\documentclass[useAMS,usenatbib]{mn2e}
\usepackage{natbib}
\usepackage{amsmath,amsfonts}
\usepackage{amssymb}
\usepackage{epsfig}
\usepackage{mathptmx}

\def\reff@jnl#1{{\rm#1\/}}

\def\aj{\reff@jnl{AJ}}                  
\def\araa{\reff@jnl{ARA\&A}}            
\def\apj{\reff@jnl{ApJ}}                        
\def\apjl{\reff@jnl{ApJ}}               
\def\apjs{\reff@jnl{ApJS}}              
\def\ao{\reff@jnl{Appl.Optics}}         
\def\apss{\reff@jnl{Ap\&SS}}            
\def\aap{\reff@jnl{A\&A}}               
\def\aapr{\reff@jnl{A\&A~Rev.}}         
\def\aaps{\reff@jnl{A\&AS}}             
\def\azh{\reff@jnl{AZh}}                        
\def\baas{\reff@jnl{BAAS}}              
\def\jrasc{\reff@jnl{JRASC}}            
\def\memras{\reff@jnl{MmRAS}}           
\def\mnras{\reff@jnl{MNRAS}}            
\def\pra{\reff@jnl{Phys. Rev. A}}         
\def\prb{\reff@jnl{Phys. Rev. B}}         
\def\prc{\reff@jnl{Phys. Rev. C}}         
\def\prd{\reff@jnl{Phys. Rev. D}}         
\def\prl{\reff@jnl{Phys. Rev. Lett}}      
\def\pasp{\reff@jnl{PASP}}              
\def\pasj{\reff@jnl{PASJ}}              
\def\qjras{\reff@jnl{QJRAS}}            
\def\rmxaa{\reff@jnl{RMxAA}}		
\def\skytel{\reff@jnl{S\&T}}            
\def\solphys{\reff@jnl{Solar~Phys.}}    
\def\sovast{\reff@jnl{Soviet~Ast.}}     
\def\ssr{\reff@jnl{Space~Sci.Rev.}}     
\def\zap{\reff@jnl{ZAp}}                        
\def\nat{\reff@jnl{Nature}}             

\def\p#1by#2{{\partial{#1} \over \partial{#2}}}
\def\pp#1by#2#3{{\partial^2{#1} \over \partial{#2}\partial{#3}}}
\def\d#1by#2{{{\rm d}{#1} \over {\rm d}{#2}}}
\def\dd#1by#2#3{{{\rm d}^2{#1} \over {\rm d}{#2}{\rm d}{#3}}}

\title[AMI YSOs with known outflows]{ AMI radio continuum observations of young stellar objects with known outflows\thanks{We request that any reference to this paper cites ``AMI Consortium: Ainsworth et~al. 2012''.}}

\label{firstpage}

\author[Ainsworth et~al.]{
 AMI Consortium:  Rachael E. Ainsworth$^1$\thanks{email: Email rainsworth@cp.dias.ie},
 Anna M. M. Scaife$^{2}$,
 Tom P. Ray$^1$,
\newauthor
 Jane V. Buckle$^{3,4}$,
 Matthew Davies$^3$,
 Thomas M. O. Franzen$^{5}$,
 Keith J. B. Grainge$^{3,4}$,
\newauthor
 Michael P. Hobson$^3$,
 Natasha Hurley-Walker$^{6}$,
 Anthony N. Lasenby$^{3,4}$,
 Malak Olamaie$^3$,
\newauthor
 Yvette C. Perrott$^3$,
 Guy G. Pooley$^3$,
 John S. Richer$^{3,4}$,
 Carmen Rodr{\'i}guez-Gonz{\'a}lvez$^7$,
\newauthor
 Richard D. E. Saunders$^{3,4}$,
 Michel P. Schammel$^3$,
 Paul F. Scott$^3$,
 Timothy Shimwell$^5$,
\newauthor
 David Titterington$^3$,
 Elizabeth Waldram$^3$.
 \vspace{0.03in}\\
$^1$ Dublin Institute for Advanced Studies, 31 Fitzwilliam Place,
     Dublin 2, Ireland\\
$^2$ School of Physics \& Astronomy, University of Southampton, Highfield, Southampton, SO17 1BJ\\
$^3$ Astrophysics Group, Cavendish Laboratory, J J Thomson Avenue,
     Cambridge, CB3 0HE\\
$^4$ Kavli Institute for Cosmology, Cambridge, Madingley Road,
     Cambridge, CB3 0HA\\
$^5$ CSIRO Astronomy \& Space Science, Australia Telescope National Facility, PO Box 76, Epping, NSW 1710, Australia\\
$^6$ International Centre for Radio Astronomy Research, Curtin Institute\\
of Radio Astronomy, 1 Turner Avenue, Technology Park, Bentley, WA 6845,
Australia \\
$^7$ Spitzer Science Center, MS 220-6, California Institute of Technology, Pasadena, CA 91125
}

\date{Accepted ---; received ---; in original form \today}

\pagerange{\pageref{firstpage}--\pageref{lastpage}}

\pubyear{2012}

\begin{document}

\maketitle

\begin{abstract}

We present 16\,GHz (1.9\,cm) deep radio continuum observations made with the Arcminute Microkelvin Imager (AMI) of a sample of low-mass young stars driving jets. We combine these new data with archival information from an extensive literature search to examine spectral energy distributions (SEDs) for each source and calculate both the radio and sub-mm spectral indices in two different scenarios: (1) fixing the dust temperature ($T_{\rm d}$) according to evolutionary class; (2) allowing $T_{\rm d}$ to vary. We use the results of this analysis to place constraints on the physical mechanisms responsible for the radio emission. From AMI data alone, as well as from model fitting to the full SED in both scenarios, we find that 80~per~cent of the objects in this sample have spectral indices consistent with free-free emission. We find an average spectral index in both $T_{\rm d}$ scenarios, consistent with free-free emission. We examine correlations of the radio luminosity with bolometric luminosity, envelope mass, and outflow force and find that these data are consistent with the strong correlation with envelope mass seen in lower luminosity samples. We examine the errors associated with determining the radio luminosity and find that the dominant source of error is the uncertainty on the opacity index, $\beta$. We examine the SEDs for variability in these young objects, and find evidence for possible radio flare events in the histories of L1551~IRS~5 and Serpens~SMM~1. 

\end{abstract}


\section{Introduction}

Young stellar objects (YSOs) are divided into a number of evolutionary classes. Class~0 protostars represent the youngest phase of protostellar evolution \citep{1993ApJ...406..122A} in which the protostar has yet to accrete most of its mass from the surrounding envelope, and therefore the central object is less massive than the envelope ($M_{\rm env}>M_{*}$). The subsequent Class~I phase \citep{1987IAUS..115....1L} occurs when a majority of the remainder of the envelope has accreted onto the protostar or its circumstellar disc ($M_{\rm env}<M_{*}$). The Class~II phase \citep{1987IAUS..115....1L}, also known as the Classical T~Tauri stage, occurs when infall is almost complete and the central object is embedded in an optically thick disc and enters the pre-main-sequence (PMS) stage of evolution. Embedded YSOs are often found to have partially ionised and collimated outflows/jets driven by the accretion from their envelopes, and are detectable from molecular gas tracers such as CO and by the shocks along the length of the outflows. The outflows driven by Class~I protostars are less powerful than those of Class~0 due to a decline in the accretion rate \citep{1996AA...311..858B}.

The target sample in this work is comprised of Class~0 and Class~I protostars. Since the mass ratio $M_{\rm env}>M_{*}$ is not directly observable, \citet{1993ApJ...406..122A, 2000prpl.conf...59A} define the following observational properties required to identify an object as Class~0. The first is the detection of a compact centimetre radio continuum source, a collimated CO outflow, or an internal heating source as indirect evidence for a central YSO. These criteria distinguish the central object from a starless prestellar core. The second requirement is centrally peaked but extended submillimeter continuum emission tracing the presence of a spheroidal circumstellar dust envelope (as opposed to just a disc). The third is a high ratio of submillimeter to bolometric luminosity, corresponding to an envelope mass greater than the stellar mass. Class~0 spectral energy distributions (SEDs) are typically well characterized by single temperature blackbodies with $15\leq T_{\rm d}\leq30$\,K and where $T_{\rm d}$ is the dust temperature. The second and third criteria differentiate Class~0 sources from the more evolved Class~I and II objects. For further distinction, bolometric temperature is also used to characterise evolutionary class, where $T_{\rm bol}<70$\,K for Class~0 objects and $70<T_{\rm bol}<650$\,K for Class~I \citep{1995ApJ...445..377C}.

The radio counterpart at centimetre wavelengths for these low-mass young stars is detected almost certainly via thermal bremsstrahlung radiation. This \textquotedblleft{free-free}\textquotedblright emission is typically observed to be compact and elongated along the outflow axis, leading to the designation of these sources as \textquotedblleft\textquoteleft{thermal radio jets}\textquotedblright\textquoteright \citep[e.g.][]{1995RMxAC...1....1R}. Free-free emission at these wavelengths typically has a flat or positive power-law spectral index $\alpha$, where the flux density $S_{\nu}\propto\nu^{\alpha}$ at frequency $\nu$. It has been shown that $\alpha=0.6$ for a standard conical jet, and $\alpha<0.6$ for an unresolved, partially opaque flow where the cross-sectional area grows more slowly than length \citep{1986ApJ...304..713R, 1998AJ....116.2953A}. However, in a number of cases non-thermal emission is also seen \citep[e.g.][]{1997Natur.385..415R, 1997ApJ...490..735C, 1998ApJ...494L.215F, 2010Sci...330.1209C}. There are some objects that have yielded spectral indices too negative to be explained by free-free emission alone, such as the Serpens~MMS~1 triple radio source \citep{1989ApJ...346L..85R, 1993ApJ...415..191C, scaife2011c}, suggesting that non-thermal processes must contribute. The radio emission can also be attributed, at least at very short wavelengths, to the discs around these young stars \citep{2008AJ....136.1852R}. 

Here we present 16\,GHz observations of a sample of classic low-mass young stars driving outflows. These sources are selected to match the target sample for the e-MERLIN (extended Multi-Element Radio Linked Interferometer Network) legacy project at 5\,GHz on the morphology and time evolution of thermal jets associated with low-mass young stars (P.I. Rodr{\'i}guez). The high spatial resolution observations provided by the e-MERLIN programme will resolve out the larger scale emission from these objects and only detect the small-scale localized emission from the central source and shocks along the length of their collimated outflows. In contrast, the data presented here will not resolve the separate components of the radio emission but will measure the integrated radio emission from these objects. This \emph{total} radio emission at 16\,GHz from YSOs has been demonstrated to show clearly defined trends with the other global properties of these systems, such as bolometric luminosity, envelope mass, and outflow force \citep{scaife2011a, scaife2011b, scaife2011c}. If such free-free emission is to be reconciled with ionization mechanisms arising from the impact of the molecular outflow or protostellar jet/wind on its surroundings then it is vital to measure it in its entirety to determine quantitatively its physical correspondence to other global properties of the protostellar system, in much the same way that radio emission from massive young stellar objects can be used to infer the total ionizing flux of the central object and hence its spectral type. Such indirect measures of protostellar activity in the early embedded phase of YSO lifetimes may provide a window into the evolution of these objects, which is otherwise obscured.

In this paper we investigate the properties of the integrated radio emission from this sample of low-mass YSOs. In general, these new data confirm trends found in previous works and improve upon the statistics. We identify outliers which may provide useful targets for further study at high spatial resolution. We combine our results for the flux densities with those found in an extensive literature search to calculate the spectral indices at radio wavelengths, and which are overwhelmingly consistent with free-free emission as the mechanism for the radio emission.

The organization of this paper is as follows. In \S2 we describe the sample of targets to be observed. In \S3 we describe the AMI telescope, the observations, and the data reduction process. In \S4 we comment on the results of the observations and \S5 contains detailed notes on the individual fields. In \S6 we discuss the detections and non-detections, the expected contamination from extragalactic sources, the spectral energy distributions, the radio spectral indices, the derivation of physical parameters, correlations between the radio luminosity and other global properties, and finally, evidence for variability. We make concluding remarks in \S7. 

\section{The Sample}

This study targets eleven YSOs driving known outflows. The coordinates of each of these objects are listed in Columns~[3] and [4] of Table~\ref{tab:srclist} along with their identifier. We follow the classification scheme described in \citet{hat2007a} for Class~0 or Class~I protostars, based on three evolutionary indicators, which for a Class~I are $T_{\rm bol}>70$\,K, $L_{\rm bol}/L_{\rm smm}>3000$, and $F_{3.6}/F_{850}>0.003$, where $F_{3.6}$ is the flux at 3.6\,cm and and $F_{850}$ is the flux at 850\,$\mu$m. There are eight Class~0 and three Class~I protostars in the target sample, listed in Table~\ref{tab:srclist}. A number of additional sources were detected within the AMI fields. We identify these additional sources assuming a generous threshold of 10\,arcsec of a known protostellar source to allow for low signal to noise, however the maximum offset is found to be $<4$\,arcsec. Additional sources are listed in Table~\ref{tab:osrclist}. When combined with the original target list, these additional sources will be refered to as the \textquotedblleft{extended sample}\textquotedblright. For the purposes of this work, all other detected sources are considered to be extra-galactic and this is discussed further in \S6. Physical properties of the sources including parent cloud association, distance, bolometric temperature, bolometric luminosity, envelope mass, and outflow force are listed in Table~\ref{tab:srcinfo}. We note that the bolometric luminosities of the target sample span an order of magnitude, and the range of envelope masses places the sample in the intermediate to high end of the range for pre-main-sequence stars.

A number of the physical properties listed in Table~\ref{tab:srcinfo} depend on the distance to the source. For example, many studies assume a distance to Perseus of 320\,pc \citep[e.g.][]{hat2007a}, but here we assume a distance of $d=250$\,pc in order to remain consistent with the \emph{Spitzer} ``Cores to Disks'' (c2d) catalogue of deeply embedded protostars \citep{2008ApJS..179..249D}. Values in the literature also differ for the distance to the Taurus and Orion molecular clouds, and the values we adopt, along with their references, are listed in Table~\ref{tab:srcinfo}. Physical quantities from the literature have been adjusted for distance where necessary. 

\begin{table*}
\begin{center}
\caption{Sample selection. Column [1] contains the source name; [2] the protostellar evolutionary class as defined by our criteria described in \S2, [3] the Right Ascension in units of hours, minutes and seconds; [4] the Declination in units of degrees, minutes, and seconds; [5] the date the observation was made, [6] AMI flux calibrator; [7] AMI phase calibrator;  [8] rms noise measured from recovered map; [9] major axis of AMI synthesized beam; [10] minor axis of AMI synthesized beam.\label{tab:srclist}}
\begin{tabular}{lccccccccc}
\hline\hline
Source & Class & RA & Dec & Date & $1^{\circ}$ & $2^{\circ}$ & $\sigma_{\rm{rms}}$ & $b_{\rm{maj}}$ & $b_{\rm{min}}$ \\
       & & (J2000) & (J2000) & (dd-mm-yyyy) & & & ($\mu$Jy\,beam$^{-1}$) & (arcsec) & (arcsec)\\
\hline
L1448~IRS~3 & 0 & 03 25 36.49 & +30 45 22.0 & 23-09-2010 & 3C48 & J0324+3410 & 24 & 41.8 & 27.4 \\
HH~7-11 & I & 03 29 03.73 & +31 16 03.8 & 24-09-2010 & 3C48 & J0324+3410 & 26 & 40.2 & 28.2 \\
L1551~IRS~5 & I & 04 31 34.15 & +18 08 04.8 & 30-09-2010 & 3C48 & J0428+1732 & 24 & 50.9 & 27.2 \\
L1527 & 0 & 04 39 53.87 & +26 03 09.9 & 25-09-2010 & 3C48 & J0440+2728 & 19 & 44.3 & 28.4 \\
HH~1-2~MMS~1 & 0 & 05 36 22.85 & -06 46 06.6 & 12-10-2010 & 3C48 & J0541-0541 & 61 & 70.5 & 24.0 \\
HH~26~IR & I & 05 46 03.90 & -00 14 52.5 & 27-08-2011 & 3C48 & J0558+0044 & 34 & 78.0 & 28.5 \\
HH~111 & 0 & 05 51 46.25 & +02 48 29.7 & 06-10-2010 & 3C48 & J0552+0313 & 31 & 68.7 & 27.4 \\
NGC~2264~G & 0 & 06 41 11.05 & +09 55 59.2 & 04-10-2010 & 3C48 & J0654+1004 & 33 & 54.5 & 26.9 \\
Serpens~MMS~1 & 0 & 18 29 49.79 & +01 15 20.8 & 11-10-2010 & 3C48 & J1824+0119 & 36 & 66.5 & 25.7 \\
L723 & 0 & 19 17 53.67 & +19 12 19.6 & 30-09-2010 & 3C48 & J1914+1636 & 23 & 48.1 & 26.3 \\
L1251~A & 0 & 22 35 24.95 & +75 17 11.4 & 25-09-2010 & 3C147 & J2236+7322 & 25 & 37.6 & 27.8 \\
\hline
\end{tabular}
\vskip .05in
\begin{minipage}{0.9\textwidth} 
{\footnotesize }
\end{minipage}
\end{center}
\end{table*}

\begin{table*}
\begin{center}
\caption{Additional detected sources. Column [1] contains the additional source number, [2] the field the source is located in, [3] - [4] the Right Ascension and Declination of the peak flux value, [5] the AMI 16\,GHz combined channel flux density, [6] the spectral index $\alpha_{\rm AMI}$ calculated from the AMI fluxes, [6] the associated source, and [7] the protostellar evolutionary class where applicable or object type.\label{tab:osrclist}}
\begin{tabular}{lcccccccc}
\hline\hline
AMI & Field & RA & Dec & $S_{16{\rm GHz}}$ & $\alpha_{\rm AMI}$ & Associated & Class\\
       Source & & (J2000) & (J2000) & (mJy) & & Source & \\
\hline
1 & L1448 & 03 25 38.9 & +30 44 02 & $0.76\pm0.09$ & $-0.67\pm0.98$ & L1448~C & 0\\
2 & HH~7-11 & 03 28 55.5 & +31 14 35 & $0.39\pm0.10$ & $1.84\pm0.16$ & NGC~1333~IRAS~2A & 0\\
3 & HH~7-11 & 03 28 57.3 & +31 14 16 & $0.50\pm0.10$ & $1.65\pm0.34$ & NGC~1333~IRAS~2B & I\\
4 & L1551 & 04 31 44.3 & +18 08 29 & $0.74\pm0.09$ & $0.61\pm0.75$ & L1551~NE & 0\\
5 & L1527 & 04 39 49.4 & +26 05 25 & $0.96\pm0.09$ & $-0.10\pm0.92$ & HH~192~VLA~2 & ext?\\
6 & L1527 & 04 39 59.8 & +26 02 06 & $0.88\pm0.09$ & $-1.25\pm0.58$ & HH~192~VLA~3 & ext?\\
7 & HH~1-2 & 05 36 25.7 & -06 47 16 & $1.00\pm0.26$ & $0.10\pm1.10$ & HH~2 & HH\\
8 & HH~26~IR & 05 46 07.4 & -00 13 42 & $1.23\pm0.10$ & $1.82\pm0.13$ & HH~25~MMS & 0 \\
9 & HH~111 & 05 51 45.5 & +02 50 06 & $0.70\pm0.15$ & $-0.23\pm1.10$ & - & ext?\\
10 & L1251 & 22 35 20.5 & +75 19 44 & $0.71\pm0.09$ & $-1.20\pm0.68$ & - & ext?\\
\hline
\end{tabular}
\vskip .05in
\begin{minipage}{0.9\textwidth} 
{\footnotesize }
\end{minipage}
\end{center}
\end{table*}

\begin{table*}
\begin{center}
\caption{Source information. Column [1] contains the source name, [2] the molecular cloud association, [3] the distance to the source, [4] distance reference, [5] bolometric temperature, [6] bolometric luminosity, [7] reference for $T_{\rm{bol}}$ and $L_{\rm{bol}}$, [8] envelope mass calculated using the 850\,$\mu$m flux from the Fundamental Map Object Catalogue (Di~Francesco et~al. 2008), [9] outflow force of the source, and [10] reference for the outflow force of the source. \label{tab:srcinfo}}
\begin{tabular}{lccccccccc}
\hline\hline
Source & Cloud & $d$ & $d$  & $T_{\rm{bol}}$ & $L_{\rm{bol}}$ & Ref. & $M_{\rm{env}}$ & $F_{\rm{out}}$ & $F_{\rm{out}}$\\
       & Association & (pc) & Ref. & (K) & (L$_{\odot}$) & & (M$_{\odot}$) & (M$_{\odot}^{-5}$ km s$^{-1}$ yr$^{-1}$) & Ref.\\
\hline
L1448~IRS~3 & Perseus & 250 & 1 & 53 & 13.7 & 6 & 16.22 & $49.9\rm{x}10^{-2}\pm5.29$ & 12\\
HH~7-11 & Perseus & 250 & 1 & 180 & 18.0 & 6 & 10.75 & $1.53\pm18.0$ & 12\\
L1551~IRS~5 & Taurus & 140 & 2 & 75 & 15.3 & 7 & 6.22 & 9.63 & 13\\
L1527 & Taurus & 140 & 2 & 59 & 2.0 & 8 & 5.89 & $0.79\pm0.13$ & 17\\
HH~1-2~MMS~1 & Orion & 420 & 3 & - & 23 & 9 & 42.46 & - & -\\
HH~26~IR & Orion & 420 & 3 & - & 25.1 & 10 & 4.68 & 36.1 & 13\\
HH~111 & Orion & 420 & 3 & 49 & 19.2 & 11 & 13.90 & 4.57 & 13\\
NGC~2264~G & Mon OB1 & 800 & 4 & 25 & 12 & 4 & 11.59 & 300 & 13\\
Serpens~MMS~1 & Serpens & 260 & 14 & 56 & $17.3^{a}$ & 15 & 28.17 & 2.8\ & 16\\
L723 & - & 300 & 4 & 50 & 3 & 4 & 5.14 & 26 & 13\\
L1251~A & Cepheus Flare & 320 & 5 & - & 2.24 & 5 & 14.83 & 8.6 & 13\\
\hline
L1448~C & Perseus & 250 & 1 & 49 & 3.9 & 6 & 5.1 & $16.3\rm{x}10^{-2}\pm1.9$ & 12\\
NGC~1333~IRAS~2A & Perseus & 250 & 1 & 58 & 13.9 & 6 & $2.81^{b}$ & $56.8\rm{x}10^{-2}\pm5.6$ & 12\\
NGC~1333~IRAS~2B & Perseus & 250 & 1 & 106 & 5.3 & 15 & $0.74^{b}$ & - & -\\
L1551~NE & Taurus & 140 & 2 & 56 & 4.2 & 11 & 5.6 & - & -\\
HH~25~MMS & Orion & 420 & 3 & 34 & 5.2 & 4 & 43.96 & - & - \\
\hline
\end{tabular}
\vskip .05in
\begin{minipage}{0.9\textwidth} 
{\footnotesize References: (1) \citealt{2008ApJS..179..249D}; (2) \citealt{1994AJ....108.1872K}; (3) \citealt{2007AA...474..515M}; (4) \citealt{2000prpl.conf...59A}; (5) \citealt{2009ApJS..185..451K}; (6) \citealt{hat2007a}; (7) \citealt{2009AA...507.1425V}; (8) \citealt{2009AA...507..861J}; (9) \citealt{2010AA...518L.122F}; (10) \citealt{2002ApJ...574..246N}; (11) \citealt{2005ApJS..156..169F}; (12) \citealt{hat2007b}; (13) \citealt{1995RMxAC...1...67A}; (14) \citealt{1996BaltA...5..125S}; (15) \citealt{2009ApJ...692..973E}; (16) \citealt{2010MNRAS.409.1412G}; (17) \citealt{1996AA...311..858B}.\\
$^{a}$ \citealt{2009ApJ...707..103E}\\
$^{b}$ \citealt{2009ApJ...692..973E}}
\end{minipage}
\end{center}
\end{table*}

\section{OBSERVATIONS}

The data in this work were taken with the Arcminute Microkelvin Imager (AMI) Large Array at the Mullard Radio Astronomy Observatory near Cambridge, UK \citep{2008MNRAS.391.1545Z}. AMI is comprised of eight 13\,m dishes, operating between 13.5 and 17.9\,GHz with eight 0.75\,GHz bandwidth channels. The channels 1-3 were not used due to a poorer response in this frequency range and interference due to geostationary satellites. The FWHM of the primary beam of the AMI is $\approx6$\,arcmin at 16\,GHz.

AMI data reduction is performed using the local software tool \textsc{reduce}. This applies both automatic and manual flags for interference, shadowing and hardware errors. \textsc{reduce} Fourier transforms the lag correlator data to synthesize frequency channels and performs phase and amplitude calibration before output to disc in \emph{uv} FITS format. Flux (primary) calibration is performed using 3C48 for all fields except for L1251 where 3C147 was used, and assumed flux densities can be found in Table~\ref{tab:cals}. Phase (secondary) calibration is carried out using the bright point sources listed in Table 1. Absolute calibration is typically accurate to 5~per~cent but we note that an increased absolute calibration error of 10~per~cent is applied to flux densities from channel 8 in order to reflect the poorer phase stability of this channel relative to the others.

\begin{table}
\begin{center}
\caption{AMI frequency channels and primary calibrator flux densities measured in Jy.\label{tab:cals}}
\begin{tabular}{lcccccc}
\hline\hline
Channel No. & 4 & 5 & 6 & 7 & 8 \\
Freq. [GHz] & 14.62 & 15.37 & 16.12 & 16.87 & 17.62 \\
\hline
3C48 & 1.75 & 1.66 & 1.58 & 1.50 & 1.43 \\
3C147 & 2.62 & 2.50 & 2.40 & 2.30 & 2.20 \\
\hline\hline
\end{tabular}
\vskip .05in
\begin{minipage}{0.9\textwidth} 
{\footnotesize }
\end{minipage}
\end{center}
\end{table}

Reduced data were imaged in \textsc{CASA}\footnotemark[1]\footnotetext[1]{http://casa.nrao.edu}. Multiple data sets were concatenated, additional baseline flagging was carried out, and the \textsc{clean} task was used for deconvolution to produce the combined frequency image as well as the separate spectral channel images. Primary beam correction and fitting Gaussian models to radio sources in order to extract flux densities were performed in \textsc{AIPS}\footnotemark[2]\footnotetext[2]{http://aips.nrao.edu}. All errors quoted are 1 $\sigma$. 

\section{Results}

Maps were made using naturally-weighted visibilities to ensure optimal signal-to-noise levels, except in the case of HH~1-2 in which uniform weighting was used to improve resolution. As HH~1-2 is the lowest-declination source, the beam would otherwise have been very extended. The dimensions of the synthesized beam for each source are listed in Table~\ref{tab:srclist}. Source detection for these data was performed in the un-primary-beam-corrected maps, where we identified all objects within the FWHM of the AMI primary beam with a peak flux density $>5\sigma_{\rm rms}$, where $\sigma_{\rm rms}$ is the rms noise determined from the maps. $\sigma_{\rm rms}$ for each target field is listed in Table~\ref{tab:srclist}. Integrated flux densities for detected sources were then extracted from the primary-beam-corrected maps using \textsc{imfit} and are listed in Table~\ref{tab:intfluxes}. The errors on the flux densities are calculated as $\sigma = \sqrt{(0.05S_{\nu})^{2} + {\sigma}^{2}_{\rm fit} + {\sigma}^{2}_{\rm rms}}$, where $\sigma_{\rm fit}$ is the fitting error returned from \textsc{imfit} and $0.05S_{\nu}$ is a conservative 5~per~cent absolute calibration error (except for channel~8 where we use a 10~per~cent absolute calibration error, $0.1S_{\nu}$). Including $\sigma_{\rm fit}$ provides an over-estimation to the total error. 

Extracted flux densities are listed in Table~\ref{tab:intfluxes}, and column [8] of Table~\ref{tab:intfluxes} lists the spectral indices $\alpha_{\rm AMI}$ calculated over the AMI channels. The uncertainties of these values are large due to the short frequency coverage of AMI. Spectral indices are discussed further in \S6. In two cases, HH~26~IR and NGC~2264, the flux density was extracted only from the 16\,GHz combined-channel image. For these objects, the flux density in the combined-channel image is $<5\sqrt{n}\sigma$, where $n$ is the number of combined channels, indicating that the sources would be detected at $<5\sigma$ in the individual channel maps. Combined-channel maps for each source may be found in Appendix~A with the exception of L1448~IRS~3 which is shown in Figure~\ref{fig:L1448}(a) for illustration purposes.

\begin{table*}
\begin{center}
\caption{Integrated flux densities. Column [1] contains the source name, [2] the combined-channel 16\,GHz flux density for that source, [3]-[7] contain the AMI integrated flux densities for each source from AMI channels 4--8, and [8] contains the spectral index $\alpha_{\rm AMI}$ calculated from the AMI fluxes.\label{tab:intfluxes}}
\begin{tabular}{lccccccc}
\hline\hline
&& \multicolumn{5}{c}{AMI Channel Number} \\
\cline{3-7}
Source & $S_{\rm 16\,GHz}$ & 4 & 5 & 6 & 7 & 8 & $\alpha_{\rm AMI}$ \\
       &          & [14.62\,GHz] & [15.37\,GHz] & [16.12\,GHz] & [16.87\,GHz] & [17.62\,GHz] \\
       & (mJy)& (mJy) & (mJy)& (mJy)& (mJy)& (mJy)\\
\hline
L1448~IRS~3 & $2.39\pm0.16$ & $2.40\pm0.25$ & $2.34\pm0.26$ & $2.41\pm0.37$ & $2.26\pm0.24$ & $2.62\pm0.37$ & $0.20\pm0.73$ \\
HH~7-11 & $3.63\pm0.24$ & $3.43\pm0.20$ & $3.10\pm0.18$ & $3.91\pm0.22$ & $3.43\pm0.19$ & $3.83\pm0.41$ & $0.70\pm0.41$ \\
L1551~IRS~5 & $4.97\pm0.27$ & $4.74\pm0.25$ & $4.37\pm0.24$ & $4.79\pm0.28$ & $4.82\pm0.29$ & $5.48\pm0.57$ & $0.79\pm0.40$ \\
L1527 & $1.33\pm0.10$ & $1.04\pm0.11$ & $1.04\pm0.14$ & $1.12\pm0.12$ & $1.20\pm0.16$ & $1.38\pm0.19$ & $1.15\pm0.57$ \\
HH~1-2~MMS~1 & $1.70\pm0.27$ & $1.30\pm0.28$ & $1.53\pm0.30$ & $1.34\pm0.18$ & $1.33\pm0.14$ & - & $-0.17\pm1.02$ \\
HH~26~IR & $0.39\pm0.08$ & - & - & - & - & - & - \\
HH~111 & $2.49\pm0.20$ & $3.11\pm0.53$ & $2.68\pm0.28$ & $2.39\pm0.26$ & $2.86\pm0.30$ & $2.51\pm0.40$ & $-0.40\pm0.85$ \\
NGC~2264~G & $0.30\pm0.22$ & - & - & - & - & - & - \\
Serpens~MMS~1 & $7.32\pm0.41$ & $8.21\pm0.77$ & $7.66\pm0.55$ & $7.04\pm0.43$ & $6.77\pm0.44$ & $7.30\pm0.83$ & $-0.73\pm0.56$ \\
L723 & $0.58\pm0.09$ & $0.58\pm0.09$ & $0.52\pm0.06$ & $0.65\pm0.08$ & $0.66\pm0.07$ & $0.52\pm0.11$ & $0.42\pm0.82$ \\
L1251~A & $1.05\pm0.12$ & $0.86\pm0.18$ & $0.98\pm0.14$ & $1.00\pm0.20$ & $0.76\pm0.16$ & $1.10\pm0.25$ & $0.06\pm0.99$ \\
\hline\hline
\end{tabular}
\vskip .05in
\begin{minipage}{0.9\textwidth} 
{\footnotesize }
\end{minipage}
\end{center}
\end{table*}

\section{Notes on Individual Fields}

\begin{figure}
\centerline{\includegraphics[width=0.4\textwidth]{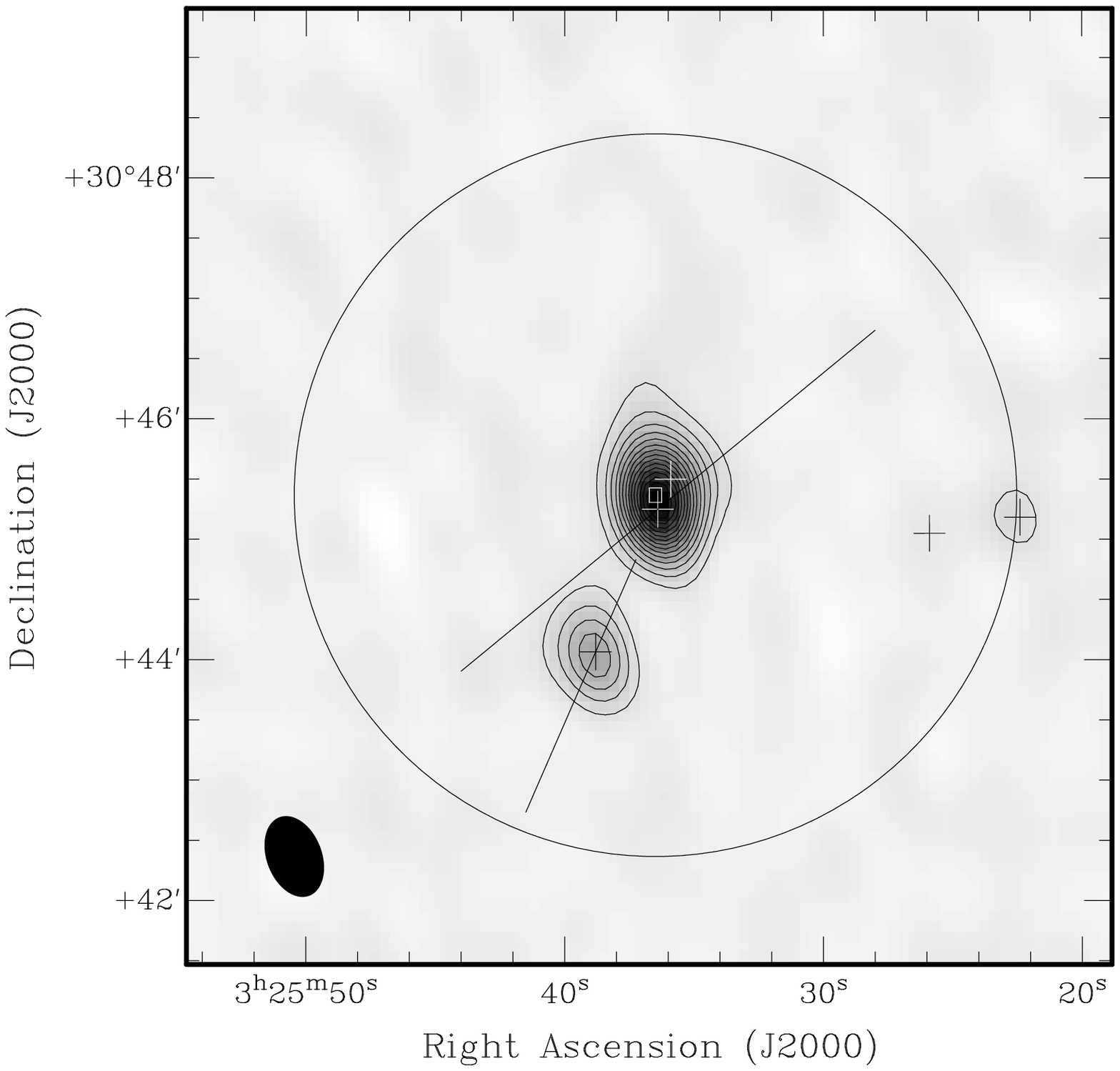}}
\centerline{(a)}
\centerline{\includegraphics[width=0.4\textwidth]{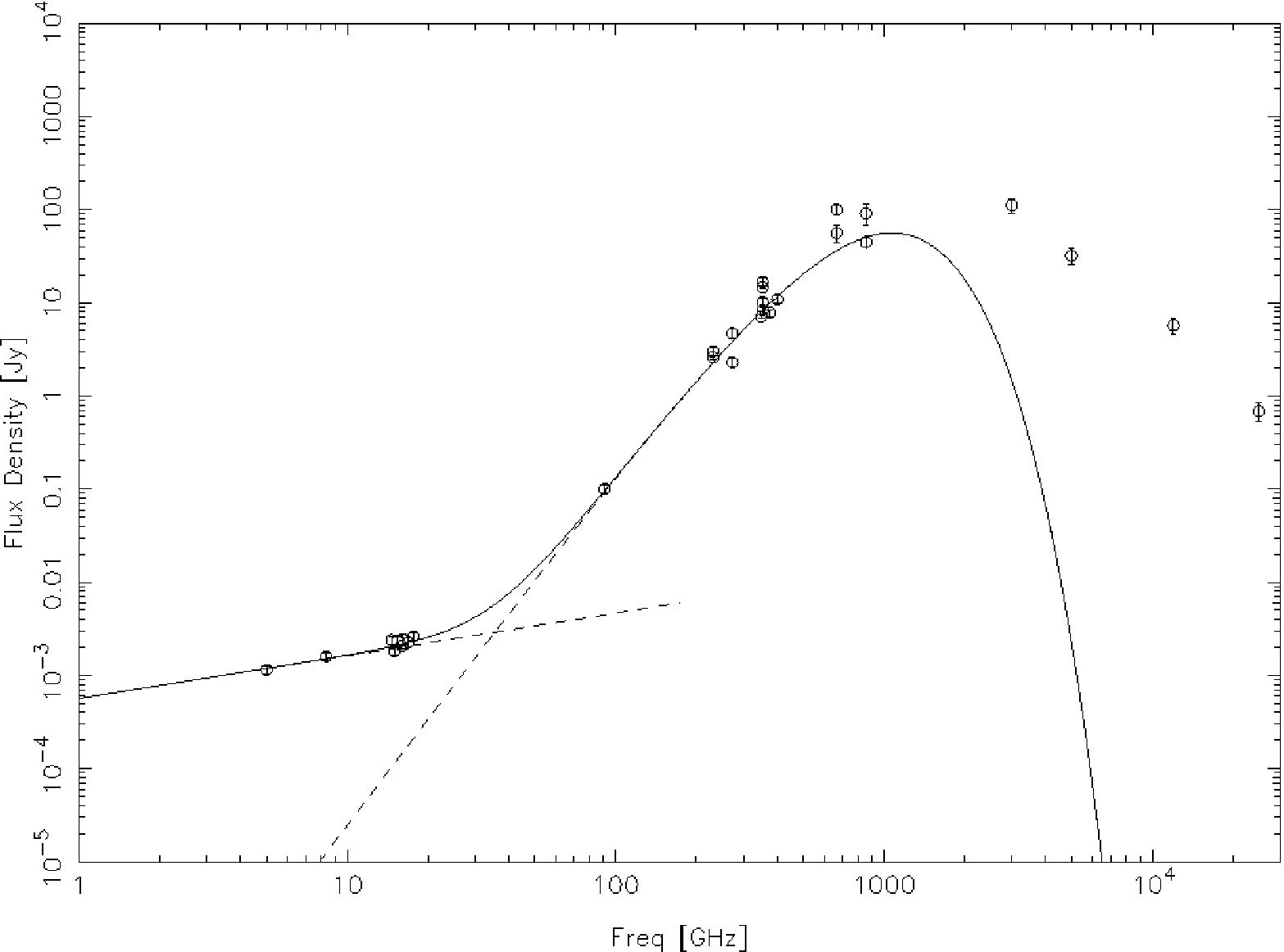}}
\centerline{(b)} 
\caption{L1448. (a) The AMI 16\,GHz combined channel map uncorrected for the primary beam response for L1448. The positions of known Class~0 and Class~I objects are indicated as crosses (+) and stars (*) respectively, and are from \citet{hat2007a}. The solid black lines indicate outflow axes for L1448~NB (northern axis) and L1448~C (southern axis) from \citet{2000AJ....120.1467W}, details in \S5. We plot the half power point of the primary beam as a solid circle ($\approx6$\,arcmin at 16\,GHz) and the FWHM of the PSF as the filled ellipse in the bottom left corner (see Table~\ref{tab:srclist}). Contours at 5, 10, 15, 20\,$\sigma_{\rm rms}$ etc, where $\sigma_{\rm rms}=24\mu$Jy\,beam$^{-1}$. (b) The SED for L1448~IRS~3 with the observed radio spectra over the AMI frequency channels 3-8 combined with flux densities from the literature.}\label{fig:L1448}
\end{figure}

\noindent{\bf L1448.} L1448~IRS~3 is made up of the L1448 North group of Class~0 protostars, NA, NB, and NW \citep{1990ApJ...365L..85C, 1997IAUS..182..507T}. AMI does not resolve the separate sources, so the data at 1.9\,cm (see Figure~\ref{fig:L1448}(a)) represents the integrated radio emission from the three protostars and appears point-like. The spectral energy distribution of this source is shown in Figure~\ref{fig:L1448}(b), and includes the observed radio spectra over the AMI frequency channels 3-8 combined with flux densities from the literature. Spectral energy distributions and spectral indices are discussed in \S6. 

L1448~NA and NB are in a $7''$ separation protobinary system \citep{1990ApJ...365L..85C} and L1448~NW lies approximately $20''$ northwest of L1448~NA \citep{1997IAUS..182..507T} in the Perseus molecular cloud. The outflows in this region have been discussed extensively by \citet{2000AJ....120.1467W}. Only redshifted gas associated with the outflow of L1448~NA is detected in their CO maps, with a PA (measured from north through east) of $\approx150^{\circ}$ and a total length of at least 0.7\,pc (at their assumed distance of 300\,pc). The outflow associated with L1448~NB has a PA $\approx129^{\circ}$ and $6'$ to the northwest along the outflow axis lie the blueshifted optical emission knots of HH~196 \citep{1997ApJ...478..603B}, the directions of these outflows are illustrated in Figure~\ref{fig:L1448}(a).

We also detect L1448~C in our map, which is the driving source of the high velocity outflow in L1448 \citep{1990AA...231..174B}. The blueshifted CO outflow lobe driven by L1448~C begins with a PA$=159^{\circ}$ \citep{1995AA...299..857B} (see Figure~\ref{fig:L1448}(a)) before being deflected through a total angle of $\approx32^{\circ}$ due to a possible collision with the core containing L1448~NA and NB \citep{1995ApJ...443L..41D, 1999ApJ...527..310C}. 

\noindent{\bf HH 7-11.} The AMI 1.9\,cm data presented here (see Appendix~A) shows emission extended in a direction almost perpendicular to the direction of the HH~7-11 outflow, but consistent with the distribution of VLA objects of \citet{1999ApJS..125..427R}, almost all of which are associated with YSOs.

HH~7-11~SVS~13 is a designated Class~I object in NGC~1333 within the Perseus molecular cloud, and is generally believed to be the driving source of the chain of HH~7-11 knots to the southeast. However, using high-resolution VLA data \citet{1997ApJ...480L.125R} propose that VLA~3, detected $\sim6''$ southwest of SVS~13 (VLA~4), is instead the driving source of the HH~7-11 outflow. VLA~3 is better aligned with the HH~7-11 knots and the elongation of the source is approximately along the axis of the HH and molecular flows. \citet{1997ApJ...480L.125R} suggest that SVS~13 is the driving source of a much smaller, nearby chain of HH objects to the east.    

We also detect NGC~1333~IRAS~2A and 2B in our map, which were first detected by \citet{1987MNRAS.226..461J} as IRAS~2, and subsequently by \citet{1994AA...285L...1S} who detected a quadrupolar outflow centered on the source suggesting that it might be a binary. The two sources, IRAS~2A and 2B, were detected at 3.6 and 6\,cm by \citet{1999ApJS..125..427R} (VLA~7 and 10 respectively) who suggest that the quadrupolar outflow may originate from IRAS~2A and which therefore itself may be a binary.

\noindent{\bf L1551.} L1551~IRS~5 is a deeply embedded multiple protostellar Class~I source in Taurus that drives a binary jet \citep{2003ApJ...583..330R}. At 1.9\,cm (see Appendix~A) this source appears point-like and it has been suggested that the emission for this source near 1\,cm arises from nearly equal contributions of dust emission from the discs and free-free emission from the ionised ejecta \citep{1998RMxAC...7...14R}.

\citet{2003ApJ...586L.137R} have shown that the emission at centimetre wavelengths is produced by a pair of closely aligned (within $\approx12^{\circ}$) bipolar ionised jets, which themselves are aligned with the larger scale bipolar molecular outflow. \citet{2006ApJ...653..425L} determined that L1551~IRS~5 is a triple protostellar system by identifying two main (N and S) components, separated by $\simeq0.3''$, each comprised of a circumstellar dust disc and bipolar ionised jet, and a third component lying $\approx0.09''$ southeast of the N component as another circumstellar dust disc.

We detect, but do not resolve, the Class~0 source L1551~NE \citep{1995ApJ...445L..55M} to the east of IRS~5 at 1.9\,cm. L1551~NE has a molecular outflow and is likely to drive the HH objects 28, 29, and 454 \citep{1999AJ....118..972D}. Radio continuum observations at 3.5\,cm identified L1551~NE as a binary source with separation $0.5''$ \citep{2002AJ....124.1045R}.

\noindent{\bf L1527.} We detect L1527 (IRAS~04368+2557) at 1.9\,cm as a point-source (see Appendix~A). This object is an embedded Class~0 \citep{2011ApJ...739L...7M} protostar in the Taurus molecular cloud  with a compact accretion disc and a $\approx0.2''$ binary companion \citep{2002ApJ...581L.109L}. The HH~192 outflow associated with this object has an east-west orientation \citep{1997AJ....114.1138G} with an edge-on disc perpendicular to the outflow axis \citep{2010ApJ...722L..12T} that is infalling onto the central source \citep{1997ApJ...475..211O}. \citet{scaife2011d} found an opacity index $\beta=0.32\pm0.04$ for this source, consistent with the presence of large dust grains, or \textit{pebbles}, in the disc. 

We also detect HH~192 VLA~2 and 3 \citep{1998RMxAA..34...13R} at 1.9\,cm; there are no other references to these objects in the literature. We find a spectral index of $\alpha_{\rm AMI}=-0.10$ for VLA~2 consistent with optically thin free-free emission and $\alpha_{\rm AMI}=-1.25$ for VLA~3 which is inconsistent with free-free emission and suggests that it could be extragalactic. For the purposes of this work, VLA~2 and 3 are considered to be extragalactic sources (see \S6) although the emission from VLA~2 could be consistent with an HII region.

\noindent{\bf HH 1-2.} At 1.9\,cm, we detect the HH~1-2 outflow driving source as a point source (see Appendix~A). HH~1 and 2 are considered to be the prototypes of the HH phenomena, displaying two bright bow shocks which expand as a part of a bipolar outflow moving away from the central source that drives a finely collimated jet \citep{2000AJ....119..882R}. The Class~0 driving source of HH~1 and 2, VLA~1, was detected as a centimetre continuum radio source by \citet{1985ApJ...293L..35P}, and subsequent observations have shown that the source powers a small thermal radio jet along the axis of the HH flow \citep{1990ApJ...352..645R}. Another source, VLA~2, was found $3''$ from VLA~1, producing the HH~144 flow which is at a large angle to the HH~1 flow axis. The AMI does not resolve the two sources VLA~1 and 2 and the combined thermal emission is approximately point-like. 

The Herbig-Haro objects HH~1 and HH~2 themselves have been previously detected at centimetre wavelengths \citep[e.g.][]{1990ApJ...352..645R, 2000AJ....119..882R} and we detect HH~2 but we do not detect HH~1 at 1.9\,cm. \citet{1990ApJ...352..645R} find a flux density of 0.78\,mJy at 2\,cm for HH~1, however we constrain an upper flux density limit of 0.34\,mJy, suggesting that this source is variable. Radio continuum emission from other HH objects has been found to be time variable \citep{2008AJ....135.2370R, 2010AJ....139.2433C}. \citet{1990ApJ...352..645R} find flux densities of 1.75\,mJy for the central source and 0.98\,mJy for HH~2, for which our data is consistent.  

\noindent{\bf HH 26 IR.} HH~26~IR is a weak source at 1.9\,cm (see Appendix~A), a Class~I protostar \citep{1997AA...324..263D} located within L1630 in Orion that drives a molecular outflow. 

L1630 is rich in star formation activity and has a complex morphology. Two embedded Class 0 protostars, HH~24MMS and HH~25MMS, are also in this star formation region and both have compact jets, while the outflow of HH~26~IR drives a more extended outflow \citep{1993AA...276..511G}. H$_{2}$ knots HH~26A, B, C, and D exist along the axis of the HH~26~IR outflow which is approximately orthogonal to the HH~25 outflow \citep{1997AA...324..263D}. We also detect HH~25~MMS within the AMI primary beam to the northeast of HH~26~IR. 

\noindent{\bf HH 111.} At 1.9\,cm, HH~111 (see Appendix~A) is significantly extended along an approximate north-south direction, perpendicular to the known outflow direction. The driving source is IRAS~05491+0247, a suspected Class~I binary source \citep{1997ApJ...475..683Y} that drives a bipolar molecular outflow. The outflow was first discovered in the optical by \citet{1989Natur.340...42R}. However, based on the classification criterion defined in \S2, this source has a bolometric temperature consistent with a Class~0 object and this is discussed further in \S6. The jet is well-collimated, aligned approximately in the E-W direction, and has knots with velocities ranging from 300 to 600\,km~s$^{-1}$ for different parts of the jet complex \citep{1992ApJ...392..145R}.  It was found that the optical jet is part of a giant HH complex extending over 7.7\,pc \citep{1997AJ....114.2708R}.

Near-infrared observations \citep{1994AA...289L..19G} revealed a second bipolar flow, HH~121, emerging from around the same position as the optical outflow and is aligned approximately in the N-S direction, consistent with the extension observed here. Such a quadrupolar outflow is indicative of a binary companion. 3.6\,cm \citep{1999AA...352L..83R} and 7\,mm \citep{2008AJ....136.1852R} VLA observations suggest a common origin within $\approx0.1''$. It has been suggested that the most viable interpretations for the structure are that the observations are of two orthogonal discs around separate protostars or a disc with a perpendicular jet \citep{2008AJ....136.1852R}.

\noindent{\bf NGC 2264.} We detect emission from NGC~2264~G at $9\sigma_{\rm rms}$ in the AMI combined-channel map (see Appendix~A). The NGC~2264~G molecular outflow was one of nine discovered in a CO survey of the Monoceros~OB1 cloud conducted by \citet{1986ApJ...309L..87M} and was found to have very high velocity emission and a well-collimated, bipolar morphology \citep{1988ApJ...333..316M}. It is centered on IRAS~06384+0958 and extends $\approx$1.6\,pc in an approximately E-W direction \citep{2006AA...449.1077C}. \citet{1996ApJ...459..638L} show that the outflow is S-shaped and symmetric with respect to the driving source, VLA~2, which was detected by \citet{1994ApJ...436..749G} using VLA ammonia observations. They obtained a spectral index for the outflow of $\alpha~\simeq~0.3\pm0.2$ which is consistent with optically thin free-free emission of a thermal jet, and their 3.6\,cm VLA images show VLA~2 to be elongated in the direction of the outflow. Subsequent observations of VLA~2 by \citet{1995MNRAS.273L..25W} using the James Clerk Maxwell Telescope (JCMT) obtained a ratio of submillimeter and bolometric luminosity indicative of a Class~0 protostar.

\noindent{\bf Serpens.} At 1.9\,cm, AMI does not resolve the triple radio source SMM~1 but measures the integrated radio emission. The emission is oriented in an approximate N-S direction (see Appendix~A), despite the outflow orientation of NW-SE which could be due to the presence of multiple objects in this region. The Serpens molecular cloud is a nearby star formation region containing clusters of low-mass protostars as well as several HH objects and outflows. The most luminous and most deeply embedded object in the Serpens cloud core is the triple radio source SMM~1, comprised of a central source, a NW component, and a SE component, and is a known Class~0 source associated with a highly collimated radio continuum jet \citep{1989ApJ...346L..85R}. 

\citet{1993ApJ...415..191C} presented high-sensitivity, multifrequency VLA radio continuum observations of the triple radio source, showing a one-sided radio jet morphology, but did not detect a counter jet. These authors show that the central source is extended at 3.6\,cm, but the difference between the direction of the major axis and the orientation of the radio jet appear to be significant, suggesting that the central source is precessing or nutating. They find that the central source has a spectral index $\alpha~\simeq~0.15\pm0.09$, consistent with the value of $\alpha\simeq0.1\pm0.1$ obtained by \citet{1989ApJ...346L..85R}, while the NW and SE components have spectral indicies $\alpha\simeq-0.05\pm0.05$ and $\alpha\simeq-0.30\pm0.04$ respectively. These results suggest a thermal origin for the emission associated with the central source and the NW component, and a borderline thermal/non-thermal origin for the SE component. \citet{scaife2011c} present evidence for radio variability in SMM~1, as previously suggested by \citet{2009ApJ...707..103E} who detect a high mass disc around the source. Here we adopt a bolometric luminosity of $L_{\rm bol}=17.3$\,L$_{\odot}$ \citep{2009ApJ...707..103E}, however we note that significantly higher luminosities have been reported for this object \citep[e.g. 46\,L$_{\odot}$ from][]{1996ApJ...460L..45H}. 

\noindent{\bf L723.} L723, also known as IRAS~19156+1906, appears point-like at 1.9\,cm (see Appendix~A). L723 is a designated low-mass, Class~0 object \citep{2002ApJ...576..294L}. The associated outflow consists of a pair of bipolar lobes aligned along the east-west direction and another pair of bipolar lobes aligned roughly in the north-south direction, with IRAS~19156+1906 located at their common centre \citep{2002ApJ...576..294L}. VLA observations at 3.6\,cm reveal two sources, VLA~1 and VLA~2, toward the centre of the outflow \citep{1991ApJ...376..615A}, although VLA~1 is likely a background source \citep{1996ApJ...473L.123A, 1997ApJ...489..734G}. VLA~2 has a jet-like morphology and partially optically thick free-free emission, characteristic of a thermal radio jet, and was first identified as the powering source of the EW pair of molecular lobes by \citet{1996ApJ...473L.123A}. Recent VLA observations by \citet{2008ApJ...676.1073C} at 3.6\,cm and 7\,mm resolve VLA~2 into several components and the two brightest sources at 3.6\,cm, VLA~2A and 2B, are separated by $\approx0.29''$.

\noindent{\bf L1251.} The emission at 1.9\,cm of L1251~A is detected as a point-source (see Appendix~A). L1251~A is a star-forming region toward Cepheus, and is centered on the bright source IRAS~22343+7501 which is associated with a large-scale ($\sim10'$) molecular outflow \citep{1989ApJ...343..773S, 1988ApJ...327..350S}. VLA observations \citep{1998AJ....115.1599M, 2001AJ....121.1556B, 2004AJ....127.1736R} show several radio sources (VLA~6, 7, and 10) within this region that could possibly drive the molecular outflow. \citet{2004AJ....127.1736R} show two sources with clear extension, VLA~7 and 10, and suggest that VLA~10 could drive the molecular outflow found by \citet{1989ApJ...343..773S}, which extends for $~10'$ in the approximate NE-SW direction. However, VLA~7 is consistent with the outflow found by \citet{1988ApJ...327..350S} that has an outflow direction in an almost perpendicular direction, and the emission we detect is consistent with this. We do not resolve these sources, but measure the integrated radio emission.

\section{Discussion}

\subsection{Detections and non-detections}

We detect 100~per~cent of the objects in the target sample with AMI at 16\,GHz, which includes eight Class~0 and three Class~I protostars based on the classification scheme described in \S2. We also detect five additional Class~0 and one additional Class~I sources within the AMI primary beam of these fields, all of which are Class~0 sources driving outflows except for NGC~1333~IRAS~2B which is Class~I. 

In the L1448 field there is one Class~0 source \citep[31;][]{hat2007a} that we do not detect, in the HH~7-11 field there is one Class~0 \citep[62;][]{hat2007a} and one Class~I \citep[50;][]{hat2007a} that we do not detect, and in the Serpens field there is one Class~0 \citep[VLA~5;][]{scaife2011c} and two Class~I \citep[VLA 9, 16;][]{scaife2011c} sources that we do not detect. 

We identify AMI~7 as the radio counterpart to HH~2, but we do not detect HH~1 at 1.9\,cm even though it has been previously detected at centimetre wavelengths. We also do not detect any HH~192 objects in the L1527 field which lie within the AMI primary beam. 

Detection statistics for the original and extended samples divided by evolutionary class are listed in Table~\ref{tab:detect}. It is not surprising that we should detect 100~per~cent of the target sample as all sources have been previously detected at centimetre wavelengths and have known radio jets. Detection rates for Class~0 in the extended sample is typical compared to \citet{scaife2011b, scaife2011c} but high for Class~I. This is likely due to the small sample size or the fact that a majority of the extended sample consists of the original sample which was chosen on the basis that these objects have been detected at these wavelengths before. The original sample is biased, and for these reasons it is difficult to draw strong statistical conclusions from this sample.

\begin{table}
\begin{center}
\caption{Summary of detection statistics. The original sample includes only objects listed in Table~\ref{tab:srclist}. The extended sample includes all objects listed in Tables~\ref{tab:srclist} and \ref{tab:osrclist}, inclusive of the original sample. \label{tab:detect}}
\begin{tabular}{lccc}
\hline\hline
Class & Present & Detected & ~per~cent \\
\hline
\textit{Original Sample}: \\
Class~0 & 8 & 8 & 100 \\
Class~I & 3 & 3 & 100 \\
\hline
\textit{Extended Sample}: \\
Class~0 & 15 & 12 & 80 \\
Class~I & 7 & 4 & 57 \\
\hline
\end{tabular}
\vskip .05in
\begin{minipage}{0.9\textwidth} 
{\footnotesize }
\end{minipage}
\end{center}
\end{table}

\subsection{Expected contamination by extragalactic radio sources}

At 16\,GHz we expect a certain number of extragalactic radio sources to be seen within each of our fields. Following \citet{scaife2011a, scaife2011b, scaife2011c}, to quantify this number we use the 15\,GHz source counts model from \citet{2005AA...431..893D} scaled to the 10C survey source counts \citep{2010arXiv1012.3659D}. The average rms noise from our datasets is $\backsimeq30$\,$\mu$Jy~beam$^{-1}$ and from this model we predict that we should see 0.043 sources\,arcmin$^{-2}$, or $\backsimeq2$ radio sources within a 6\,arcmin FWHM primary beam above a 5$\sigma_{\rm rms}$ flux density of 150\,$\mu$Jy. Accounting for the radial attenuation of the primary beam, we would therefore expect to see $\approx10\pm3$ extragalactic radio sources within our target fields. Making the assumption that all sources which cannot be identified with a previously known protostellar object are extragalactic we find 4 radio sources, lower than predicted but consistent at the 2$\sigma$ level. The low number of extragalactic sources detected in these fields compared to the model may represent an over-estimation from the 15\,GHz source counts, which are extrapolated from a completeness limit of 0.5\,mJy and is high compared with our detection limit of 0.15\,mJy. One break in the observed source counts is already known at mJy flux density levels \citep{2010arXiv1012.3659D} and it is possible that a further break exists below 0.5\,mJy. 

\subsection{Spectral Energy Distributions}

An extensive literature search was conducted for unresolved, integrated flux densities to include in the spectral energy distributions. It should be noted that \citet{1995A&AS..109..177W} was a useful reference. High resolution data that were highly discrepant due to flux loss or data with high uncertainties \citep[e.g. 450\,$\mu$m data from][]{2008ApJS..175..277D} were not included. Spectral energy distributions are shown in Figures~\ref{fig:L1448}(b), ~\ref{fig:L1551NE}, and Appendix~B, with maximum likelihood models (see Table~\ref{tab:srcalpha2}) overlaid. The list of archival data used in the spectral energy distributions can be found in Appendix~C. Where uncertainties were not provided, an error of 10~per~cent was used in the model fittings and this is indicated by a $^{\dag}$ in Appendix~C. Only data $\nu<3$\,THz ($\lambda>100\,\mu$m) were included in the fit, but IRAS data $\nu>3$\,THz are included in the plots for illustration.

\subsubsection{Maximum likelihood SED fitting}

Power-law spectral indices, $\alpha_{\rm AMI}$, were fitted to the AMI channel data alone for each object, see Tables~\ref{tab:osrclist} and \ref{tab:intfluxes}, using the Markov Chain Monte Carlo based Maximum Likelihood algorithm \textsc{METRO} \citep{hob04}. The fit is of the form:
\begin{equation}
S_{\nu} \propto \nu^{\alpha_{\rm AMI}}
\end{equation}
where $S_{\nu}$ is the flux density measured at frequency $\nu$.

{\sc METRO} was then used to fit a combined radio power law, with spectral index $\alpha'$, and blackbody model to the larger dataset for each source. This fit utilized data at wavelengths longer than 100\,$\mu$m and had the form: 
\begin{equation}
S_{\rm total} = S_{1} + S_{2} = K_{1}\left(\frac{\nu}{\nu_{1}}\right)^{\alpha'} + K_{2}\frac{\nu^{\beta}B_{\nu}(T_{\rm{d}})}{\nu_{2}^{\beta}B_{\nu_{2}}(T_{\rm{d}})},
\end{equation}
where $\beta$ is the dust opacity index, $B_{\nu}$ is the Planck function for a dust temperature $T_{\rm{d}}$, $K_{1}$ is the normalized flux density at $\nu_{1}=16$\,GHz and $K_{2}$ is the normalized flux density at $\nu_{2}=300$\,GHz. Fitting was performed in two scenarios to obtain the maximum likelihood values for the spectral index $\alpha'$ of each source. These two scenarios are:
\begin{enumerate}
\item fixing the dust temperature based on evolutionary class,
\item allowing the dust temperature to vary. 
\end{enumerate}
It can be seen that when $\nu=\nu_{1}$, $S_{1}$ equals $K_{1}$, the normalized flux density at 16\,GHz and when $\nu=\nu_{2}$, $S_{2}$ equals $K_{2}$, the normalized flux density at 300\,GHz, and we define these parameters as $S_{16}^{\rm norm}$ and $S_{300}^{\rm norm}$ respectively. We use uniform and separable priors for all parameters, with ranges
\begin{equation}
\Pi=\Pi_{\alpha}(-2,2)\Pi_{\beta}(0,3)\Pi_{T_{\rm d}}(5,45).
\end{equation}

\subsubsection{The greybody contribution}

When $\nu=\nu_{1}=16$\,GHz, the free-free component should dominate the total flux density, however at 16\,GHz there is also expected to be a small contribution to the radio flux density of protostars due to the long wavelength tail of the thermal dust emission from the envelopes around these objects \citep[e.g.][]{scaife2011d}. 

It is important to separate the greybody component from the free-free emission and we calculate this predicted contribution to the 16\,GHz flux density in the following way:
\begin{equation}
S_{\rm gb,16} = f(K_{2},\beta,T_{\rm d}) = K_{2}\frac{\nu_{1}^{\beta}B_{16}(T_{\rm{d}})}{\nu_{2}^{\beta}B_{300}(T_{\rm d})},
\end{equation}
\noindent
with an associated error
\begin{equation}
\sigma_{S,16}^{2} = \left(\frac{\partial f}{\partial K_{2}}\sigma_{K_{2}}\right)^{2} +\left(\frac{\partial f}{\partial \beta}\sigma_{\beta}\right)^{2} + \left(\frac{\partial f}{\partial T_{\rm d}}\sigma_{T_{\rm d}}\right)^{2},
\end{equation}
where $\nu=16$\,GHz, $\nu_{2}=300$\,GHz, and $\sigma_{K_{2}}$, $\sigma_{\beta}$ and $\sigma_{T_{\rm d}}$ are the uncertainties on $K_{2}$, $\beta$ and $T_{\rm d}$ respectively and listed in Tables~\ref{tab:srcalpha1} and \ref{tab:srcalpha2}. In all cases, we find that the error, $\sigma_{S,16}$, is dominated by the uncertainty on $\beta$. For example, in the case of L1448: $\sigma_{S,16}^{2}=4.39^{2}\pm75.8^{2}\pm14.3^{2}$\,$\mu$Jy. 

We subtract the value of $S_{\rm gb,16}$ from the measured 16\,GHz flux density to remove the contribution of thermal dust emission and obtain values of only the radio emission due to the outflow component at 16\,GHz, $S_{\rm rad,16}$. This is to ensure that the values used do not include contributions from the thermal dust tail which varies greatly between sources and which might therefore influence any conclusions being drawn from the correlations examined in a later section of this paper. We compute the radio luminosity $S_{\rm rad,16}d^{2}$ using the greybody-subtracted radio flux densities and the distance values listed in Table~\ref{tab:srcinfo} to use in the correlations in \S6.6. If the greybody subtracted flux density is not distinct from zero by $>3\sigma$, we consider the radio luminosity consistent with zero (see Tables~\ref{tab:srcalpha1} and \ref{tab:srcalpha2}).

\begin{table*}
\begin{center}
\caption{Model results for fixed $T_{\rm d}$. Column [1] contains the source name, [2] the spectral index, [3] the opacity index, [4] the dust temperature, where 15\,K was used for Class~0 sources and 20\,K for Class~I, [5] the normalized flux density at 16\,GHz, [6] the normalized flux density at 300\,GHz, [7] the predicted greybody contribution at 16\,GHz, and [8] the radio luminosity measured at 16\,GHz with the predicted thermal dust contribution subtracted. \textquoteleft{UC}\textquoteright is used to indicate that the parameter is unconstrained. \label{tab:srcalpha1}}
\begin{tabular}{lccccccc}
\hline\hline
Source & $\alpha'$ & $\beta$ & $T_{\rm d}$ & $S_{16}^{\rm norm}$ & $S_{300}^{\rm norm}$ & $S_{\rm gb,16}$ & $S_{\rm rad,16}d^{2}$ \\
       & & & (K) & (mJy) & (Jy) & ($\mu$Jy) & (mJy kpc$^{2}$) \\
\hline
L1448~IRS~3 & $0.37\pm0.09$ & $1.48\pm0.05$ & 15 & $1.88\pm0.09$ & $4.96\pm0.15$ & $297.47\pm52.50$ & 0.131 \\
HH~7-11 & $1.12\pm0.04$ & $1.92\pm0.02$ & 20 & $3.03\pm0.08$ & $5.30\pm0.19$ & $77.96\pm4.83$ & 0.222 \\
L1551~IRS~5 & $0.10\pm0.04$ & $1.52\pm0.14$ & 20 & $3.90\pm0.17$ & $4.47\pm0.43$ & $214.58\pm130.38$ & 0.093 \\
L1527 & $0.17\pm0.08$ & $0.96\pm0.10$ & 15 & $0.85\pm0.05$ & $0.49\pm0.02$ & $134.94\pm52.75$ & 0.023 \\
HH~1-2~MMS~1 & $0.23\pm0.04$ & $2.38\pm0.08$ & 15 & $1.49\pm0.07$ & $0.99\pm0.02$ & $4.33\pm1.25$ & 0.299 \\
HH~26~IR & UC & $0.62\pm0.07$ & 20 & $0.04\pm0.06$ & $0.56\pm0.03$ & $372.00\pm101.97$ & $-$ \\
HH~111 & $0.88\pm0.07$ & $1.66\pm0.07$ & 15 & $2.16\pm0.09$ & $0.90\pm0.02$ & $31.95\pm8.81$ & 0.434 \\
NGC~2264~G & $-0.29\pm0.24$ & $1.44\pm0.39$ & 15 & $0.50\pm0.10$ & $0.54\pm0.65$ & $37.06\pm192.75$ & $-$ \\
Serpens~MMS~1 & $-0.06\pm0.02$ & $1.47\pm0.05$ & 15 & $6.89\pm0.14$ & $5.22\pm0.20$ & $327.71\pm56.62$ & 0.473 \\
L723 & $-0.26\pm0.09$ & $1.83\pm0.07$ & 15 & $0.58\pm0.03$ & $0.93\pm0.05$ & $20.04\pm5.15$ & 0.050 \\
L1251~A & $1.79\pm0.15$ & $2.92\pm0.27$ & 15 & $0.91\pm0.07$ & $0.32\pm0.07$ & $0.29\pm0.51$ & 0.107 \\
\hline
L1448~C & $1.12\pm0.12$ & $2.10\pm0.04$ & 15 & $0.53\pm0.02$ & $1.64\pm0.04$ & $16.25\pm2.41$ & 0.046 \\
NGC~1333~IRAS~2A & $2.42\pm0.07$ & $2.01\pm0.05$ & 15 & $0.43\pm0.02$ & $1.52\pm0.10$ & $19.33\pm3.67$ & 0.023 \\
NGC~1333~IRAS~2B & $1.31\pm0.05$ & $1.95\pm0.09$ & 20 & $1.02\pm0.04$ & $0.53\pm0.04$ & $7.16\pm2.69$ & 0.031 \\
L1551~NE & $-0.13\pm0.31$ & $1.41\pm0.03$ & 15 & $0.60\pm0.09$ & $1.83\pm0.02$ & $136.09\pm11.08$ & 0.012 \\
HH~25~MMS & $2.13\pm0.04$ & $1.68\pm0.16$ & 15 & $0.97\pm0.07$ & $0.83\pm0.10$ & $27.82\pm22.30$ & 0.212 \\
\hline
\end{tabular}
\vskip .05in
\begin{minipage}{0.9\textwidth} 
{\footnotesize }
\end{minipage}
\end{center}
\end{table*}

\begin{table*}
\begin{center}
\caption{Model results for variable $T_{\rm d}$. Column [1] contains the source name, [2] the spectral index, [3] the opacity index, [4] the dust temperature, [5] the normalized flux density at 16\,GHz, [6] the normalized flux density at 300\,GHz, [7] the predicted greybody contribution at 16\,GHz, and [8] the radio luminosity measured at 16\,GHz with the predicted thermal dust contribution subtracted. \textquoteleft{UC}\textquoteright is used to indicate that the parameter is unconstrained. \label{tab:srcalpha2}}
\begin{tabular}{lccccccc}
\hline\hline
Source & $\alpha'$ & $\beta$ & $T_{\rm d}$ & $S_{16}^{\rm norm}$ & $S_{300}^{\rm norm}$ & $S_{\rm gb,16}$ & $S_{\rm rad,16}d^{2}$ \\
       & & & (K) & (mJy) & (Jy) & ($\mu$Jy) & (mJy kpc$^{2}$) \\
\hline
L1448~IRS~3 & $0.46\pm0.09$ & $1.81\pm0.12$ & $10.77\pm0.95$ & $2.02\pm0.09$ & $5.13\pm0.12$ & $145.74\pm77.31$ & 0.140 \\
HH~7-11 & $1.28\pm0.04$ & $3.00\pm0.08$ & $12.53\pm0.90$ & $3.24\pm0.07$ & $6.21\pm0.27$ & $4.92\pm1.59$ & 0.227 \\
L1551~IRS~5 & $0.09\pm0.03$ & $1.46\pm0.05$ & $26.74\pm2.07$ & $3.87\pm0.0.09$ & $4.54\pm0.17$ & $233.58\pm43.54$ & 0.093 \\
L1527 & $0.20\pm0.08$ & $0.81\pm0.11$ & $44.86\pm1.57$ & $0.86\pm0.06$ & $0.44\pm0.03$ & $138.81\pm60.00$ & 0.023 \\
HH~1-2~MMS~1 & $0.21\pm0.04$ & $2.18\pm0.19$ & $16.93\pm1.80$ & $1.46\pm0.07$ & $1.01\pm0.02$ & $7.35\pm7.38$ & 0.299 \\
HH~26~IR & UC & $0.66\pm0.16$ & $17.36\pm5.25$ & $0.04\pm0.09$ & $0.59\pm0.04$ & $368.59\pm300.38$ & $-$ \\
HH~111 & $0.84\pm0.07$ & $1.55\pm0.17$ & $17.33\pm3.68$ & $2.11\pm0.10$ & $0.90\pm0.02$ & $41.31\pm35.89$ & 0.366 \\
NGC~2264~G & $-0.31\pm0.11$ & $1.71\pm0.13$ & $18.30\pm1.88$ & $0.49\pm0.06$ & $0.36\pm0.02$ & $10.02\pm5.59$ & 0.186 \\
Serpens & $-0.07\pm0.02$ & $1.25\pm0.04$ & $28.54\pm1.54$ & $6.74\pm0.14$ & $4.85\pm0.18$ & $451.87\pm71.01$ & 0.464 \\
L723 & $-0.29\pm0.09$ & $1.59\pm0.13$ & $17.56\pm1.31$ & $0.56\pm0.04$ & $0.92\pm0.04$ & $37.73\pm20.70$ & 0.049 \\
L1251~A & $1.70\pm0.14$ & $2.27\pm0.41$ & $20.85\pm4.36$ & $0.86\pm0.07$ & $0.38\pm0.06$ & $1.98\pm7.84$ & 0.107 \\
\hline
L1448~C & $1.17\pm0.15$ & $2.22\pm0.11$ & $12.70\pm1.25$ & $0.54\pm0.03$ & $1.69\pm0.04$ & $12.88\pm5.99$ & 0.047 \\
NGC~1333~IRAS~2A & $2.56\pm0.03$ & $2.76\pm0.19$ & $9.85\pm1.10$ & $0.46\pm0.02$ & $1.17\pm0.07$ & $2.22\pm2.20$ & 0.024 \\
NGC~1333~IRAS~2B & $1.40\pm0.05$ & $2.89\pm0.26$ & $9.55\pm1.90$ & $1.06\pm0.03$ & $0.57\pm0.06$ & $0.76\pm1.28$ & 0.031 \\
L1551~NE & UC & $0.78\pm0.05$ & $40.64\pm3.29$ & $0.17\pm0.07$ & $1.69\pm0.02$ & $585.75\pm0.05$ & $-$ \\
HH~25~MMS & UC & $0.55\pm0.09$ & $41.27\pm3.65$ & $0.004\pm0.18$ & $1.51\pm0.09$ & $1021.76\pm352.26$ & $-$ \\
\hline
\end{tabular}
\vskip .05in
\begin{minipage}{0.9\textwidth} 
{\footnotesize }
\end{minipage}
\end{center}
\end{table*}

\subsubsection{Comparison of the two scenarios}

In Scenario (i), the algorithm was run by fixing $T_{\rm d}$ to 15\,K for Class~0 objects and to 20\,K for Class~I objects, and allowing the other parameters to vary to determine their maximum likelihood values. Overall, fixing the dust temperature provided reasonable fits to the spectral energy distributions, with the exception of L1527, NGC~2264~G, Serpens~MMS~1, and L1551~NE where it provided a poor fit to the greybody peaks. The results for the maximum likelihood values of $\alpha'$, $\beta$, $S_{16}^{\rm norm}$ and $S_{300}^{\rm norm}$ in each case are listed in Table~\ref{tab:srcalpha1}, along with the dust temperature used [Columns~2-6], the greybody contribution at 16\,GHz [Column~7], and the radio luminosity at 16\,GHz with the greybody contribution subtracted out [Column~8]. We find a weighted average spectral index of $\overline{\alpha'}=0.20\pm0.41$ for the original sample, consistent with free-free emission, and a dust opacity index of $\overline{\beta}=1.75\pm0.36$. Quoted errors represent the standard deviation of the distribution of values, rather than the standard error on the mean.

In Scenario (ii), we do not fix $T_{\rm d}$ and allow the algorithm to determine the maximum likelihood value for this parameter along with $\alpha'$, $\beta$, $S_{16}^{\rm norm}$ and $S_{300}^{\rm norm}$. The results for these parameters along with $S_{\rm gb,16}$ and $S_{\rm rad,16}d^{2}$ are listed in Table~\ref{tab:srcalpha2}. We again find a weighted average spectral index of $\overline{\alpha'}=0.20\pm0.46$ for the original sample consistent with free-free emission. We find an opacity index of $\overline{\beta}=1.52\pm0.56$ and a dust temperature $\overline{T_{\rm d}}=18.46\pm9.77$ for the original sample.  

In the first scenario, with fixed $T_{\rm d}$, 80~per~cent of the sources in the original sample for which the spectral indices could be constrained had maximum likelihood values of $2>\alpha'>-0.1$, consistent with free-free emission. In the second scenario, with $T_{\rm d}$ varying, again 80~per~cent of the sources in the original sample had $\alpha'$ consistent with free-free emission. In both cases, NGC~2264 and L723 have spectral indices that are inconsistent with free-free emission and $\alpha' <-0.1$, suggesting non-thermal emission. The one source for which an accurate value of $\alpha'$ is unconstrained in the original sample is HH~26~IR, due to insufficient low-frequency data. Unconstrained parameters are indicated in Tables~\ref{tab:srcalpha1} and \ref{tab:srcalpha2} by \textquoteleft{UC}\textquoteright. Since $\alpha'$ is unconstrained for HH~26~IR, the algorithm attributes all the flux density at 16\,GHz to thermal dust emission, resulting in a value of $S_{\rm rad,16}d^{2}\approx0$ when subtracted from the AMI flux density. Consequently it does not appear in the correlations in \S6.6.  

A spectral index of $\alpha>2$ is generally considered un-physical for free-free emission, and is attributed to dust emission. In the cases where the likelihood distribution pushed the upper limit of the prior range, we increased the prior range to $\Pi_{\alpha}(-2,3)$. In the extended sample, NGC~1333~IRAS~2A has $\alpha'>2$ in both model scenarios, suggesting that the lower frequency data is an extension of the greybody emission and that there is no free-free component. This will not be reflected accurately in the calculation of the greybody emission at 16\,GHz since the algorithm constrains a separate $\alpha'$ component, and this will in turn affect the radio luminosity calculated for this source and the correlations discussed in \S6.6. When $T_{\rm d}$ is fixed (Scenario (i)), HH~25~MMS also has $\alpha'>2$, suggesting that it is purely greybody emission. This will have the same consequences as for NGC~1333~IRAS~2A. However, $\alpha'$ is unconstrained for HH~25~MMS when $T_{\rm d}$ is allowed to vary (Scenario (ii)) and this will have the reverse affect on the radio luminosity. Examining the SED plot (see Appendix~B), it is possible that HH~25~MMS has a free-free component based on one or two data points, but the model is unable to constrain it without more low-frequency data. 

\begin{figure}
\centerline{\includegraphics[width=0.4\textwidth]{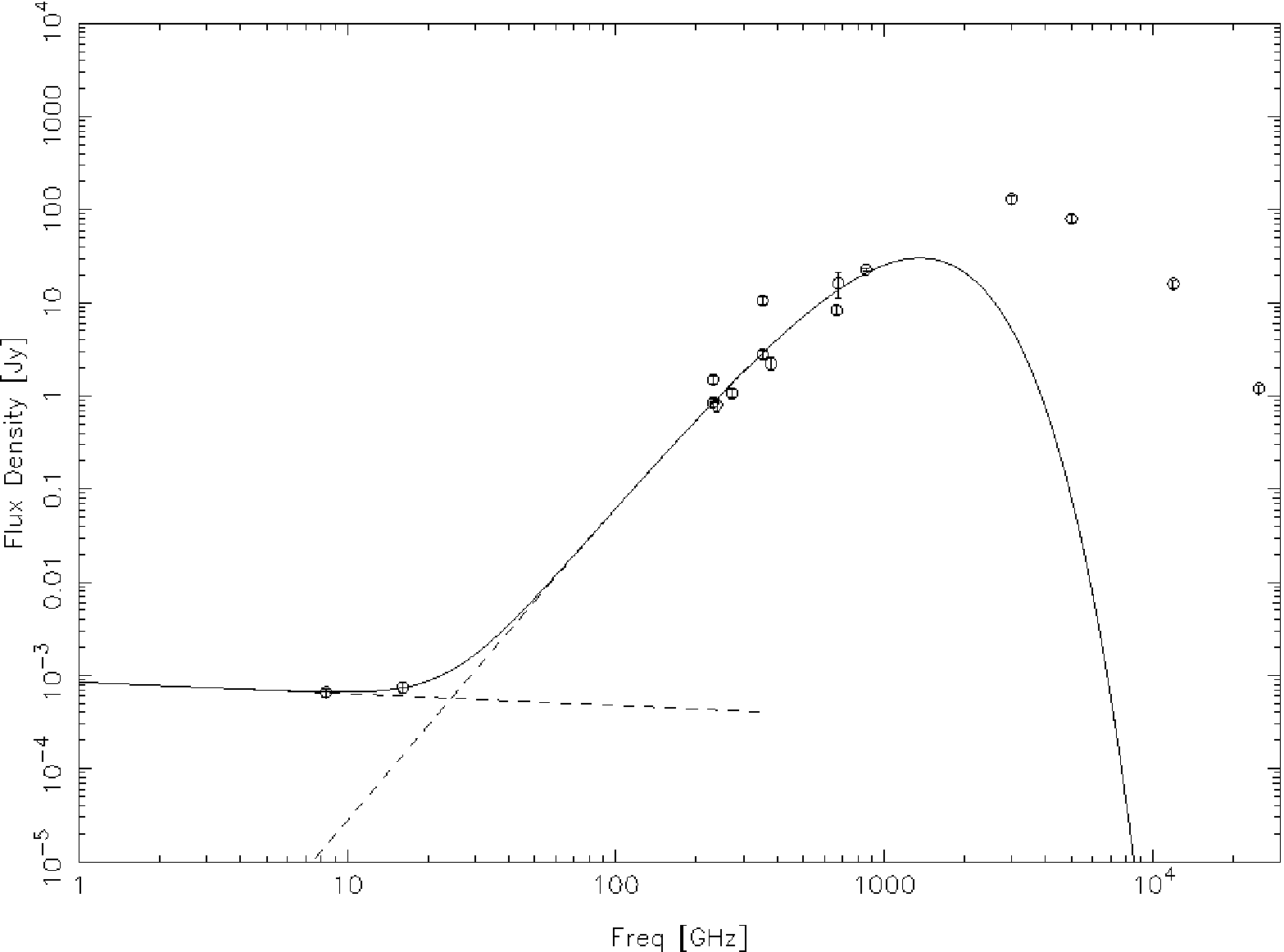}}
\centerline{(a)}
\centerline{\includegraphics[width=0.4\textwidth]{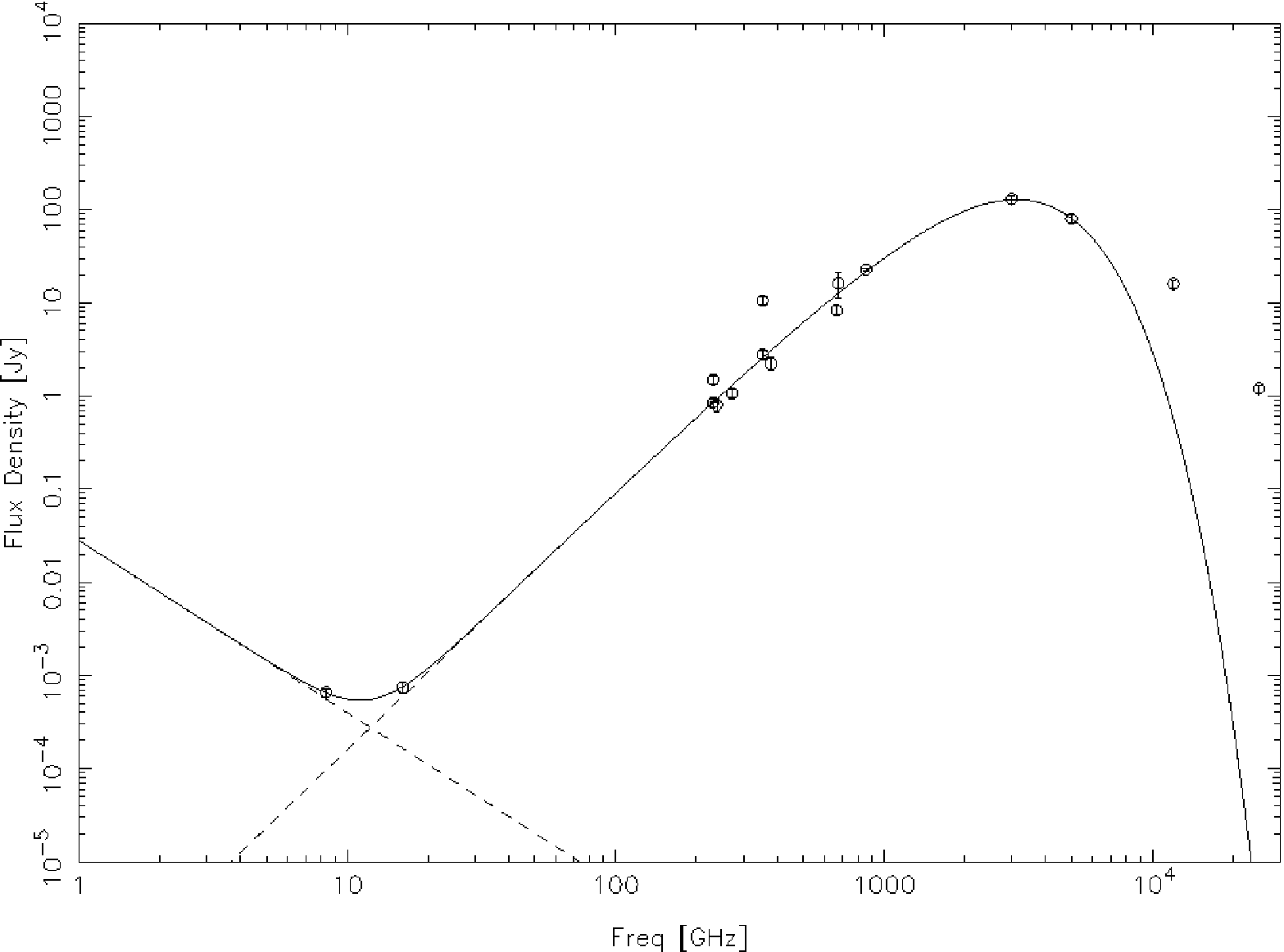}}
\centerline{(b)} 
\caption{The spectral energy distribution fits for L1551~NE when (a) $T_{\rm d}=15$\,K, and (b) when $T_{\rm d}$ is allowed to vary. In the first case, the method determines $\alpha'$ for the two low-frequency data points, but in the second case the AMI data point is fit as greybody emission and $\alpha'$ is unconstrained. }\label{fig:L1551NE}
\end{figure}

For L1551~NE, $\alpha'$ is constrained in Scenario (i) but not in Scenario (ii). Consequently, the calculated radio luminosities differ significantly between the two scenarios and this will affect the correlations in \S6.6. Specifically, $S_{\rm rad,16}d^{2}$ is a factor of 4.3 higher when $T_{\rm d}$ is allowed to vary than when it is fixed to 15\,K (see Tables~\ref{tab:srcalpha1} and \ref{tab:srcalpha2}). When $T_{\rm d}$ is fixed for this source, the peak of the greybody is constrained by the given dust temperature (and does not fit the higher frequency data), allowing the algorithm to determine $\alpha'$ using the two low-frequency data points (see Figure~\ref{fig:L1551NE}(a)). However, when $T_{\rm d}$ is allowed to vary, the AMI data point at 16\,GHz is fit as part of the greybody emission and $\alpha'$ is left unconstrained, as there is only one remaining low-frequency data point (see Figure~\ref{fig:L1551NE}(b)). This results in a more significant greybody contribution at 16\,GHz. Therefore the radio luminosity will be significantly lower when $S_{\rm gb,16}$ is subtracted in this scenario compared to that when the dust temperature is fixed, and this will drastically change its position on the correlation plots between the two scenarios. The same situation is also found in the case of HH~25~MMS. 

\subsection{Dust Temperatures}

In Scenario (ii), where $T_{\rm d}$ varies, our results for $T_{\rm d}$ range between 9--45\,K. Values in the literature for the dust temperatures associated with Class~0 and I sources are found to range between 10--40\,K \citep{1993AA...273..221R, 1998MNRAS.301.1049D, 2002ApJ...575..337S, 2003ApJS..145..111Y, hat2007a}. \citet{1998MNRAS.301.1049D} find that their sample of Class~0 protostars can be fitted with optically thin emission, a $\beta$ index between 1--1.5, and a dust temperature between 20--30\,K. These authors derive for their Class~I sources $\beta$ between 1-2 and a dust temperature of $~30$\,K, of which our results are consistent. 

For Class~0 sources, we find weighted average values of $\overline{\beta}=1.35\pm0.29$ and $\overline{T_{\rm d}}=20.1\pm10.89$, and for Class~I sources we find $\overline{\beta}=1.81\pm0.74$ and $\overline{T_{\rm d}}=14.85\pm5.15$. These results are reversed from what we expect based on the first SED scenario, but this is likely to be due to the low number of Class~I sources in our sample compared to Class~0. \citet{scaife2011d} find $\beta=0.3$ for L1527 by fitting data from 5\,GHz to 350\,GHz jointly using the sum of two power-laws, which is slightly shallower than the value of $\beta=0.81\pm0.11$ found in this work. However, both are consistent with the value of $\beta\leq1$, indicating the presence of large dust grains.

\subsection{Radio Spectral Indices}

Typically, the radio emission detected at centimetre wavelengths for Class~0 and I protostars is observed to possess a flat or rising spectrum, indicating that it occurs as a consequence of free-free radiation from ionised gas. Unresolved sources can exhibit partially opaque spectra with spectral indices in the range $-0.1\leq\alpha\leq2$, where a value of $-0.1$ indicates optically thin free-free emission and a value of 2 indicates optically thick. For the case of a standard, canonical jet with constant opening angle, the spectral index $\alpha=0.6$. Shallower spectral indices ($\alpha<0.6$) are expected for unresolved, well-collimated, ionised outflows, and steeper spectra still ($\alpha>0.9$) are expected for fully ionised, adiabatic jets \citep{1986ApJ...304..713R}. In this section we compare our results for $\alpha_{\rm AMI}$ (Tables~\ref{tab:osrclist} and \ref{tab:intfluxes}) and $\alpha'$ from the spectral energy distribution scenario where $T_{\rm d}$ is allowed to vary (Table~\ref{tab:srcalpha2}).

Across the AMI band, 7/9 (78~per~cent) of the target sources, where the spectral index $\alpha_{\rm AMI}$ could be constrained, are within the range $-0.1\leq\alpha_{\rm AMI}\leq2$ indicating free-free emission (see Table~\ref{tab:intfluxes}). When including the additional protostellar sources detected in these fields, 11/14 (79~per~cent) of the sources have $\alpha_{\rm AMI}$ within this range. The three sources that lie outside this range, HH~111, Serpens and L1448~C, have $\alpha_{\rm AMI}<-0.1$, typically indicating that another emission mechanism is at work. However, spectral indices inconsistent with free-free emission have been found previously for Serpens~MMS~1 \citep{1989ApJ...346L..85R, 1993ApJ...415..191C}. The weighted average spectral index of the original sample calculated from the AMI fluxes is $\overline{\alpha_{\rm AMI}}=0.42\pm0.59$, consistent with a canonical jet. The flux densities for HH~26~IR and NGC~2264 could only be extracted from the 16\,GHz combined-channel map, so it was not possible to measure $\alpha_{\rm AMI}$ for these sources. For the extended sample, $\overline{\alpha_{\rm AMI}}=1.52\pm0.64$, still consistent with free-free emission. The large uncertainties of $\alpha_{\rm AMI}$ in Table~\ref{tab:intfluxes} are due to the short frequency coverage of AMI. 

For comparison, 8/10 (80~per~cent) of the original sample have $-0.1\leq\alpha'\leq2$ when including archival data (again, excluding HH~26~IR, see Table~\ref{tab:srcalpha2}). For the extended sample, 10/13 (77~per~cent) of the sources (excluding L1551~NE and HH~25~MMS) have $\alpha'$ within this range. The values for $\alpha'$ that lie outside this range are NGC~2264 and L723 ($\alpha'<-0.1$) and NGC~1333~IRAS~2A ($\alpha'>2$). The original sample has a weighted average spectral index $\overline{\alpha'}=0.20\pm0.46$ and the extended sample has $\overline{\alpha'}=0.37\pm0.63$, both consistent with free-free emission and a well-collimated flow.

In the case of NGC~2264~G, there are two known radio continuum sources in this outflow region: VLA~1, which was proposed as the driving source by \citet{1989RMxAA..17..115R}, and VLA~2, which is suggested to be the driving source source by \citet{1994ApJ...436..749G}. \citet{1994ApJ...436..749G} find a spectral index $\alpha\simeq-1.3\pm0.4$ for VLA~1 and a significantly lower flux density than \citet{1989RMxAA..17..115R}, suggesting that it appears to be highly variable and an extragalactic nature cannot be ruled out. For VLA~2, these authors find a spectral index $\alpha\simeq0.3\pm0.2$, consistent with free-free emission. The negative spectral index $\alpha'=-0.31\pm0.11$ obtained here for NGC~2264~G is likely contaminated by VLA~1 as the sources are unresolved by AMI. Additional high resolution observations are needed to confirm the non-thermal component.

L723 is in a similar situation to NGC~2264~G, where there is a nearby source, VLA~1, with spectral index $\alpha\leq-0.1$ that could be contaminating the results in the SED for the prospective driving source, VLA~2. \citet{1996ApJ...473L.123A} find a spectral index $\alpha=-0.3\pm0.2$ for VLA~1, and discuss the possibility that the source is a radio-emitting, optically obscured T~Tauri star. These authors find a spectral index $\alpha=0.5\pm0.2$ for VLA~2, consistent with free-free emission, and the results of \citet{2008ApJ...676.1073C} for the spectral index of VLA~2 are also indicative of free-free emission. We find $\alpha_{\rm AMI}=0.42\pm0.82$ for L723, consistent with free-free emission from a canonical jet even though we do not resolve these sources. The negative spectral index $\alpha'=-0.29\pm0.09$ we obtain is likely affected by both VLA~1 and insufficient low-frequency data to better constrain $\alpha'$.

From the data presented here, L1527 is found to have a spectral index $\alpha_{\rm AMI}=1.15\pm0.57$, which agrees with the value of $\alpha_{\rm AMI}=1.17\pm0.42$ found by \citet{scaife2011d}. A spectral index of $\alpha>0.8$ can be easily produced by velocity gradients or the recombination that might be expected for rapidly diverging flows \citep{1986ApJ...304..713R}. For consistency purposes, we have used the same archival data as in \citet{scaife2011d} for the SED, who find $\alpha'=-0.11\pm0.09$. However, we find $\alpha'=0.70\pm0.06$ in this work. We attribute this difference to the fact that the data is fit with multiple greybody components in \citet{scaife2011d}. L1527 has a known disc \citep{2002ApJ...581L.109L, 2010ApJ...722L..12T} which dominates the SED at sub-mm frequencies, while the envelope contribution dominates at far-infrared frequencies. The single greybody model used in this work does not provide the best fit to both the envelope and disc. However, in both scenarios the SED model here predicts $\beta\leq1$, consistent with the presence of large dust grains in the disc \citep{scaife2011d}.

Overall, the values for the spectral indices are consistent as all sources listed in Table~\ref{tab:intfluxes} and the protostellar sources listed in Table~\ref{tab:srcalpha2} have known outflows. In the cases of NGC~2264~G and L723, higher spatial resolution data can help constrain the emission mechanism for the driving sources of these outflows and identify non-thermal components.

\subsection{Correlations}

Radio luminosity has been shown to correlate with global properties such as bolometric luminosity, envelope mass, and outflow force \citep{scaife2011a, scaife2011b, scaife2011c}. We do not repeat these discussions here, but combine our data with the previous work to redraw the correlations, see Figures~\ref{fig:corr1} and \ref{fig:corr2}. We use the radio luminosity ($S_{\rm rad,16}d^{2}$) results calculated by subtracting the predicted greybody contribution at 16\,GHz ($S_{\rm gb,16}$) from the measured flux density ($S_{\rm 16\,GHz}$), in the correlations that follow to ensure that the values used are representative of only the outflow component of the radio emission. We compare the correlations between both SED scenarios and examine the relation of this sample to existing correlations. 

\subsubsection{Note on Classification}

\citet{scaife2011b} found that when comparing bolometric temperatures determined using IRAS data \citep{hat2007a} and those found from fluxes calcuated using \emph{Spitzer} SEDs \citep{2008ApJS..179..249D}, the IRAS temperatures were, on average, a factor of 1.6 higher than the corresponding \emph{Spitzer} values. Bolometric temperature is one of the indicators for protostellar class and within this sample, one of the sources with $70\leq~T_{\rm{bol}}\leq150$ (HH~111) will change from Class~I to 0 when the \emph{Spitzer} temperature is used.

\subsubsection{Correlation with bolometric luminosity}

\begin{figure}
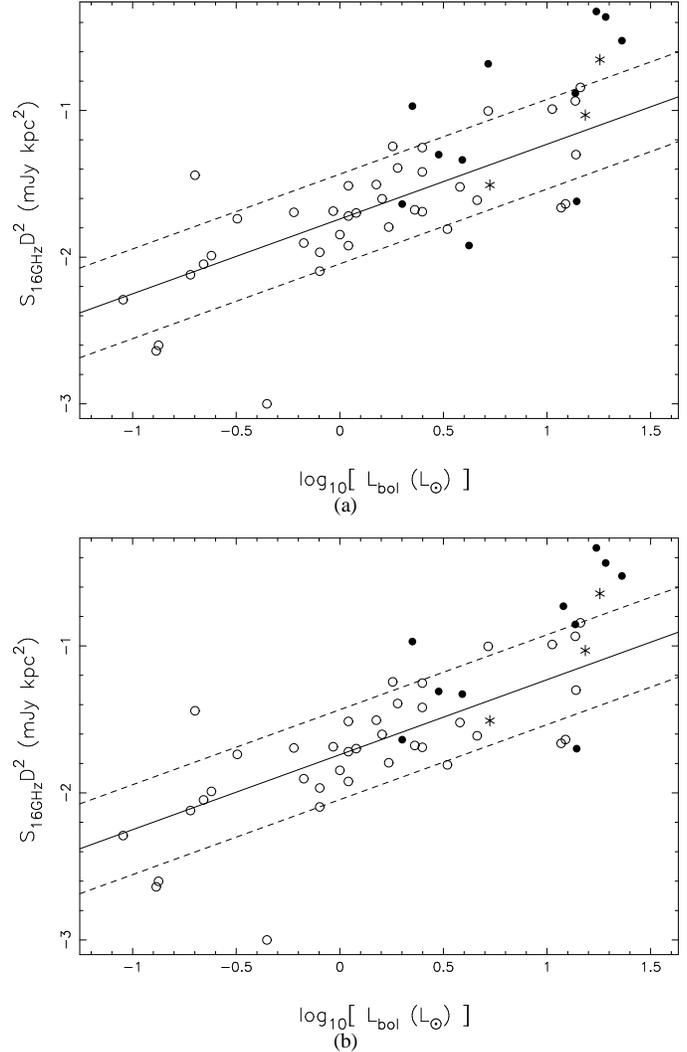

\centerline{\includegraphics[angle=-90,width=0.5\textwidth]{./lbol_corr1.ps}}
\centerline{(a)}
\centerline{\includegraphics[angle=-90,width=0.5\textwidth]{./lbol_corr2.ps}}
\centerline{(b)}
\caption{(a) Correlation of bolometric luminosity with 1.9\,cm radio luminosity when $T_{\rm d}$ is fixed, and (b) correlation of bolometric luminosity with 1.9\,cm radio luminosity when $T_{\rm d}$ varies. Filled circles represent Class~0 sources from this sample, stars are Class~I, and unfilled circles are previous data. Best fitting correlations from \citet{scaife2011b} are shown as solid lines. \label{fig:corr1}}
\end{figure}

Figure~\ref{fig:corr1} shows the correlation of radio luminosity with bolometric luminosity for (a) Scenario (i), where $T_{\rm d}$ is fixed, and (b) Scenario (ii), where $T_{\rm d}$ varies. These new data extend the trend into higher luminosities. We note that the bolometric luminosity for the Serpens~MMS~1 source used here is $L_{\rm bol}=17.3$\,L$_{\odot}$ \citep{2009ApJ...707..103E}, however significantly higher luminosities have been reported for this object in the literature which could explain the offset from the trend (see \S5.9). The value of $\alpha'$ is unconstrained for HH~26~IR due to insufficient low-frequency data. As a result, our method calculates a significant value for $S_{\rm gb,16}$, resulting in a value of $S_{\rm rad,16}d^{2}$ consistent with zero due to the low flux density detected by AMI. HH~26~IR is therefore not included in the plots. Data below 8\,GHz is needed to constrain the spectral index of this source and establish an accurate measure of the radio luminosity.

Two sources, L1551~NE and HH~25~MMS, have their location change significantly between Figures~\ref{fig:corr1}(a) and (b), due to the large difference in their radio luminosities between the two SED scenarios (as discussed extensively in \S6). Larger radio luminosities were calculated for both sources in the case where $T_{\rm d}$ is fixed. However, the lower radio luminosity calculated from the scenario where $T_{\rm d}$ is allowed to vary for HH~25~MMS is in better agreement with the trend than the higher value from the scenario where $T_{\rm d}$ is fixed. The opposite is true for L1551~NE, and the radio luminosity calculated for this source from Scenario (i) is in better agreement with the trend than the one calculated from Scenario (ii). Like HH~26~IR, L1551~NE deviates from the trend due to a radio luminosity $S_{\rm rad,16}d^{2}\approx0$ when $T_{\rm d}$ varies, and more observations at these frequencies are needed to constrain the spectral index and establish the radio luminosity. 

We use a two sample Kolmogorov-Smirnov (KS) test to determine whether the sample from this paper is consistent with the known correlation. We reject the null hypothesis of no difference between the datasets if the probability is less than the standard cutoff value of p=0.05. We find p-values of p=0.021 and =0.037 for Scenarios (i) and (ii) respectively, indicating that the sample from this work is statistically different from the previous sample in both scenarios. 

We attribute this discrepancy to the values of the bolometric luminosity, rather than the radio luminosity, as no similar effect is seen in the correlation with envelope mass (see \S~\ref{sec:masscorr}). There are a number of possible explanations. The data set from previous papers (open circles in Fig.~\ref{fig:corr1}) has bolometric luminosity values dominated by a single catalogue \citep{hat2007a}, whereas the current sample has values from a much wider range of references. We might therefore expect a larger degree of scatter in the current dataset due to differences in methodology. Alternatively, it is possible that the bolometric luminosities derived in \citet{hat2007a} which dominate the archival dataset have themselves a systematic offset from those found more generally in the literature, however this seems unlikely as it has not been noted in previous comparisons. A further possibility is that the current sample is (on average) biased towards sources with high ratios of bolometric luminosity to envelope mass. Since these sources are preferentially selected to be well known radio protostars it is likely that this effect is present at some level. We therefore suggest that the underlying cause of the discrepancy is a combination of both this last possibility and the first.

\subsubsection{Correlation with envelope mass}
\label{sec:masscorr}

\begin{figure}
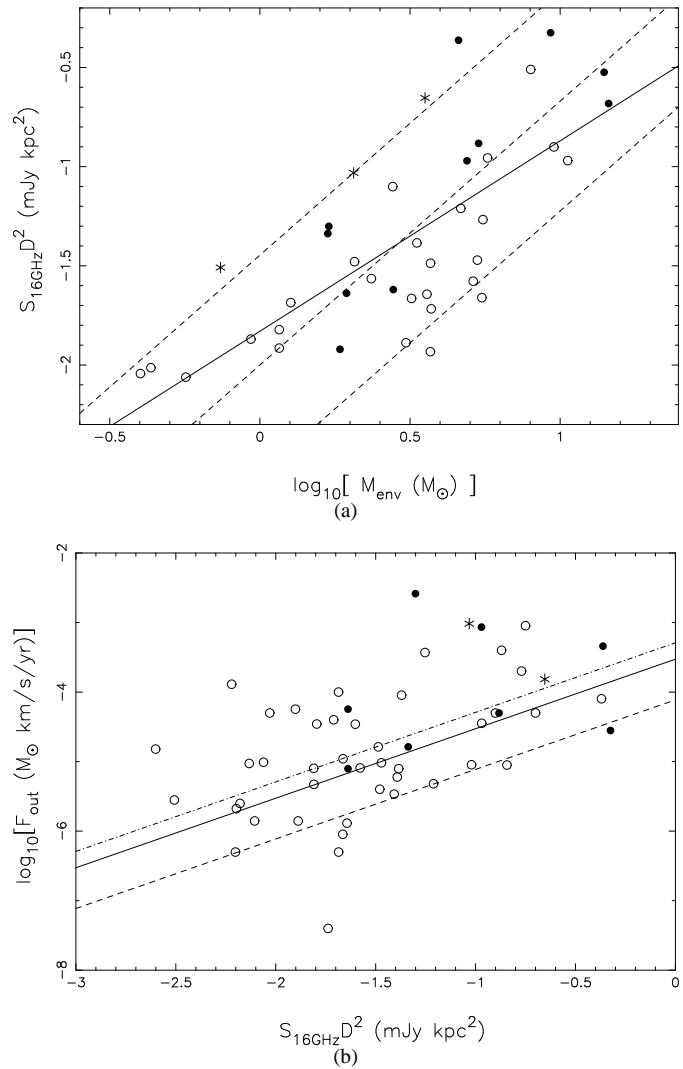

\centerline{\includegraphics[angle=-90,width=0.5\textwidth]{./menv_corr.ps}}
\centerline{(a)}
\centerline{\includegraphics[angle=-90,width=0.5\textwidth]{./fout_corr.ps}}
\centerline{(b)}
\caption{(a) Correlation of envelope mass with 1.9\,cm radio luminosity, where the solid line indicates the best fitting correlation using the combined data set, see \S~\ref{sec:masscorr}, the dashed lines indicate a slope of 1.33 and the 1$\sigma$ limits on the distribution of data relative to this slope (see \citet{scaife2011b}); (b) correlation of 1.9\,cm radio luminosity with outflow force. Filled circles represent Class~0 sources from this sample, stars are Class~I, and unfilled circles are previous data. The solid line indicates the shock ionisation model of \citet{1989ApL&C..27..299C} with a temperature of $T_{\rm e}=10^{4}$\,K. The upper (lower) dashed line indicates the same model with a temperature of $T_{\rm e}=3000$\,K ($10^{5}$\,K; \citet{scaife2011b}). \label{fig:corr2}}
\end{figure}

Figure~\ref{fig:corr2}a shows the correlation between radio luminosity and envelope mass. The envelope masses listed in Column [8] of Table~\ref{tab:srcinfo} were determined using the 850\,$\mu$m flux density from the Fundamental Map Object Catalogue \citep{2008ApJS..175..277D} in the following way:
\begin{equation}
M = \frac{S_{\nu}d^2}{\kappa_{\nu}B_{\nu}(T_{\rm d})}
\end{equation}
where $\kappa_{\nu}$ is the opacity index, $B_{\nu}(T_{\rm d})$ is the Planck function at a dust temperature $T_{\rm d}$, and the scaling factor $\big[\kappa_{\nu}B_{\nu}(T_{\rm{d}})\big]^{-1}$ uses the scaled value from \citet{hat2007a}, applied in \citet{scaife2011c}. This fixes us on the same scale as \citet{2009ApJ...707..103E} and allows for comparisons between the datasets. Distances used are listed in Column [3] of Table~\ref{tab:srcinfo}. \citet{2005ApJS..156..169F} find that lower dust temperature of 15\,K e.g., as suggested by \citet{2001AA...365..440M} and \citet{hat2007a}, would lead to envelope masses that are a factor of 1.6 higher, however this should not affect our results as we are maintaining consistency within our sample. The SCUBA \citep{2008ApJS..175..277D} 850\,$\mu$m flux density is unresolved for NGC~1333~IRAS~2A and B, and so we use the values from \citet{2009ApJ...692..973E} for these objects.

The data from this paper are combined with those from \citet{scaife2011a, scaife2011b} to examine the correlation of radio luminosity with envelope mass. It can be seen from Figure~\ref{fig:corr2}(a) that the data from this work are consistent with those from the literature. A best-fitting correlation using the combined data-set is found to be,
\begin{eqnarray}
\nonumber \log[L_{\rm rad} {(\rm mJy\,kpc^2)}] = &&(-1.83\pm0.25)\\
\nonumber && +(0.96\pm0.41)\log[M_{\rm env} {(M_{\odot})}].
\end{eqnarray}

Following \citet{scaife2011b} we also compare the data to a power-law with index 1.33, shown as a dashed line in Figure~\ref{fig:corr2}(a). We note that the fitted correlation is consistent with this index at the 1\,$\sigma$ level.

Class~I sources are expected to have smaller envelope masses than Class~0 by definition (see \S1) but from these data there is no evident deviation from the trend based on evolutionary class. In fact, the Class~I sources in this sample are all found along the +1~$\sigma$ level line of the correlation.

From the KS test we find p-values of p=0.374 and p=0.54 for Scenarios (i) and (ii) respectively, indicating that the sample from this work is consistent with the correlations found from the previous sample in both scenarios. .

\noindent{\bf Observational bias in the $L_{\rm rad}-M_{\rm env}$ correlation} We note that the correlation fitted directly to the data shown in Figure~\ref{fig:corr2} will be subject to an observational bias due to the low significance end of the data. This form of Eddington bias causes the slope of the fitted correlation to be artificially shallow due to the increased data scattered above the correlation, relative to that scattered below the correlation which will be missed due to the detection limit. Such a bias can be corrected statistically, and following the method of \citet{2006spit.conf..541H} we use a maximum likelihood estimator for the slope, $m$, to recover a value of $\hat{m} = 1.01$. 

The correlation with bolometric luminosity will also be subject to a similar bias. However, we do not correct the fit in this case as the detection limit is not the only criterion for detection of a radio counterpart in this case, as shown in \citet{scaife2011b}.

\subsubsection{Correlation with outflow force}

The 1.9\,cm radio luminosity data presented here are consistent with the loose correlation with outflow force measured in \citet{scaife2011a, scaife2011b}, see Figure~\ref{fig:corr2}(b). As discussed in \citet{scaife2011a} the lack of a well defined correlation in this case may by due to the errors inherent in measurements of outflow force. Class~0 objects are generally considered to have more energetic outflows than Class~I, but there is no clear evolutionary division in the correlation presented here. 

Many studies increase their outflow forces by a factor of 10 to account for optical depth and inclination effects, following \citet{1996AA...311..858B}. From the current sample, only two the values for outflow force from the literature have been corrected in this way: L1527 and Serpens~MMS (see Table~\ref{tab:srcinfo}). As the majority of the data included in Figure~\ref{fig:corr2}(b) has this correction implemented, we apply the same adjustment to the values in Table~\ref{tab:srcinfo} for the purposes of the plot.

\subsection{Variability}

Radio variability might be expected for Class~0 and I sources due to episodic accretion rates and mass ejection in the outflows \citep{1995ApJS..101..117K, 2005ApJ...627..293Y}. However, only two sources, L1551~IRS~5 and Serpens~SMM~1, show strong signs of variability based on examination of their spectral energy distributions (see Appendix~B). This could be due to the fact that there is a comparatively larger amount of archival data for these sources, and it is likely that a majority of the sources in this sample possess variable natures but more observations at multiple epochs are needed to confirm this. In the case of NGC~2264~G, evidence for variability has been shown \citep{1994ApJ...436..749G} for an object in the same region as the driving source of this outflow which will affect the results of unresolved data.

\noindent{\bf L1551.} \citet{2003ApJ...583..330R} have compiled measurements of L1551~IRS~5 at 2\,cm from 1983 to 1998 to determine the proper motion of the multiple sources, and by comparing different epochs they found that the southern component underwent a major ejection event during the late 1980s and showed considerable elongation in the E-W direction. By comparing data at 5 and 15\,GHz taken at different epochs (see Appendix~C) there is clear evidence of variability. The flux densities measured at 5\,GHz range by a factor of $\approx1.7$ over all epochs, while the flux densities measured at 15\,GHz range by a factor of $\approx2.2$. In addition, the VLA observations of \citet{1990ApJ...355..635K} were measured in September 1988, and their flux density at 15\,GHz is a factor of $\approx1.7$ higher than the flux density measured in this work at a similar frequency. \citet{2006ApJ...653..425L} detect three circumstellar dust disks in this region, and therefore luminosity variations might be expected. These authors have discussed that the flux of the N jet (see \S5 in this work) is typically higher than the flux of the S jet, except during the epoch of the major ejection event of the southern jet.

\noindent{\bf Serpens.} Evidence for variability has been shown for Serpens~SMM~1 by \citet{scaife2011c} and references therein. \citet{1993ApJ...415..191C} suggest that the knotty structure in the NW and SE components is the result of a discrete ejection of material (\textquotedblleft{bullets}\textquotedblright) from the source, which could account for the radio flare event in the archival data \citet{scaife2011c} that appears to have faded. The observations at 1.9\,cm for SMM~1 presented in this work were taken 11~October~2011 and the flux density (see Table~\ref{tab:intfluxes}) is a factor of $\approx1.5$ higher than that presented in \citet{scaife2011c} who measure a flux density of $4.736\pm0.237$ at 1.8\,cm from observations taken between December~2010 and January~2011. Our measured flux density is within the uncertainty limits of the 15\,GHz flux density of $10\pm3$\,mJy found by \citet{1986ApJ...303..683S}. \citet{2009ApJ...707..103E} detect a high mass disc around this source which should be unstable and therefore exhibit luminosity variations. 

\noindent{\bf HH~1} The Herbig-Haro object HH~1 has been previously detected at centimetre wavelengths \citep[e.g.][]{1990ApJ...352..645R, 2000AJ....119..882R} however, here we do not detect HH~1 at 1.9\,cm. We calculate a flux density upper limit of 0.34\,mJy for this object, inconsistent with the value of 0.78\,mJy at 2\,cm found by \citet{1990ApJ...352..645R}. However, \citet{1985ApJ...293L..35P} also failed to detect HH~1 in the radio, suggesting that this source is variable. Indeed multi-wavelength observations of this object have been discussed by \citet{2011RMxAA..47..425R}, and HH~1 has been shown to be variable at radio wavelengths. Future observations with the AMI and eMERLIN at the same epoch will provide a better picture of the variability of these objects. 

\section{Conclusions}

We have presented 16\,GHz observations made with AMI of a sample of low-mass young stars driving jets. 

\begin{itemize}

\item The data presented in this work provides support for free-free emission as the dominant mechanism for producing the radio luminosity from these YSOs. Across the AMI channel bands 80~per~cent of the detected protostellar sources in the target sample have spectral indices $-0.1\leq\alpha_{\rm AMI}\leq2$ consistent with free-free emission and we find an average spectral index $\overline{\alpha_{\rm AMI}}=0.42\pm0.59$. Although the flux densities integrated over the whole of the source are in general indicative of free-free emission, it could be possible that subcomponents within the source show non-thermal (i.e. negative) spectra.

\item We have also presented and examined spectral energy distributions for each detected protostellar source, combining the AMI data with available archival data from an extensive literature search. We tested two scenarios to obtain maximum likelihood values for the spectral indices, dust opacity indices, and dust temperatures. In Scenario (i) we fix $T_{\rm d}$ for each source based on protostellar evolutionary class and find that 80~per~cent of the objects in the target sample have spectral indices consistent with free-free emission, and we calculate an average spectral index $\overline{\alpha'}=0.20\pm0.41$ consistent with the value for a well collimated outflow. In Scenario (ii) we do not fix $T_{\rm d}$, and we find the same average value for the spectral index. These values for the spectral indices are consistent with expectations as all target sources have known outflows. 

\item We examine the errors associated with determining the radio luminosity and find that the dominant source of error is the uncertainty on $\beta$.

\item In both scenarios we find that NGC~2264~G and L723 have $\alpha'\leq-0.1$, inconsistent with free-free emission. For these two sources there are other objects within close proximity (i.e. within the synthesized beam) that have non-thermal spectral indices that affect the integrated flux densities. Higher spatial resolution data can help constrain the emission mechanism for the driving sources of these outflows. We also find $\alpha'>2$ in some cases (e.g. NGC~1333~IRAS~2A)  suggesting that the lower frequency data is an extension of the greybody emission and that there is no free-free component detectable in these data.

\item We examine correlations between the radio luminosity and bolometric luminosity, envelope mass, and outflow force and find that these data are broadly consistent with correlations found by previous samples. We find a slight bias in the correlation with bolometric luminosity towards higher values of $L_{\rm bol}$.

\item Evidence for variability has been shown for L1551~IRS~5 and Serpens~SMM~1 by close inspection of their SEDs. It is apparent from their histories that they have undergone radio flare events, supporting non-steady accretion accretion models of protostellar evolution. We also find variability in the Herbig-Haro object HH~1.
\end{itemize}

\section{ACKNOWLEDGEMENTS}

We thank the staff of the Lord's Bridge Observatory for their invaluable assistance in the commissioning and operation of the Arcminute Microkelvin Imager. The AMI is supported by Cambridge University and the STFC. We thank the anonymous referee for their careful and constructive reading of our paper. RA would like to acknowledge support from Science Foundation Ireland under grant 11/RFP/AST3331.

\bibliographystyle{mn2e}
\bibliography{AMILAbib}

\begin{thebibliography}{163}
\expandafter\ifx\csname natexlab\endcsname\relax\def\natexlab#1{#1}\fi

\bibitem[{{Altenhoff} {et~al}\mbox{.}(1994){Altenhoff}, {Thum}, \&
  {Wendker}}]{1994AA...281..161A}
{Altenhoff} W.~J., {Thum} C., {Wendker} H.~J., 1994, \aap, 281, 161

\bibitem[{{AMI Consortium: Davies} {et~al}\mbox{.}(2011){AMI Consortium:
  Davies}, {Franzen}, {Waldram}, {Grainge}, {Hobson}, {Hurley-Walker},
  {Lasenby}, {Olamaie}, {Pooley}, {Riley}, {Rodriguez-Gonzalvez}, {Saunders},
  {Scaife}, {Schammel}, {Scott}, {Shimwell}, {Titterington}, \&
  {Zwart}}]{2010arXiv1012.3659D}
{AMI Consortium: Davies} M.~L. {et~al.}, 2011, \mnras, 415, 2708

\bibitem[{{AMI Consortium: Scaife} {et~al}\mbox{.}(2012{\natexlab{a}}){AMI
  Consortium: Scaife}, {Buckle}, {Ainsworth}, {Davies}, {Franzen}, {Grainge},
  {Hobson}, {Hurley-Walker}, {Lasenby}, {Olamaie}, {Perrott}, {Pooley}, {Ray},
  {Richer}, {Rodr{\'{\i}}guez-Gonz{\'a}lvez}, {Saunders}, {Schammel}, {Scott},
  {Shimwell}, {Titterington}, \& {Waldram}}]{scaife2011d}
{AMI Consortium: Scaife} A.~M.~M. {et~al.}, 2012{\natexlab{a}}, \mnras, 420,
  3334

\bibitem[{{AMI Consortium: Scaife} {et~al}\mbox{.}(2011{\natexlab{a}}){AMI
  Consortium: Scaife}, {Curtis}, {Davies}, {Franzen}, {Grainge}, {Hobson},
  {Hurley-Walker}, {Lasenby}, {Olamaie}, {Pooley},
  {Rodr{\'{\i}}guez-Gonz{\'a}lvez}, {Saunders}, {Schammel}, {Scott},
  {Shimwell}, {Titterington}, {Waldram}, \& {Zwart}}]{scaife2011a}
{AMI Consortium: Scaife} A.~M.~M. {et~al.}, 2011{\natexlab{a}}, \mnras, 410,
  2662

\bibitem[{{AMI Consortium: Scaife} {et~al}\mbox{.}(2012{\natexlab{b}}){AMI
  Consortium: Scaife}, {Hatchell}, {Ainsworth}, {Buckle}, {Davies}, {Franzen},
  {Grainge}, {Hobson}, {Hurley-Walker}, {Lasenby}, {Olamaie}, {Perrott},
  {Pooley}, {Richer}, {Rodr{\'{\i}}guez-Gonz{\'a}lvez}, {Saunders}, {Schammel},
  {Scott}, {Shimwell}, {Titterington}, \& {Waldram}}]{scaife2011c}
{AMI Consortium: Scaife} A.~M.~M. {et~al.}, 2012{\natexlab{b}}, \mnras, 420,
  1019

\bibitem[{{AMI Consortium: Scaife} {et~al}\mbox{.}(2011{\natexlab{b}}){AMI
  Consortium: Scaife}, {Hatchell}, {Davies}, {Franzen}, {Grainge}, {Hobson},
  {Hurley-Walker}, {Lasenby}, {Olamaie}, {Perrott}, {Pooley},
  {Rodr{\'{\i}}guez-Gonz{\'a}lvez}, {Saunders}, {Schammel}, {Scott},
  {Shimwell}, {Titterington}, \& {Waldram}}]{scaife2011b}
{AMI Consortium: Scaife} A.~M.~M. {et~al.}, 2011{\natexlab{b}}, \mnras, 415,
  893

\bibitem[{{AMI Consortium: Zwart} {et~al}\mbox{.}(2008){AMI Consortium: Zwart},
  {Barker}, {Biddulph}, {Bly}, {Boysen}, {Brown}, {Clementson}, {Crofts},
  {Culverhouse}, {Czeres}, {Dace}, {Davies}, {D'Alessandro}, {Doherty},
  {Duggan}, {Ely}, {Felvus}, {Feroz}, {Flynn}, {Franzen}, {Geisb{\"u}sch},
  {G{\'e}nova-Santos}, {Grainge}, {Grainger}, {Hammett}, {Hills}, {Hobson},
  {Holler}, {Hurley-Walker}, {Jilley}, {Jones}, {Kaneko}, {Kneissl},
  {Lancaster}, {Lasenby}, {Marshall}, {Newton}, {Norris}, {Northrop}, {Odell},
  {Petencin}, {Pober}, {Pooley}, {Pospieszalski}, {Quy},
  {Rodr{\'{\i}}guez-Gonz{\'a}lvez}, {Saunders}, {Scaife}, {Schofield}, {Scott},
  {Shaw}, {Shimwell}, {Smith}, {Taylor}, {Titterington}, {Veli{\'c}},
  {Waldram}, {West}, {Wood}, {Yassin}, \& {AMI
  Consortium}}]{2008MNRAS.391.1545Z}
{AMI Consortium: Zwart} J.~T.~L. {et~al.}, 2008, \mnras, 391, 1545

\bibitem[{{Andr{\'e}} {et~al}\mbox{.}(1993){Andr{\'e}}, {Ward-Thompson}, \&
  {Barsony}}]{1993ApJ...406..122A}
{Andr{\'e}} P., {Ward-Thompson} D., {Barsony} M., 1993, \apj, 406, 122

\bibitem[{{Andr{\'e}} {et~al}\mbox{.}(2000){Andr{\'e}}, {Ward-Thompson}, \&
  {Barsony}}]{2000prpl.conf...59A}
{Andr{\'e}} P., {Ward-Thompson} D., {Barsony} M., 2000, Protostars and Planets
  IV, 59

\bibitem[{{Andrews} \& {Williams}(2005)}]{2005ApJ...631.1134A}
{Andrews} S.~M., {Williams} J.~P., 2005, \apj, 631, 1134

\bibitem[{{Anglada}(1995)}]{1995RMxAC...1...67A}
{Anglada} G., 1995, in Revista Mexicana de Astronomia y Astrofisica, vol. 27,
  Vol.~1, Revista Mexicana de Astronomia y Astrofisica Conference Series,
  {S.~Lizano \& J.~M.~Torrelles}, ed., p.~67

\bibitem[{{Anglada} {et~al}\mbox{.}(1991){Anglada}, {Estalella},
  {Rodr{\'{\i}}guez}, {Torrelles}, {Lopez}, \& {Canto}}]{1991ApJ...376..615A}
{Anglada} G., {Estalella} R., {Rodr{\'{\i}}guez} L.~F., {Torrelles} J.~M.,
  {Lopez} R., {Canto} J., 1991, \apj, 376, 615

\bibitem[{{Anglada} {et~al}\mbox{.}(2004){Anglada}, {Rodr{\'{\i}}guez},
  {Osorio}, {Torrelles}, {Estalella}, {Beltr{\'a}n}, \&
  {Ho}}]{2004ApJ...605L.137A}
{Anglada} G., {Rodr{\'{\i}}guez} L.~F., {Osorio} M., {Torrelles} J.~M.,
  {Estalella} R., {Beltr{\'a}n} M.~T., {Ho} P.~T.~P., 2004, \apjl, 605, L137

\bibitem[{{Anglada} {et~al}\mbox{.}(1996){Anglada}, {Rodr{\'{\i}}guez}, \&
  {Torrelles}}]{1996ApJ...473L.123A}
{Anglada} G., {Rodr{\'{\i}}guez} L.~F., {Torrelles} J.~M., 1996, \apjl, 473,
  L123

\bibitem[{{Anglada} {et~al}\mbox{.}(1998){Anglada}, {Villuendas}, {Estalella},
  {Beltr{\'a}n}, {Rodr{\'{\i}}guez}, {Torrelles}, \&
  {Curiel}}]{1998AJ....116.2953A}
{Anglada} G., {Villuendas} E., {Estalella} R., {Beltr{\'a}n} M.~T.,
  {Rodr{\'{\i}}guez} L.~F., {Torrelles} J.~M., {Curiel} S., 1998, \aj, 116,
  2953

\bibitem[{{Antoniucci} {et~al}\mbox{.}(2008){Antoniucci}, {Nisini}, {Giannini},
  \& {Lorenzetti}}]{2008AA...479..503A}
{Antoniucci} S., {Nisini} B., {Giannini} T., {Lorenzetti} D., 2008, \aap, 479,
  503

\bibitem[{{Bachiller} {et~al}\mbox{.}(1991){Bachiller}, {Andre}, \&
  {Cabrit}}]{1991AA...241L..43B}
{Bachiller} R., {Andre} P., {Cabrit} S., 1991, \aap, 241, L43

\bibitem[{{Bachiller} {et~al}\mbox{.}(1995){Bachiller}, {Guilloteau}, {Dutrey},
  {Planesas}, \& {Martin-Pintado}}]{1995AA...299..857B}
{Bachiller} R., {Guilloteau} S., {Dutrey} A., {Planesas} P., {Martin-Pintado}
  J., 1995, \aap, 299, 857

\bibitem[{{Bachiller} {et~al}\mbox{.}(1990){Bachiller}, {Martin-Pintado},
  {Tafalla}, {Cernicharo}, \& {Lazareff}}]{1990AA...231..174B}
{Bachiller} R., {Martin-Pintado} J., {Tafalla} M., {Cernicharo} J., {Lazareff}
  B., 1990, \aap, 231, 174

\bibitem[{{Bally} {et~al}\mbox{.}(1997){Bally}, {Devine}, {Alten}, \&
  {Sutherland}}]{1997ApJ...478..603B}
{Bally} J., {Devine} D., {Alten} V., {Sutherland} R.~S., 1997, \apj, 478, 603

\bibitem[{{Barsony} \& {Chandler}(1993)}]{1993ApJ...406L..71B}
{Barsony} M., {Chandler} C.~J., 1993, \apjl, 406, L71

\bibitem[{{Barsony} {et~al}\mbox{.}(1998){Barsony}, {Ward-Thompson},
  {Andr{\'e}}, \& {O'Linger}}]{1998ApJ...509..733B}
{Barsony} M., {Ward-Thompson} D., {Andr{\'e}} P., {O'Linger} J., 1998, \apj,
  509, 733

\bibitem[{{Beltr{\'a}n} {et~al}\mbox{.}(2001){Beltr{\'a}n}, {Estalella},
  {Anglada}, {Rodr{\'{\i}}guez}, \& {Torrelles}}]{2001AJ....121.1556B}
{Beltr{\'a}n} M.~T., {Estalella} R., {Anglada} G., {Rodr{\'{\i}}guez} L.~F.,
  {Torrelles} J.~M., 2001, \aj, 121, 1556

\bibitem[{{Bieging} \& {Cohen}(1985)}]{1985ApJ...289L...5B}
{Bieging} J.~H., {Cohen} M., 1985, \apjl, 289, L5

\bibitem[{{Bieging} {et~al}\mbox{.}(1984){Bieging}, {Cohen}, \&
  {Schwartz}}]{1984ApJ...282..699B}
{Bieging} J.~H., {Cohen} M., {Schwartz} P.~R., 1984, \apj, 282, 699

\bibitem[{{Bontemps} {et~al}\mbox{.}(1996){Bontemps}, {Andr{\'e}}, {Terebey},
  \& {Cabrit}}]{1996AA...311..858B}
{Bontemps} S., {Andr{\'e}} P., {Terebey} S., {Cabrit} S., 1996, \aap, 311, 858

\bibitem[{{Bontemps} {et~al}\mbox{.}(1995){Bontemps}, {Andre}, \&
  {Ward-Thompson}}]{1995AA...297...98B}
{Bontemps} S., {Andre} P., {Ward-Thompson} D., 1995, \aap, 297, 98

\bibitem[{{Cabrit} \& {Andre}(1991)}]{1991ApJ...379L..25C}
{Cabrit} S., {Andre} P., 1991, \apjl, 379, L25

\bibitem[{{Caratti o Garatti} {et~al}\mbox{.}(2006){Caratti o Garatti},
  {Giannini}, {Nisini}, \& {Lorenzetti}}]{2006AA...449.1077C}
{Caratti o Garatti} A., {Giannini} T., {Nisini} B., {Lorenzetti} D., 2006,
  \aap, 449, 1077

\bibitem[{{Carkner} {et~al}\mbox{.}(1997){Carkner}, {Mamajek}, {Feigelson},
  {Neuhauser}, {Wichmann}, \& {Krautter}}]{1997ApJ...490..735C}
{Carkner} L., {Mamajek} E., {Feigelson} E., {Neuhauser} R., {Wichmann} R.,
  {Krautter} J., 1997, \apj, 490, 735

\bibitem[{{Carrasco-Gonz{\'a}lez} {et~al}\mbox{.}(2008){Carrasco-Gonz{\'a}lez},
  {Anglada}, {Rodr{\'{\i}}guez}, {Torrelles}, {Osorio}, \&
  {Girart}}]{2008ApJ...676.1073C}
{Carrasco-Gonz{\'a}lez} C., {Anglada} G., {Rodr{\'{\i}}guez} L.~F., {Torrelles}
  J.~M., {Osorio} M., {Girart} J.~M., 2008, \apj, 676, 1073

\bibitem[{{Carrasco-Gonz{\'a}lez}
  {et~al}\mbox{.}(2010{\natexlab{a}}){Carrasco-Gonz{\'a}lez},
  {Rodr{\'{\i}}guez}, {Anglada}, {Mart{\'{\i}}}, {Torrelles}, \&
  {Osorio}}]{2010Sci...330.1209C}
{Carrasco-Gonz{\'a}lez} C., {Rodr{\'{\i}}guez} L.~F., {Anglada} G.,
  {Mart{\'{\i}}} J., {Torrelles} J.~M., {Osorio} M., 2010{\natexlab{a}},
  Science, 330, 1209

\bibitem[{{Carrasco-Gonz{\'a}lez}
  {et~al}\mbox{.}(2010{\natexlab{b}}){Carrasco-Gonz{\'a}lez},
  {Rodr{\'{\i}}guez}, {Torrelles}, {Anglada}, \&
  {Gonz{\'a}lez-Mart{\'{\i}}n}}]{2010AJ....139.2433C}
{Carrasco-Gonz{\'a}lez} C., {Rodr{\'{\i}}guez} L.~F., {Torrelles} J.~M.,
  {Anglada} G., {Gonz{\'a}lez-Mart{\'{\i}}n} O., 2010{\natexlab{b}}, \aj, 139,
  2433

\bibitem[{{Casali} {et~al}\mbox{.}(1993){Casali}, {Eiroa}, \&
  {Duncan}}]{1993AA...275..195C}
{Casali} M.~M., {Eiroa} C., {Duncan} W.~D., 1993, \aap, 275, 195

\bibitem[{{Chandler} \& {Richer}(2000)}]{2000ApJ...530..851C}
{Chandler} C.~J., {Richer} J.~S., 2000, \apj, 530, 851

\bibitem[{{Chen} {et~al}\mbox{.}(1995){Chen}, {Myers}, {Ladd}, \&
  {Wood}}]{1995ApJ...445..377C}
{Chen} H., {Myers} P.~C., {Ladd} E.~F., {Wood} D.~O.~S., 1995, \apj, 445, 377

\bibitem[{{Chen} {et~al}\mbox{.}(2009){Chen}, {Launhardt}, \&
  {Henning}}]{2009ApJ...691.1729C}
{Chen} X., {Launhardt} R., {Henning} T., 2009, \apj, 691, 1729

\bibitem[{{Choi}(2009)}]{2009ApJ...705.1730C}
{Choi} M., 2009, \apj, 705, 1730

\bibitem[{{Cohen} {et~al}\mbox{.}(1982){Cohen}, {Bieging}, \&
  {Schwartz}}]{1982ApJ...253..707C}
{Cohen} M., {Bieging} J.~H., {Schwartz} P.~R., 1982, \apj, 253, 707

\bibitem[{{Condon} {et~al}\mbox{.}(1998){Condon}, {Cotton}, {Greisen}, {Yin},
  {Perley}, {Taylor}, \& {Broderick}}]{1998AJ....115.1693C}
{Condon} J.~J., {Cotton} W.~D., {Greisen} E.~W., {Yin} Q.~F., {Perley} R.~A.,
  {Taylor} G.~B., {Broderick} J.~J., 1998, \aj, 115, 1693

\bibitem[{{Curiel} {et~al}\mbox{.}(1990){Curiel}, {Raymond}, {Moran},
  {Rodr{\'{\i}}guez}, \& {Canto}}]{1990ApJ...365L..85C}
{Curiel} S., {Raymond} J.~C., {Moran} J.~M., {Rodr{\'{\i}}guez} L.~F., {Canto}
  J., 1990, \apjl, 365, L85

\bibitem[{{Curiel} {et~al}\mbox{.}(1989){Curiel}, {Rodriguez}, {Bohigas},
  {Roth}, {Canto}, \& {Torrelles}}]{1989ApL&C..27..299C}
{Curiel} S., {Rodriguez} L.~F., {Bohigas} J., {Roth} M., {Canto} J.,
  {Torrelles} J.~M., 1989, Astrophysical Letters and Communications, 27, 299

\bibitem[{{Curiel} {et~al}\mbox{.}(1993){Curiel}, {Rodr{\'{\i}}guez}, {Moran},
  \& {Canto}}]{1993ApJ...415..191C}
{Curiel} S., {Rodr{\'{\i}}guez} L.~F., {Moran} J.~M., {Canto} J., 1993, \apj,
  415, 191

\bibitem[{{Curiel} {et~al}\mbox{.}(1999){Curiel}, {Torrelles},
  {Rodr{\'{\i}}guez}, {G{\'o}mez}, \& {Anglada}}]{1999ApJ...527..310C}
{Curiel} S., {Torrelles} J.~M., {Rodr{\'{\i}}guez} L.~F., {G{\'o}mez} J.~F.,
  {Anglada} G., 1999, \apj, 527, 310

\bibitem[{{Davis} {et~al}\mbox{.}(1999){Davis}, {Matthews}, {Ray}, {Dent}, \&
  {Richer}}]{1999MNRAS.309..141D}
{Davis} C.~J., {Matthews} H.~E., {Ray} T.~P., {Dent} W.~R.~F., {Richer} J.~S.,
  1999, \mnras, 309, 141

\bibitem[{{Davis} {et~al}\mbox{.}(1997){Davis}, {Ray}, {Eisloeffel}, \&
  {Corcoran}}]{1997AA...324..263D}
{Davis} C.~J., {Ray} T.~P., {Eisloeffel} J., {Corcoran} D., 1997, \aap, 324,
  263

\bibitem[{{Davis} \& {Smith}(1995)}]{1995ApJ...443L..41D}
{Davis} C.~J., {Smith} M.~D., 1995, \apjl, 443, L41

\bibitem[{{de Zotti} {et~al}\mbox{.}(2005){de Zotti}, {Ricci}, {Mesa}, {Silva},
  {Mazzotta}, {Toffolatti}, \& {Gonz{\'a}lez-Nuevo}}]{2005AA...431..893D}
{de Zotti} G., {Ricci} R., {Mesa} D., {Silva} L., {Mazzotta} P., {Toffolatti}
  L., {Gonz{\'a}lez-Nuevo} J., 2005, \aap, 431, 893

\bibitem[{{Dent} {et~al}\mbox{.}(1998){Dent}, {Matthews}, \&
  {Ward-Thompson}}]{1998MNRAS.301.1049D}
{Dent} W.~R.~F., {Matthews} H.~E., {Ward-Thompson} D., 1998, \mnras, 301, 1049

\bibitem[{{Devine} {et~al}\mbox{.}(1999){Devine}, {Reipurth}, \&
  {Bally}}]{1999AJ....118..972D}
{Devine} D., {Reipurth} B., {Bally} J., 1999, \aj, 118, 972

\bibitem[{{di Francesco} {et~al}\mbox{.}(2008){di Francesco}, {Johnstone},
  {Kirk}, {MacKenzie}, \& {Ledwosinska}}]{2008ApJS..175..277D}
{di Francesco} J., {Johnstone} D., {Kirk} H., {MacKenzie} T., {Ledwosinska} E.,
  2008, \apjs, 175, 277

\bibitem[{{Duarte-Cabral} {et~al}\mbox{.}(2010){Duarte-Cabral}, {Fuller},
  {Peretto}, {Hatchell}, {Ladd}, {Buckle}, {Richer}, \&
  {Graves}}]{2010AA...519A..27D}
{Duarte-Cabral} A., {Fuller} G.~A., {Peretto} N., {Hatchell} J., {Ladd} E.~F.,
  {Buckle} J., {Richer} J., {Graves} S.~F., 2010, \aap, 519, A27

\bibitem[{{Duncan} {et~al}\mbox{.}(1987){Duncan}, {Forster}, {Gardner}, \&
  {Whiteoak}}]{1987MNRAS.224..721D}
{Duncan} R.~A., {Forster} J.~R., {Gardner} F.~F., {Whiteoak} J.~B., 1987,
  \mnras, 224, 721

\bibitem[{{Dunham} {et~al}\mbox{.}(2008){Dunham}, {Crapsi}, {Evans}, {Bourke},
  {Huard}, {Myers}, \& {Kauffmann}}]{2008ApJS..179..249D}
{Dunham} M.~M., {Crapsi} A., {Evans}, II N.~J., {Bourke} T.~L., {Huard} T.~L.,
  {Myers} P.~C., {Kauffmann} J., 2008, \apjs, 179, 249

\bibitem[{{Eiroa} {et~al}\mbox{.}(2005){Eiroa}, {Torrelles}, {Curiel}, \&
  {Djupvik}}]{2005AJ....130..643E}
{Eiroa} C., {Torrelles} J.~M., {Curiel} S., {Djupvik} A.~A., 2005, \aj, 130,
  643

\bibitem[{{Emerson} {et~al}\mbox{.}(1984){Emerson}, {Harris}, {Jennings},
  {Beichman}, {Baud}, {Beintema}, {Wesselius}, \&
  {Marsden}}]{1984ApJ...278L..49E}
{Emerson} J.~P., {Harris} S., {Jennings} R.~E., {Beichman} C.~A., {Baud} B.,
  {Beintema} D.~A., {Wesselius} P.~R., {Marsden} P.~L., 1984, \apjl, 278, L49

\bibitem[{{Enoch} {et~al}\mbox{.}(2009{\natexlab{a}}){Enoch}, {Corder},
  {Dunham}, \& {Duch{\^e}ne}}]{2009ApJ...707..103E}
{Enoch} M.~L., {Corder} S., {Dunham} M.~M., {Duch{\^e}ne} G.,
  2009{\natexlab{a}}, \apj, 707, 103

\bibitem[{{Enoch} {et~al}\mbox{.}(2009{\natexlab{b}}){Enoch}, {Evans},
  {Sargent}, \& {Glenn}}]{2009ApJ...692..973E}
{Enoch} M.~L., {Evans}, II N.~J., {Sargent} A.~I., {Glenn} J.,
  2009{\natexlab{b}}, \apj, 692, 973

\bibitem[{{Enoch} {et~al}\mbox{.}(2006){Enoch}, {Young}, {Glenn}, {Evans},
  {Golwala}, {Sargent}, {Harvey}, {Aguirre}, {Goldin}, {Haig}, {Huard},
  {Lange}, {Laurent}, {Maloney}, {Mauskopf}, {Rossinot}, \&
  {Sayers}}]{2006ApJ...638..293E}
{Enoch} M.~L. {et~al.}, 2006, \apj, 638, 293

\bibitem[{{Evans} {et~al}\mbox{.}(1987){Evans}, {Levreault}, {Beckwith}, \&
  {Skrutskie}}]{1987ApJ...320..364E}
{Evans}, II N.~J., {Levreault} R.~M., {Beckwith} S., {Skrutskie} M., 1987,
  \apj, 320, 364

\bibitem[{{Feigelson} {et~al}\mbox{.}(1998){Feigelson}, {Carkner}, \&
  {Wilking}}]{1998ApJ...494L.215F}
{Feigelson} E.~D., {Carkner} L., {Wilking} B.~A., 1998, \apjl, 494, L215+

\bibitem[{{Froebrich}(2005)}]{2005ApJS..156..169F}
{Froebrich} D., 2005, \apjs, 156, 169

\bibitem[{{Gezari} {et~al}\mbox{.}(1999){Gezari}, {Pitts}, \&
  {Schmitz}}]{1999yCat.2225....0G}
{Gezari} D.~Y., {Pitts} P.~S., {Schmitz} M., 1999, VizieR Online Data Catalog,
  2225, 0

\bibitem[{{Gibb}(1999)}]{1999MNRAS.304....1G}
{Gibb} A.~G., 1999, \mnras, 304, 1

\bibitem[{{Gibb} \& {Heaton}(1993)}]{1993AA...276..511G}
{Gibb} A.~G., {Heaton} B.~D., 1993, \aap, 276, 511

\bibitem[{{Girart} {et~al}\mbox{.}(1997){Girart}, {Estalella}, {Anglada},
  {Torrelles}, {Ho}, \& {Rodr{\'{\i}}guez}}]{1997ApJ...489..734G}
{Girart} J.~M., {Estalella} R., {Anglada} G., {Torrelles} J.~M., {Ho} P.~T.~P.,
  {Rodr{\'{\i}}guez} L.~F., 1997, \apj, 489, 734

\bibitem[{{G{\'o}mez} {et~al}\mbox{.}(1994){G{\'o}mez}, {Curiel}, {Torrelles},
  {Rodr{\'{\i}}guez}, {Anglada}, \& {Girart}}]{1994ApJ...436..749G}
{G{\'o}mez} J.~F., {Curiel} S., {Torrelles} J.~M., {Rodr{\'{\i}}guez} L.~F.,
  {Anglada} G., {Girart} J.~M., 1994, \apj, 436, 749

\bibitem[{{Gomez} {et~al}\mbox{.}(1993){Gomez}, {Torrelles}, {Ho},
  {Rodr{\'{\i}}guez}, \& {Canto}}]{1993ApJ...414..333G}
{Gomez} J.~F., {Torrelles} J.~M., {Ho} P.~T.~P., {Rodr{\'{\i}}guez} L.~F.,
  {Canto} J., 1993, \apj, 414, 333

\bibitem[{{Gomez} {et~al}\mbox{.}(1997){Gomez}, {Whitney}, \&
  {Kenyon}}]{1997AJ....114.1138G}
{Gomez} M., {Whitney} B.~A., {Kenyon} S.~J., 1997, \aj, 114, 1138

\bibitem[{{Gramajo} {et~al}\mbox{.}(2010){Gramajo}, {Whitney}, {G{\'o}mez}, \&
  {Robitaille}}]{2010AJ....139.2504G}
{Gramajo} L.~V., {Whitney} B.~A., {G{\'o}mez} M., {Robitaille} T.~P., 2010,
  \aj, 139, 2504

\bibitem[{{Gredel} \& {Reipurth}(1994)}]{1994AA...289L..19G}
{Gredel} R., {Reipurth} B., 1994, \aap, 289, L19

\bibitem[{{Guilloteau} {et~al}\mbox{.}(1992){Guilloteau}, {Bachiller},
  {Fuente}, \& {Lucas}}]{1992AA...265L..49G}
{Guilloteau} S., {Bachiller} R., {Fuente} A., {Lucas} R., 1992, \aap, 265, L49

\bibitem[{{Hatchell} {et~al}\mbox{.}(2007{\natexlab{a}}){Hatchell}, {Fuller},
  \& {Richer}}]{hat2007b}
{Hatchell} J., {Fuller} G.~A., {Richer} J.~S., 2007{\natexlab{a}}, \aap, 472,
  187

\bibitem[{{Hatchell} {et~al}\mbox{.}(2007{\natexlab{b}}){Hatchell}, {Fuller},
  {Richer}, {Harries}, \& {Ladd}}]{hat2007a}
{Hatchell} J., {Fuller} G.~A., {Richer} J.~S., {Harries} T.~J., {Ladd} E.~F.,
  2007{\natexlab{b}}, \aap, 468, 1009

\bibitem[{{Herranz} {et~al}\mbox{.}(2006){Herranz}, {Sanz}, {Lopez-Caniego}, \&
  {Gonzalez-Nuevo}}]{2006spit.conf..541H}
{Herranz} D., {Sanz} J.~L., {Lopez-Caniego} M., {Gonzalez-Nuevo} J., 2006, in
  2006 IEEE International Symposium on Signal Processing and Information=
  Technology, vol. 1, p. 541-544, pp. 541--544

\bibitem[{{Hobson} \& {Baldwin}(2004)}]{hob04}
{Hobson} M.~P., {Baldwin} J.~E., 2004, \ao, 43, 2651

\bibitem[{{Hogerheijde} {et~al}\mbox{.}(1999){Hogerheijde}, {van Dishoeck},
  {Salverda}, \& {Blake}}]{1999ApJ...513..350H}
{Hogerheijde} M.~R., {van Dishoeck} E.~F., {Salverda} J.~M., {Blake} G.~A.,
  1999, \apj, 513, 350

\bibitem[{{Hurt} \& {Barsony}(1996)}]{1996ApJ...460L..45H}
{Hurt} R.~L., {Barsony} M., 1996, \apjl, 460, L45

\bibitem[{{Jennings} {et~al}\mbox{.}(1987){Jennings}, {Cameron}, {Cudlip}, \&
  {Hirst}}]{1987MNRAS.226..461J}
{Jennings} R.~E., {Cameron} D.~H.~M., {Cudlip} W., {Hirst} C.~J., 1987, \mnras,
  226, 461

\bibitem[{{J{\o}rgensen} {et~al}\mbox{.}(2004){J{\o}rgensen}, {Hogerheijde},
  {van Dishoeck}, {Blake}, \& {Sch{\"o}ier}}]{2004AA...413..993J}
{J{\o}rgensen} J.~K., {Hogerheijde} M.~R., {van Dishoeck} E.~F., {Blake} G.~A.,
  {Sch{\"o}ier} F.~L., 2004, \aap, 413, 993

\bibitem[{{J{\o}rgensen} {et~al}\mbox{.}(2009){J{\o}rgensen}, {van Dishoeck},
  {Visser}, {Bourke}, {Wilner}, {Lommen}, {Hogerheijde}, \&
  {Myers}}]{2009AA...507..861J}
{J{\o}rgensen} J.~K., {van Dishoeck} E.~F., {Visser} R., {Bourke} T.~L.,
  {Wilner} D.~J., {Lommen} D., {Hogerheijde} M.~R., {Myers} P.~C., 2009, \aap,
  507, 861

\bibitem[{{Keene} \& {Masson}(1990)}]{1990ApJ...355..635K}
{Keene} J., {Masson} C.~R., 1990, \apj, 355, 635

\bibitem[{{Keene} \& {Masson}(1986)}]{1986BAAS...18..973K}
{Keene} J.~B., {Masson} C.~R., 1986, in Bulletin of the American Astronomical
  Society, Vol.~18, Bulletin of the American Astronomical Society, p. 973

\bibitem[{{Kenyon} {et~al}\mbox{.}(1994){Kenyon}, {Dobrzycka}, \&
  {Hartmann}}]{1994AJ....108.1872K}
{Kenyon} S.~J., {Dobrzycka} D., {Hartmann} L., 1994, \aj, 108, 1872

\bibitem[{{Kenyon} \& {Hartmann}(1995)}]{1995ApJS..101..117K}
{Kenyon} S.~J., {Hartmann} L., 1995, \apjs, 101, 117

\bibitem[{{Kirk} {et~al}\mbox{.}(2007){Kirk}, {Johnstone}, \&
  {Tafalla}}]{2007ApJ...668.1042K}
{Kirk} H., {Johnstone} D., {Tafalla} M., 2007, \apj, 668, 1042

\bibitem[{{Kun} {et~al}\mbox{.}(2009){Kun}, {Balog}, {Kenyon}, {Mamajek}, \&
  {Gutermuth}}]{2009ApJS..185..451K}
{Kun} M., {Balog} Z., {Kenyon} S.~J., {Mamajek} E.~E., {Gutermuth} R.~A., 2009,
  \apjs, 185, 451

\bibitem[{{Lada}(1987)}]{1987IAUS..115....1L}
{Lada} C.~J., 1987, in IAU Symposium, Vol. 115, Star Forming Regions,
  {M.~Peimbert \& J.~Jugaku}, ed., pp. 1--17

\bibitem[{{Lada} \& {Fich}(1996)}]{1996ApJ...459..638L}
{Lada} C.~J., {Fich} M., 1996, \apj, 459, 638

\bibitem[{{Lee} {et~al}\mbox{.}(2002){Lee}, {Mundy}, {Stone}, \&
  {Ostriker}}]{2002ApJ...576..294L}
{Lee} C.-F., {Mundy} L.~G., {Stone} J.~M., {Ostriker} E.~C., 2002, \apj, 576,
  294

\bibitem[{{Lim} \& {Takakuwa}(2006)}]{2006ApJ...653..425L}
{Lim} J., {Takakuwa} S., 2006, \apj, 653, 425

\bibitem[{{Lis} {et~al}\mbox{.}(1999){Lis}, {Menten}, \&
  {Zylka}}]{1999ApJ...527..856L}
{Lis} D.~C., {Menten} K.~M., {Zylka} R., 1999, \apj, 527, 856

\bibitem[{{Loinard} {et~al}\mbox{.}(2002){Loinard}, {Rodr{\'{\i}}guez},
  {D'Alessio}, {Wilner}, \& {Ho}}]{2002ApJ...581L.109L}
{Loinard} L., {Rodr{\'{\i}}guez} L.~F., {D'Alessio} P., {Wilner} D.~J., {Ho}
  P.~T.~P., 2002, \apjl, 581, L109

\bibitem[{{Looney} {et~al}\mbox{.}(2000){Looney}, {Mundy}, \&
  {Welch}}]{2000ApJ...529..477L}
{Looney} L.~W., {Mundy} L.~G., {Welch} W.~J., 2000, \apj, 529, 477

\bibitem[{{Margulis} \& {Lada}(1986)}]{1986ApJ...309L..87M}
{Margulis} M., {Lada} C.~J., 1986, \apjl, 309, L87

\bibitem[{{Margulis} {et~al}\mbox{.}(1988){Margulis}, {Lada}, \&
  {Snell}}]{1988ApJ...333..316M}
{Margulis} M., {Lada} C.~J., {Snell} R.~L., 1988, \apj, 333, 316

\bibitem[{{McMullin} {et~al}\mbox{.}(1994){McMullin}, {Mundy}, {Wilking},
  {Hezel}, \& {Blake}}]{1994ApJ...424..222M}
{McMullin} J.~P., {Mundy} L.~G., {Wilking} B.~A., {Hezel} T., {Blake} G.~A.,
  1994, \apj, 424, 222

\bibitem[{{Meehan} {et~al}\mbox{.}(1998){Meehan}, {Wilking}, {Claussen},
  {Mundy}, \& {Wootten}}]{1998AJ....115.1599M}
{Meehan} L.~S.~G., {Wilking} B.~A., {Claussen} M.~J., {Mundy} L.~G., {Wootten}
  A., 1998, \aj, 115, 1599

\bibitem[{{Melis} {et~al}\mbox{.}(2011){Melis}, {Duch{\^e}ne}, {Chomiuk},
  {Palmer}, {Perrin}, {Maddison}, {M{\'e}nard}, {Stapelfeldt}, {Pinte}, \&
  {Duvert}}]{2011ApJ...739L...7M}
{Melis} C. {et~al.}, 2011, \apjl, 739, L7

\bibitem[{{Menten} {et~al}\mbox{.}(2007){Menten}, {Reid}, {Forbrich}, \&
  {Brunthaler}}]{2007AA...474..515M}
{Menten} K.~M., {Reid} M.~J., {Forbrich} J., {Brunthaler} A., 2007, \aap, 474,
  515

\bibitem[{{Morgan} {et~al}\mbox{.}(1990){Morgan}, {Snell}, \&
  {Strom}}]{1990ApJ...362..274M}
{Morgan} J.~A., {Snell} R.~L., {Strom} K.~M., 1990, \apj, 362, 274

\bibitem[{{Moriarty-Schieven} {et~al}\mbox{.}(1995){Moriarty-Schieven},
  {Butner}, \& {Wannier}}]{1995ApJ...445L..55M}
{Moriarty-Schieven} G.~H., {Butner} H.~M., {Wannier} P.~G., 1995, \apjl, 445,
  L55

\bibitem[{{Moriarty-Schieven} {et~al}\mbox{.}(2006){Moriarty-Schieven},
  {Johnstone}, {Bally}, \& {Jenness}}]{2006ApJ...645..357M}
{Moriarty-Schieven} G.~H., {Johnstone} D., {Bally} J., {Jenness} T., 2006,
  \apj, 645, 357

\bibitem[{{Moriarty-Schieven} {et~al}\mbox{.}(2000){Moriarty-Schieven},
  {Powers}, {Butner}, {Wannier}, \& {Keene}}]{2000ApJ...533L.143M}
{Moriarty-Schieven} G.~H., {Powers} J.~A., {Butner} H.~M., {Wannier} P.~G.,
  {Keene} J., 2000, \apjl, 533, L143

\bibitem[{{Motte} \& {Andr{\'e}}(2001)}]{2001AA...365..440M}
{Motte} F., {Andr{\'e}} P., 2001, \aap, 365, 440

\bibitem[{{Nisini} {et~al}\mbox{.}(2002){Nisini}, {Giannini}, \&
  {Lorenzetti}}]{2002ApJ...574..246N}
{Nisini} B., {Giannini} T., {Lorenzetti} D., 2002, \apj, 574, 246

\bibitem[{{Nutter} \& {Ward-Thompson}(2007)}]{2007MNRAS.374.1413N}
{Nutter} D., {Ward-Thompson} D., 2007, \mnras, 374, 1413

\bibitem[{{Ohashi} {et~al}\mbox{.}(1997){Ohashi}, {Hayashi}, {Ho}, \&
  {Momose}}]{1997ApJ...475..211O}
{Ohashi} N., {Hayashi} M., {Ho} P.~T.~P., {Momose} M., 1997, \apj, 475, 211

\bibitem[{{Ohashi} {et~al}\mbox{.}(1991){Ohashi}, {Kawabe}, {Ishiguro}, \&
  {Hayashi}}]{1991AJ....102.2054O}
{Ohashi} N., {Kawabe} R., {Ishiguro} M., {Hayashi} M., 1991, \aj, 102, 2054

\bibitem[{{Phillips} {et~al}\mbox{.}(2001){Phillips}, {Gibb}, \&
  {Little}}]{2001MNRAS.326..927P}
{Phillips} R.~R., {Gibb} A.~G., {Little} L.~T., 2001, \mnras, 326, 927

\bibitem[{{Pravdo} {et~al}\mbox{.}(1985){Pravdo}, {Rodr{\'{\i}}guez}, {Curiel},
  {Canto}, {Torrelles}, {Becker}, \& {Sellgren}}]{1985ApJ...293L..35P}
{Pravdo} S.~H., {Rodr{\'{\i}}guez} L.~F., {Curiel} S., {Canto} J., {Torrelles}
  J.~M., {Becker} R.~H., {Sellgren} K., 1985, \apjl, 293, L35

\bibitem[{{Raga} {et~al}\mbox{.}(2011){Raga}, {Reipurth}, {Cant{\'o}},
  {Sierra-Flores}, \& {Guzm{\'a}n}}]{2011RMxAA..47..425R}
{Raga} A.~C., {Reipurth} B., {Cant{\'o}} J., {Sierra-Flores} M.~M.,
  {Guzm{\'a}n} M.~V., 2011, \rmxaa, 47, 425

\bibitem[{{Ray} {et~al}\mbox{.}(1997){Ray}, {Muxlow}, {Axon}, {Brown},
  {Corcoran}, {Dyson}, \& {Mundt}}]{1997Natur.385..415R}
{Ray} T.~P., {Muxlow} T.~W.~B., {Axon} D.~J., {Brown} A., {Corcoran} D.,
  {Dyson} J., {Mundt} R., 1997, \nat, 385, 415

\bibitem[{{Reipurth}(1989)}]{1989Natur.340...42R}
{Reipurth} B., 1989, \nat, 340, 42

\bibitem[{{Reipurth} {et~al}\mbox{.}(1997){Reipurth}, {Bally}, \&
  {Devine}}]{1997AJ....114.2708R}
{Reipurth} B., {Bally} J., {Devine} D., 1997, \aj, 114, 2708

\bibitem[{{Reipurth} {et~al}\mbox{.}(1993){Reipurth}, {Chini}, {Krugel},
  {Kreysa}, \& {Sievers}}]{1993AA...273..221R}
{Reipurth} B., {Chini} R., {Krugel} E., {Kreysa} E., {Sievers} A., 1993, \aap,
  273, 221

\bibitem[{{Reipurth} {et~al}\mbox{.}(1992){Reipurth}, {Raga}, \&
  {Heathcote}}]{1992ApJ...392..145R}
{Reipurth} B., {Raga} A.~C., {Heathcote} S., 1992, \apj, 392, 145

\bibitem[{{Reipurth} {et~al}\mbox{.}(2002){Reipurth}, {Rodr{\'{\i}}guez},
  {Anglada}, \& {Bally}}]{2002AJ....124.1045R}
{Reipurth} B., {Rodr{\'{\i}}guez} L.~F., {Anglada} G., {Bally} J., 2002, \aj,
  124, 1045

\bibitem[{{Reipurth} {et~al}\mbox{.}(2004{\natexlab{a}}){Reipurth},
  {Rodr{\'{\i}}guez}, {Anglada}, \& {Bally}}]{2004AJ....127.1736R}
{Reipurth} B., {Rodr{\'{\i}}guez} L.~F., {Anglada} G., {Bally} J.,
  2004{\natexlab{a}}, \aj, 127, 1736

\bibitem[{{Reipurth} {et~al}\mbox{.}(2004{\natexlab{b}}){Reipurth},
  {Rodr{\'{\i}}guez}, {Anglada}, \& {Bally}}]{2004AJ...127.1736R}
{Reipurth} B., {Rodr{\'{\i}}guez} L.~F., {Anglada} G., {Bally} J.,
  2004{\natexlab{b}}, \aj, 127, 1736

\bibitem[{{Reipurth} {et~al}\mbox{.}(1999){Reipurth}, {Yu}, {Rodr{\'{\i}}guez},
  {Heathcote}, \& {Bally}}]{1999AA...352L..83R}
{Reipurth} B., {Yu} K.~C., {Rodr{\'{\i}}guez} L.~F., {Heathcote} S., {Bally}
  J., 1999, \aap, 352, L83

\bibitem[{{Reynolds}(1986)}]{1986ApJ...304..713R}
{Reynolds} S.~P., 1986, \apj, 304, 713

\bibitem[{{Rodr{\'{\i}}guez}(1995)}]{1995RMxAC...1....1R}
{Rodr{\'{\i}}guez} L.~F., 1995, in Revista Mexicana de Astronomia y
  Astrofisica, vol. 27, Vol.~1, Revista Mexicana de Astronomia y Astrofisica
  Conference Series, {S.~Lizano \& J.~M.~Torrelles}, ed., p.~1

\bibitem[{{Rodr{\'{\i}}guez}(1998)}]{1998RMxAC...7...14R}
{Rodr{\'{\i}}guez} L.~F., 1998, in Revista Mexicana de Astronomia y
  Astrofisica, vol. 27, Vol.~7, Revista Mexicana de Astronomia y Astrofisica
  Conference Series, {R.~J.~Dufour \& S.~Torres-Peimbert}, ed., p.~14

\bibitem[{{Rodr{\'{\i}}guez} {et~al}\mbox{.}(1997){Rodr{\'{\i}}guez},
  {Anglada}, \& {Curiel}}]{1997ApJ...480L.125R}
{Rodr{\'{\i}}guez} L.~F., {Anglada} G., {Curiel} S., 1997, \apjl, 480, L125

\bibitem[{{Rodr{\'{\i}}guez} {et~al}\mbox{.}(1999){Rodr{\'{\i}}guez},
  {Anglada}, \& {Curiel}}]{1999ApJS..125..427R}
{Rodr{\'{\i}}guez} L.~F., {Anglada} G., {Curiel} S., 1999, \apjs, 125, 427

\bibitem[{{Rodr{\'{\i}}guez}
  {et~al}\mbox{.}(1989{\natexlab{a}}){Rodr{\'{\i}}guez}, {Canto}, {Mirabel}, \&
  {Ruiz}}]{1989ApJ...337..712R}
{Rodr{\'{\i}}guez} L.~F., {Canto} J., {Mirabel} I.~F., {Ruiz} A.,
  1989{\natexlab{a}}, \apj, 337, 712

\bibitem[{{Rodr{\'{\i}}guez} {et~al}\mbox{.}(1986){Rodr{\'{\i}}guez}, {Canto},
  {Torrelles}, \& {Ho}}]{1986ApJ...301L..25R}
{Rodr{\'{\i}}guez} L.~F., {Canto} J., {Torrelles} J.~M., {Ho} P.~T.~P., 1986,
  \apjl, 301, L25

\bibitem[{{Rodr{\'{\i}}guez} \& {Curiel}(1989)}]{1989RMxAA..17..115R}
{Rodr{\'{\i}}guez} L.~F., {Curiel} S., 1989, \rmxaa, 17, 115

\bibitem[{{Rodr{\'{\i}}guez}
  {et~al}\mbox{.}(2003{\natexlab{a}}){Rodr{\'{\i}}guez}, {Curiel}, {Cant{\'o}},
  {Loinard}, {Raga}, \& {Torrelles}}]{2003ApJ...583..330R}
{Rodr{\'{\i}}guez} L.~F., {Curiel} S., {Cant{\'o}} J., {Loinard} L., {Raga}
  A.~C., {Torrelles} J.~M., 2003{\natexlab{a}}, \apj, 583, 330

\bibitem[{{Rodr{\'{\i}}guez}
  {et~al}\mbox{.}(1989{\natexlab{b}}){Rodr{\'{\i}}guez}, {Curiel}, {Moran},
  {Mirabel}, {Roth}, \& {Garay}}]{1989ApJ...346L..85R}
{Rodr{\'{\i}}guez} L.~F., {Curiel} S., {Moran} J.~M., {Mirabel} I.~F., {Roth}
  M., {Garay} G., 1989{\natexlab{b}}, \apjl, 346, L85

\bibitem[{{Rodr{\'{\i}}guez} {et~al}\mbox{.}(2000){Rodr{\'{\i}}guez},
  {Delgado-Arellano}, {G{\'o}mez}, {Reipurth}, {Torrelles}, {Noriega-Crespo},
  {Raga}, \& {Cant{\'o}}}]{2000AJ....119..882R}
{Rodr{\'{\i}}guez} L.~F., {Delgado-Arellano} V.~G., {G{\'o}mez} Y., {Reipurth}
  B., {Torrelles} J.~M., {Noriega-Crespo} A., {Raga} A.~C., {Cant{\'o}} J.,
  2000, \aj, 119, 882

\bibitem[{{Rodr{\'{\i}}guez} {et~al}\mbox{.}(1990){Rodr{\'{\i}}guez}, {Ho},
  {Torrelles}, {Curiel}, \& {Canto}}]{1990ApJ...352..645R}
{Rodr{\'{\i}}guez} L.~F., {Ho} P.~T.~P., {Torrelles} J.~M., {Curiel} S.,
  {Canto} J., 1990, \apj, 352, 645

\bibitem[{{Rodr{\'{\i}}guez}
  {et~al}\mbox{.}(2008{\natexlab{a}}){Rodr{\'{\i}}guez}, {Moran},
  {Franco-Hern{\'a}ndez}, {Garay}, {Brooks}, \&
  {Mardones}}]{2008AJ....135.2370R}
{Rodr{\'{\i}}guez} L.~F., {Moran} J.~M., {Franco-Hern{\'a}ndez} R., {Garay} G.,
  {Brooks} K.~J., {Mardones} D., 2008{\natexlab{a}}, \aj, 135, 2370

\bibitem[{{Rodr{\'{\i}}guez}
  {et~al}\mbox{.}(2003{\natexlab{b}}){Rodr{\'{\i}}guez}, {Porras}, {Claussen},
  {Curiel}, {Wilner}, \& {Ho}}]{2003ApJ...586L.137R}
{Rodr{\'{\i}}guez} L.~F., {Porras} A., {Claussen} M.~J., {Curiel} S., {Wilner}
  D.~J., {Ho} P.~T.~P., 2003{\natexlab{b}}, \apjl, 586, L137

\bibitem[{{Rodriguez} \& {Reipurth}(1994)}]{1994AA...281..882R}
{Rodriguez} L.~F., {Reipurth} B., 1994, \aap, 281, 882

\bibitem[{{Rodr{\'{\i}}guez} \& {Reipurth}(1998)}]{1998RMxAA..34...13R}
{Rodr{\'{\i}}guez} L.~F., {Reipurth} B., 1998, \rmxaa, 34, 13

\bibitem[{{Rodr{\'{\i}}guez}
  {et~al}\mbox{.}(2008{\natexlab{b}}){Rodr{\'{\i}}guez}, {Torrelles},
  {Anglada}, \& {Reipurth}}]{2008AJ....136.1852R}
{Rodr{\'{\i}}guez} L.~F., {Torrelles} J.~M., {Anglada} G., {Reipurth} B.,
  2008{\natexlab{b}}, \aj, 136, 1852

\bibitem[{{Rosvick} \& {Davidge}(1995)}]{1995PASP..107...49R}
{Rosvick} J.~M., {Davidge} T.~J., 1995, \pasp, 107, 49

\bibitem[{{S.~F. Graves et~al.}(2010)}]{2010MNRAS.409.1412G}
{S.~F. Graves et~al.}, 2010, \mnras, 409, 1412

\bibitem[{{Sandell} {et~al}\mbox{.}(1994){Sandell}, {Knee}, {Aspin}, {Robson},
  \& {Russell}}]{1994AA...285L...1S}
{Sandell} G., {Knee} L.~B.~G., {Aspin} C., {Robson} I.~E., {Russell} A.~P.~G.,
  1994, \aap, 285, L1

\bibitem[{{Sato} \& {Fukui}(1989)}]{1989ApJ...343..773S}
{Sato} F., {Fukui} Y., 1989, \apj, 343, 773

\bibitem[{{Schwartz} {et~al}\mbox{.}(1988){Schwartz}, {Gee}, \&
  {Huang}}]{1988ApJ...327..350S}
{Schwartz} P.~R., {Gee} G., {Huang} Y.-L., 1988, \apj, 327, 350

\bibitem[{{Shirley} {et~al}\mbox{.}(2002){Shirley}, {Evans}, \&
  {Rawlings}}]{2002ApJ...575..337S}
{Shirley} Y.~L., {Evans}, II N.~J., {Rawlings} J.~M.~C., 2002, \apj, 575, 337

\bibitem[{{Shirley} {et~al}\mbox{.}(2000){Shirley}, {Evans}, {Rawlings}, \&
  {Gregersen}}]{2000ApJS..131..249S}
{Shirley} Y.~L., {Evans}, II N.~J., {Rawlings} J.~M.~C., {Gregersen} E.~M.,
  2000, \apjs, 131, 249

\bibitem[{{Shirley} {et~al}\mbox{.}(2011){Shirley}, {Mason}, {Mangum}, {Bolin},
  {Devlin}, {Dicker}, \& {Korngut}}]{2011AJ....141...39S}
{Shirley} Y.~L., {Mason} B.~S., {Mangum} J.~G., {Bolin} D.~E., {Devlin} M.~J.,
  {Dicker} S.~R., {Korngut} P.~M., 2011, \aj, 141, 39

\bibitem[{{Snell} \& {Bally}(1986)}]{1986ApJ...303..683S}
{Snell} R.~L., {Bally} J., 1986, \apj, 303, 683

\bibitem[{{Snell} {et~al}\mbox{.}(1985){Snell}, {Bally}, {Strom}, \&
  {Strom}}]{1985ApJ...290..587S}
{Snell} R.~L., {Bally} J., {Strom} S.~E., {Strom} K.~M., 1985, \apj, 290, 587

\bibitem[{{Stapelfeldt} \& {Scoville}(1993)}]{1993ApJ...408..239S}
{Stapelfeldt} K.~R., {Scoville} N.~Z., 1993, \apj, 408, 239

\bibitem[{{Strai\v{z}ys} {et~al}\mbox{.}(1996){Strai\v{z}ys}, {\v{C}ernis}, \&
  {Barta\v{s}i\={u}t\.{e}}}]{1996BaltA...5..125S}
{Strai\v{z}ys} V., {\v{C}ernis} K., {Barta\v{s}i\={u}t\.{e}} S., 1996, Baltic
  Astronomy, 5, 125

\bibitem[{{Terebey} \& {Padgett}(1997)}]{1997IAUS..182..507T}
{Terebey} S., {Padgett} D.~L., 1997, in IAU Symposium, Vol. 182, Herbig-Haro
  Flows and the Birth of Stars, {B.~Reipurth \& C.~Bertout}, ed., pp. 507--514

\bibitem[{{Tobin} {et~al}\mbox{.}(2010){Tobin}, {Hartmann}, \&
  {Loinard}}]{2010ApJ...722L..12T}
{Tobin} J.~J., {Hartmann} L., {Loinard} L., 2010, \apjl, 722, L12

\bibitem[{{Torrelles} {et~al}\mbox{.}(1985){Torrelles}, {Ho},
  {Rodr{\'{\i}}guez}, \& {Canto}}]{1985ApJ...288..595T}
{Torrelles} J.~M., {Ho} P.~T.~P., {Rodr{\'{\i}}guez} L.~F., {Canto} J., 1985,
  \apj, 288, 595

\bibitem[{{van Kempen} {et~al}\mbox{.}(2009){van Kempen}, {van Dishoeck},
  {G{\"u}sten}, {Kristensen}, {Schilke}, {Hogerheijde}, {Boland}, {Menten}, \&
  {Wyrowski}}]{2009AA...507.1425V}
{van Kempen} T.~A. {et~al.}, 2009, \aap, 507, 1425

\bibitem[{{W.~J. Fischer et~al.}(2010)}]{2010AA...518L.122F}
{W.~J. Fischer et~al.}, 2010, \aap, 518, L122+

\bibitem[{{Ward-Thompson} {et~al}\mbox{.}(1995){Ward-Thompson}, {Eiroa}, \&
  {Casali}}]{1995MNRAS.273L..25W}
{Ward-Thompson} D., {Eiroa} C., {Casali} M.~M., 1995, \mnras, 273, L25

\bibitem[{{Wendker}(1995)}]{1995A&AS..109..177W}
{Wendker} H.~J., 1995, \aaps, 109, 177

\bibitem[{{Wilking} {et~al}\mbox{.}(1989){Wilking}, {Blackwell}, {Mundy}, \&
  {Howe}}]{1989ApJ...345..257W}
{Wilking} B.~A., {Blackwell} J.~H., {Mundy} L.~G., {Howe} J.~E., 1989, \apj,
  345, 257

\bibitem[{{Wolf-Chase} {et~al}\mbox{.}(2000){Wolf-Chase}, {Barsony}, \&
  {O'Linger}}]{2000AJ....120.1467W}
{Wolf-Chase} G.~A., {Barsony} M., {O'Linger} J., 2000, \aj, 120, 1467

\bibitem[{{Wu} {et~al}\mbox{.}(2007){Wu}, {Dunham}, {Evans}, {Bourke}, \&
  {Young}}]{2007AJ....133.1560W}
{Wu} J., {Dunham} M.~M., {Evans}, II N.~J., {Bourke} T.~L., {Young} C.~H.,
  2007, \aj, 133, 1560

\bibitem[{{Yang} {et~al}\mbox{.}(1997){Yang}, {Ohashi}, {Yan}, {Liu}, {Kaifu},
  \& {Kimura}}]{1997ApJ...475..683Y}
{Yang} J., {Ohashi} N., {Yan} J., {Liu} C., {Kaifu} N., {Kimura} H., 1997,
  \apj, 475, 683

\bibitem[{{Young} \& {Evans}(2005)}]{2005ApJ...627..293Y}
{Young} C.~H., {Evans}, II N.~J., 2005, \apj, 627, 293

\bibitem[{{Young} {et~al}\mbox{.}(2003){Young}, {Shirley}, {Evans}, \&
  {Rawlings}}]{2003ApJS..145..111Y}
{Young} C.~H., {Shirley} Y.~L., {Evans}, II N.~J., {Rawlings} J.~M.~C., 2003,
  \apjs, 145, 111

\end{thebibliography}

\appendix
\section{AMI Maps}

Maps of each field are shown. AMI data is shown uncorrected for the primary beam response. The positions of known Class~0 and Class~I objects are indicated as crosses (+) and stars (*), respectively. The positions for those found in Perseus (L1448 and HH~7-11) are from \citet{hat2007a}, HH~1-2 from \citet{2010AA...518L.122F}, and Serpens from \citet{2010AA...519A..27D}. Crosses (x) represent known sources of unknown evolutionary class. We plot the half power point of the primary beam as a solid circle ($\approx6$\,arcmin at 16\,GHz) and the FWHM of the PSF as the filled ellipse in the bottom left corner (see Table~\ref{tab:srclist}). Contours at 5, 10, 15, 20\,$\sigma_{\rm rms}$ etc, where $\sigma_{\rm rms}$ values for each source are listed in Table 1.

\begin{figure*}
\caption{The AMI 16\,GHz combined-channel map for each source. \label{fig:amimap1}}
\centerline{\includegraphics[width=0.4\textwidth]{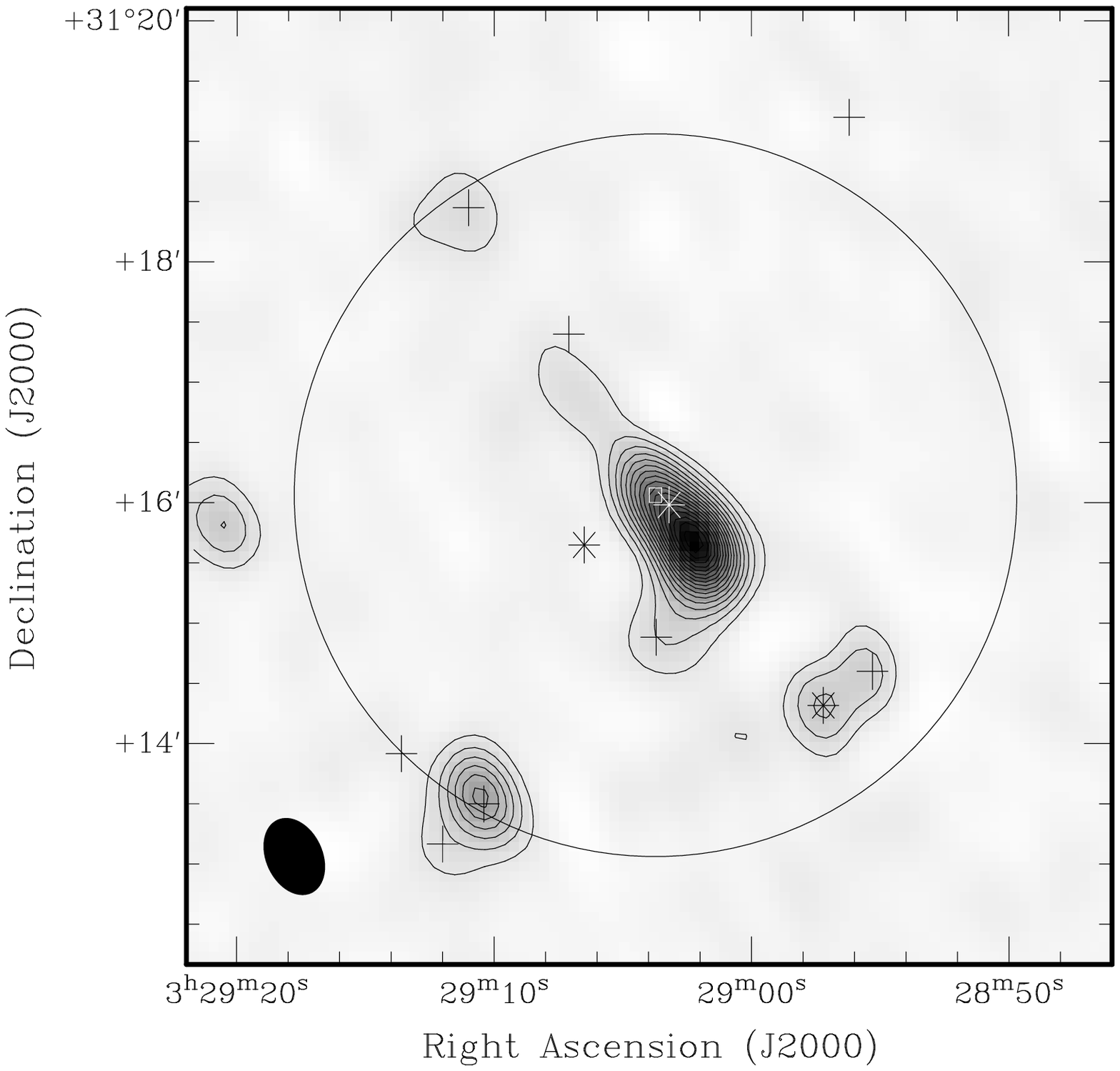}\qquad \includegraphics[width=0.4\textwidth]{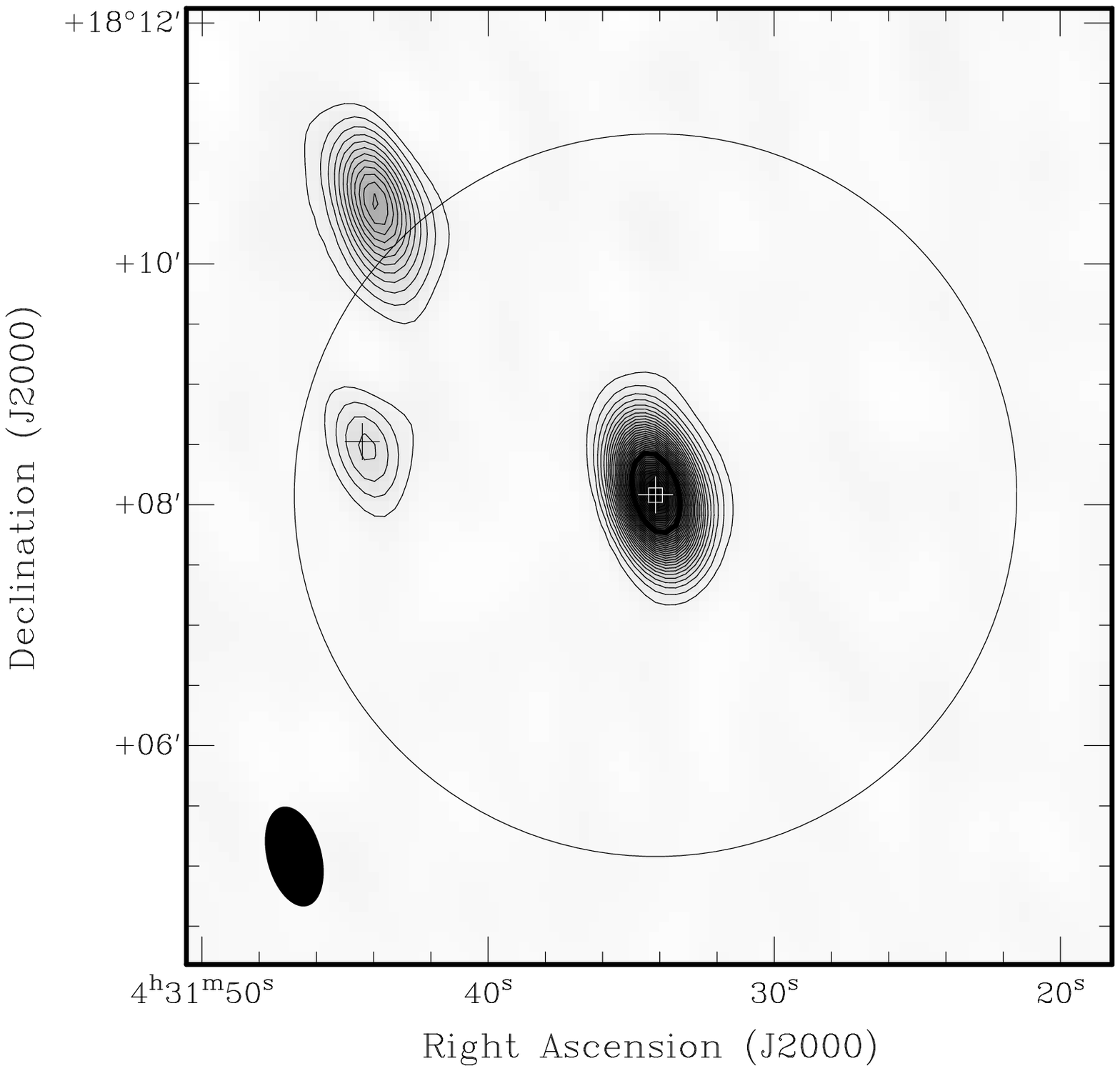}}
\centerline{HH~7-11\hspace{0.4\textwidth}L1551}
\centerline{\includegraphics[width=0.4\textwidth]{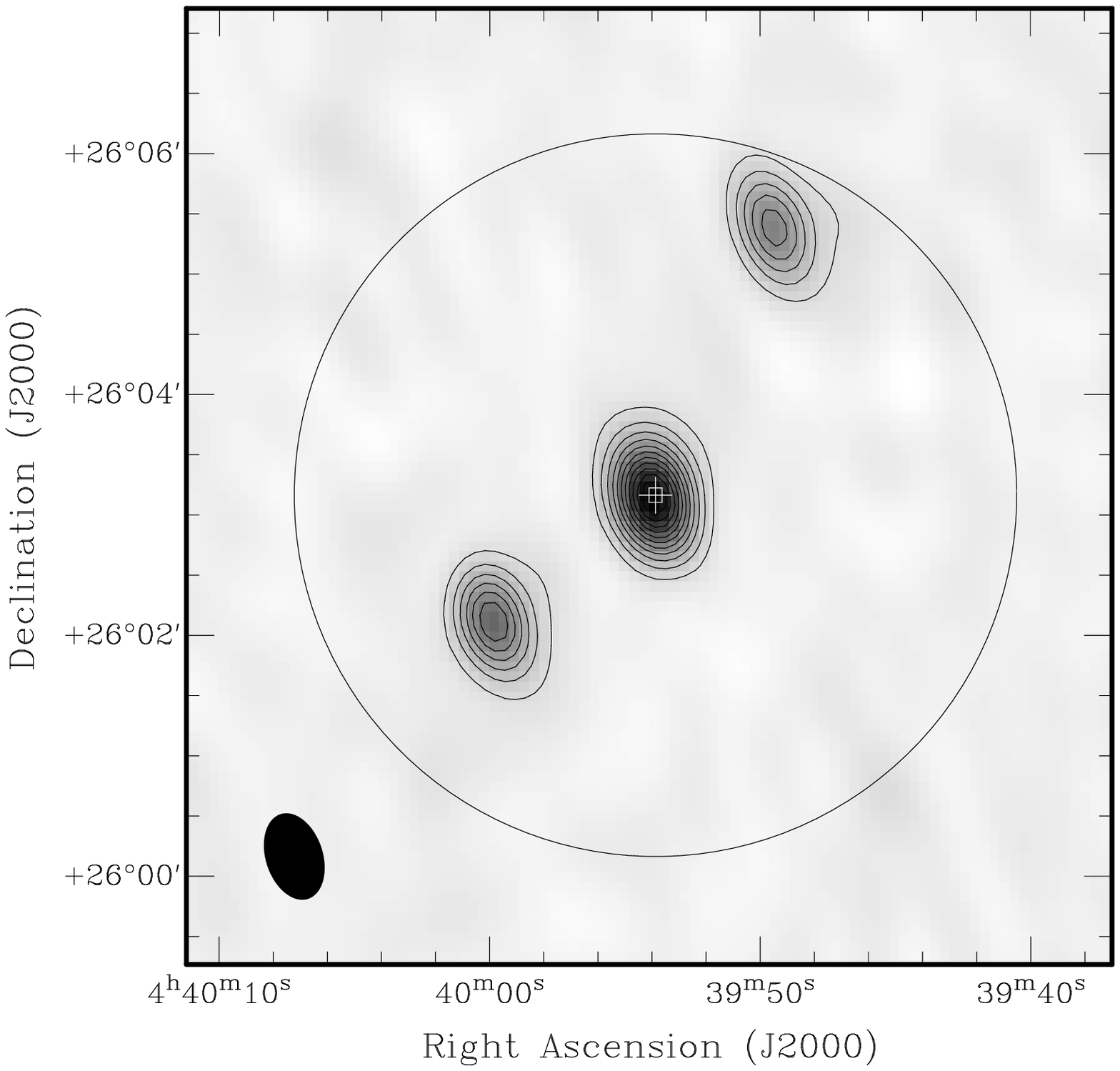}\qquad \includegraphics[width=0.4\textwidth]{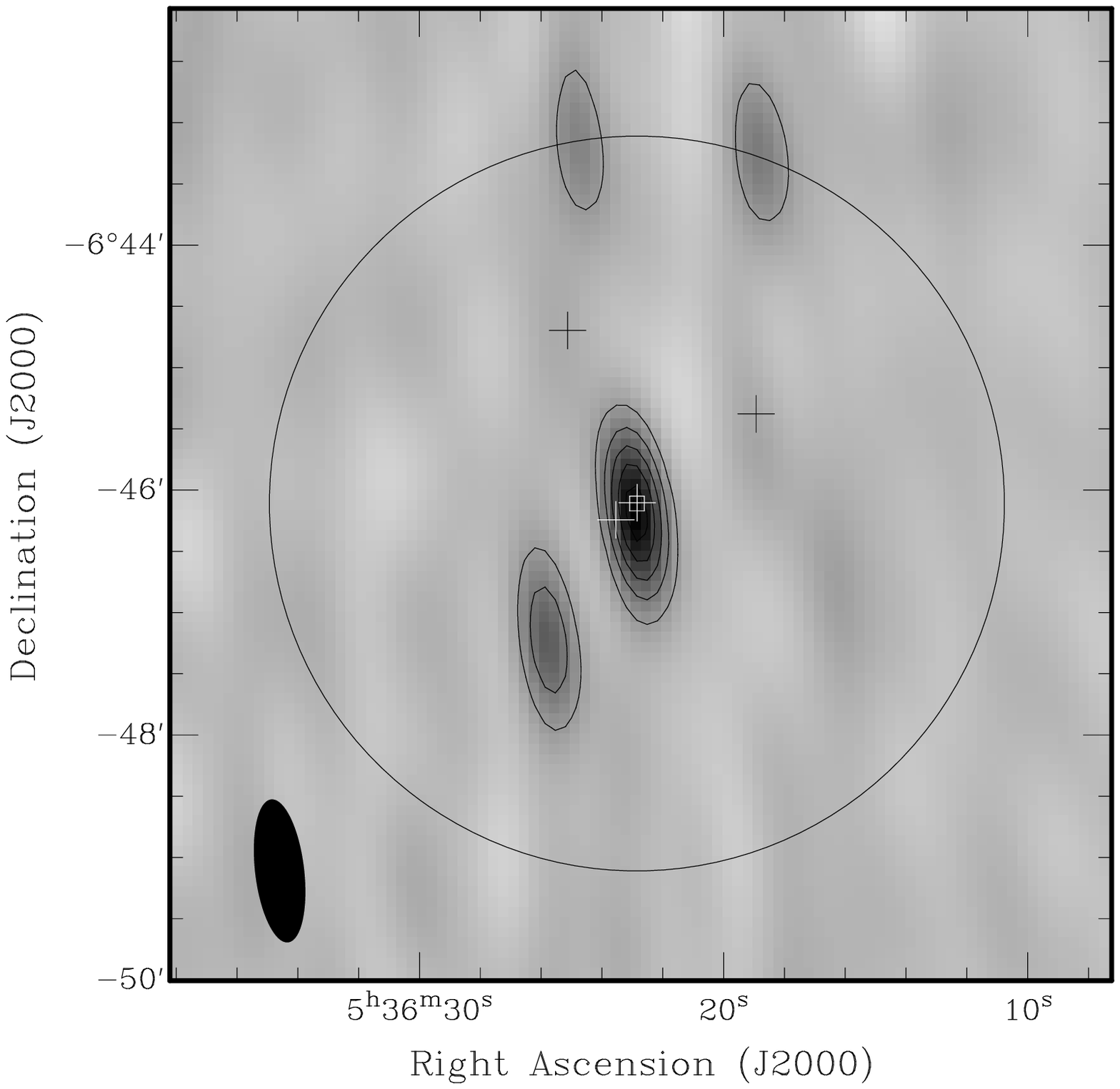}}
\centerline{L1527\hspace{0.4\textwidth}HH~1-2}
\centerline{\includegraphics[width=0.4\textwidth]{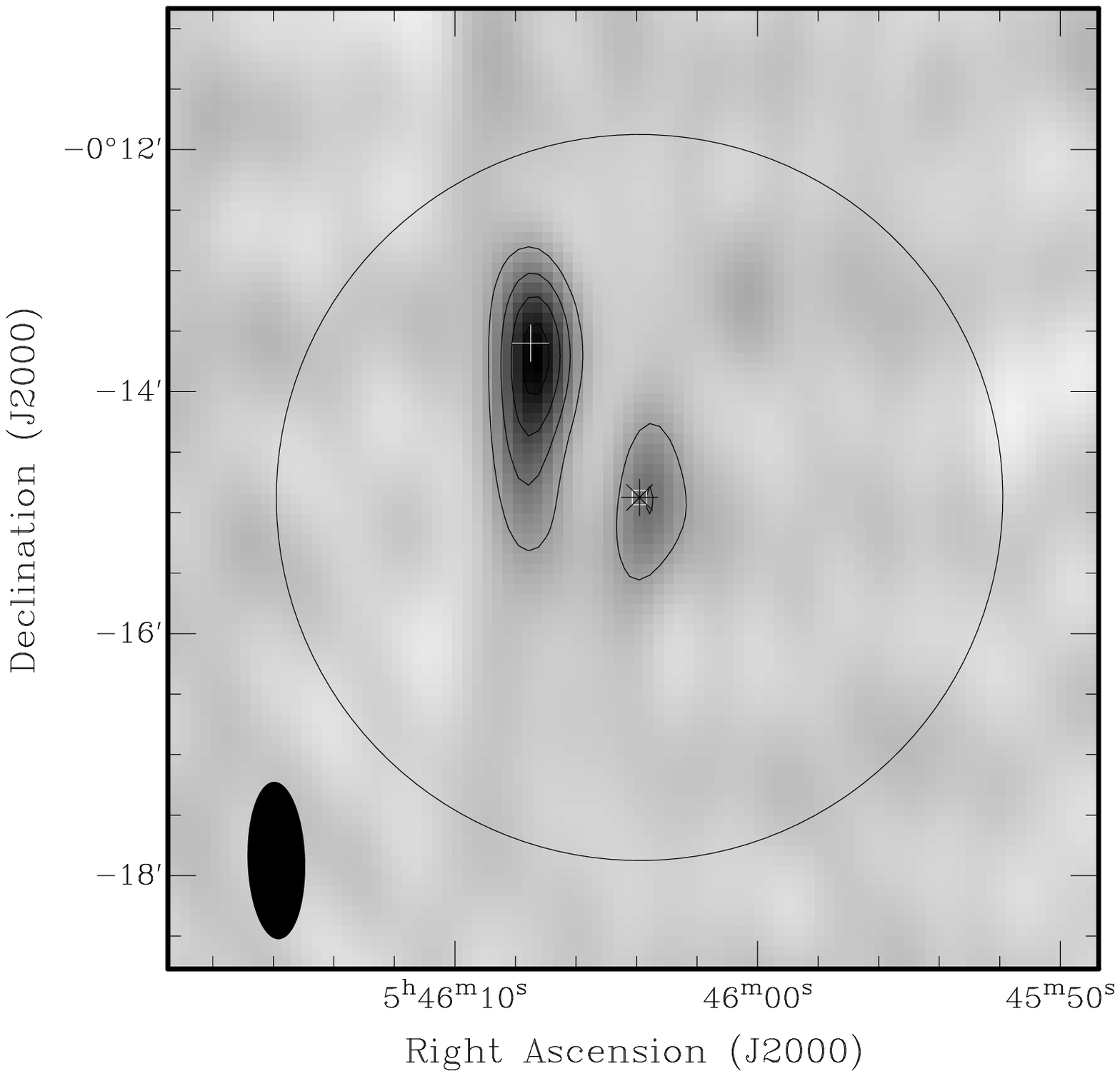}\qquad \includegraphics[width=0.4\textwidth]{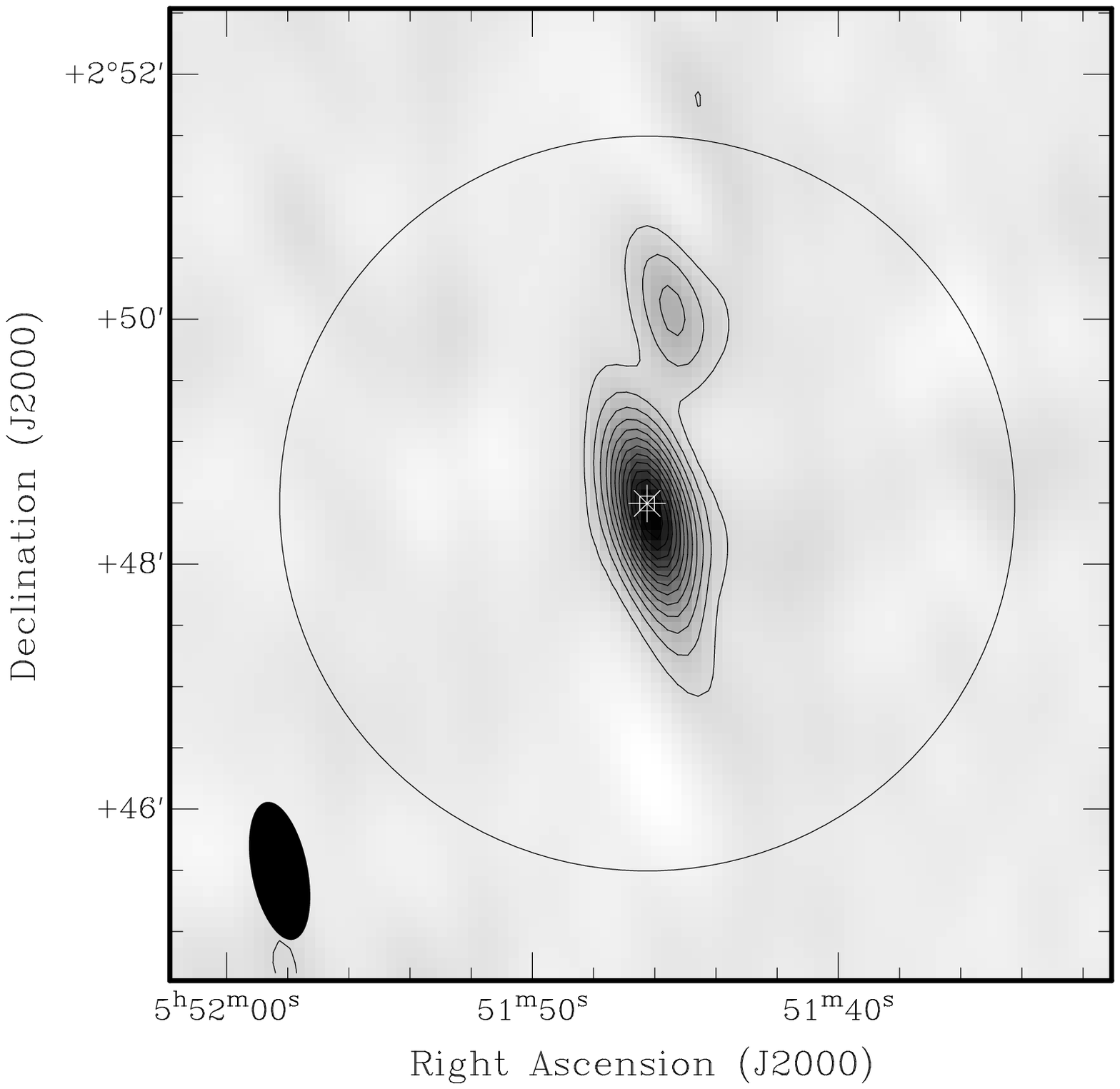}}
\centerline{HH~26IR\hspace{0.4\textwidth}HH~111}
\end{figure*}

\begin{figure*}
\caption{The AMI 16\,GHz combined-channel map for each source - continued. \label{fig:amimap1}}
\centerline{\includegraphics[width=0.4\textwidth]{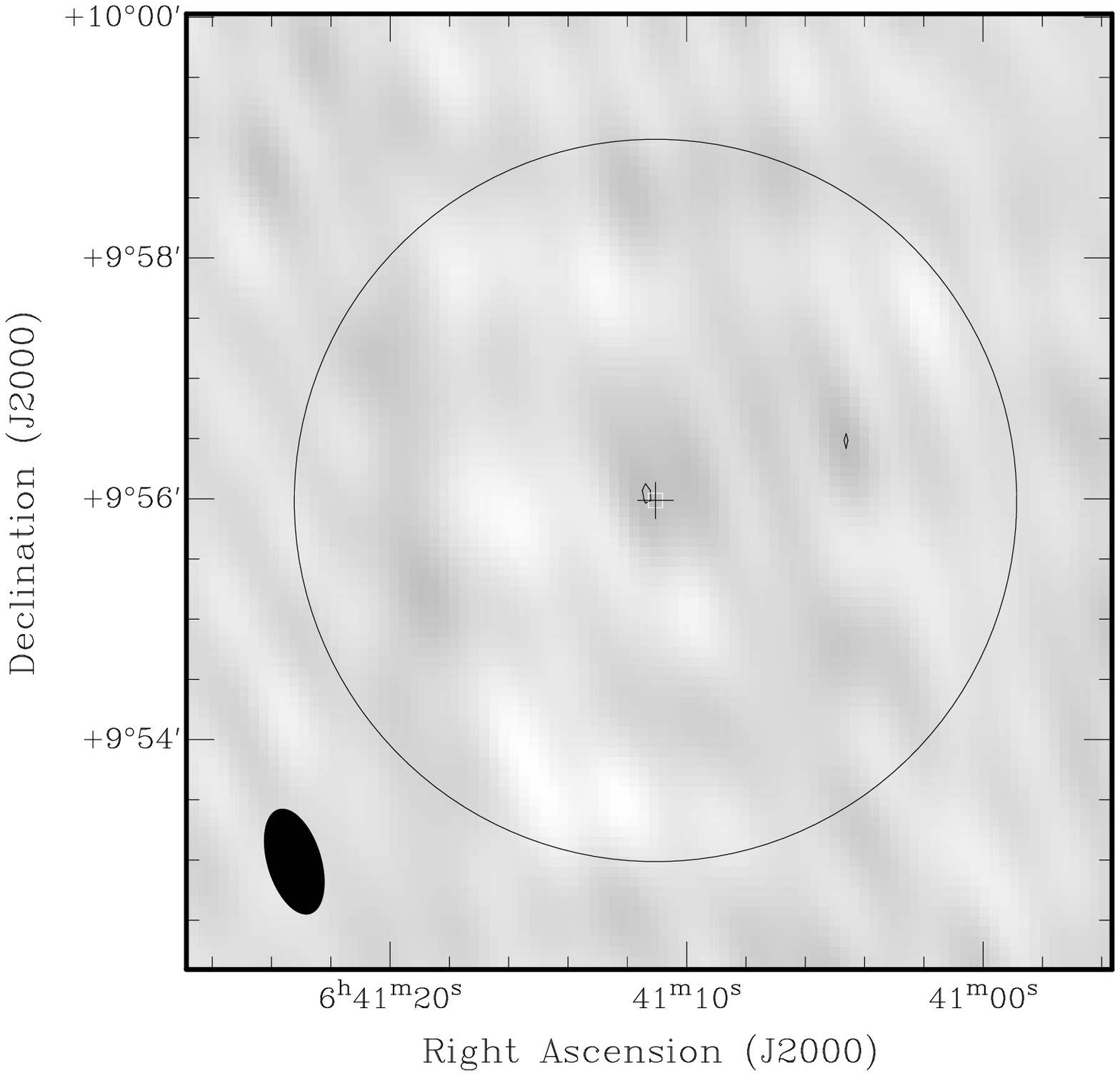}\qquad \includegraphics[width=0.4\textwidth]{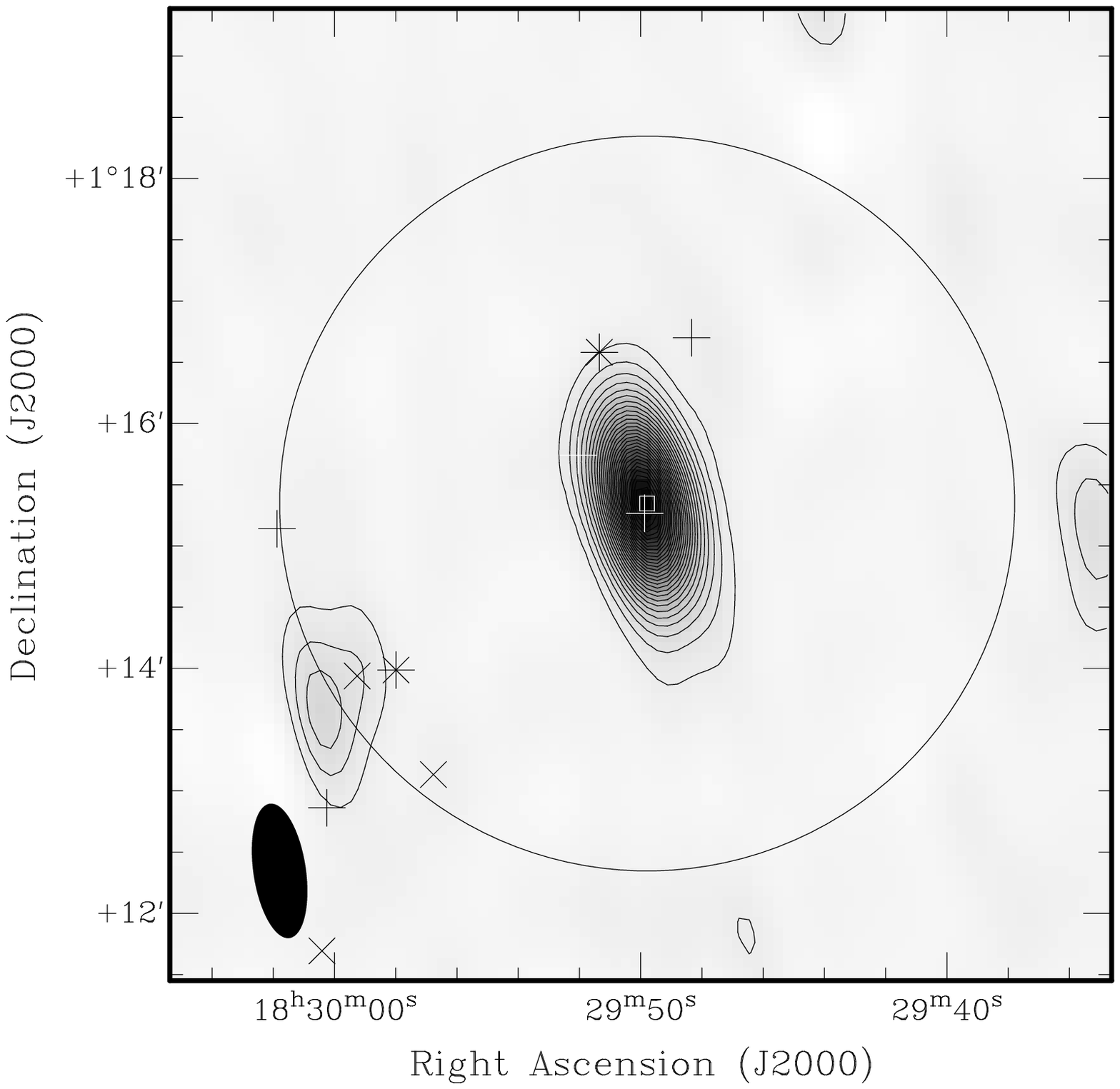}}
\centerline{NGC~2264\hspace{0.4\textwidth}Serpens}
\centerline{\includegraphics[width=0.4\textwidth]{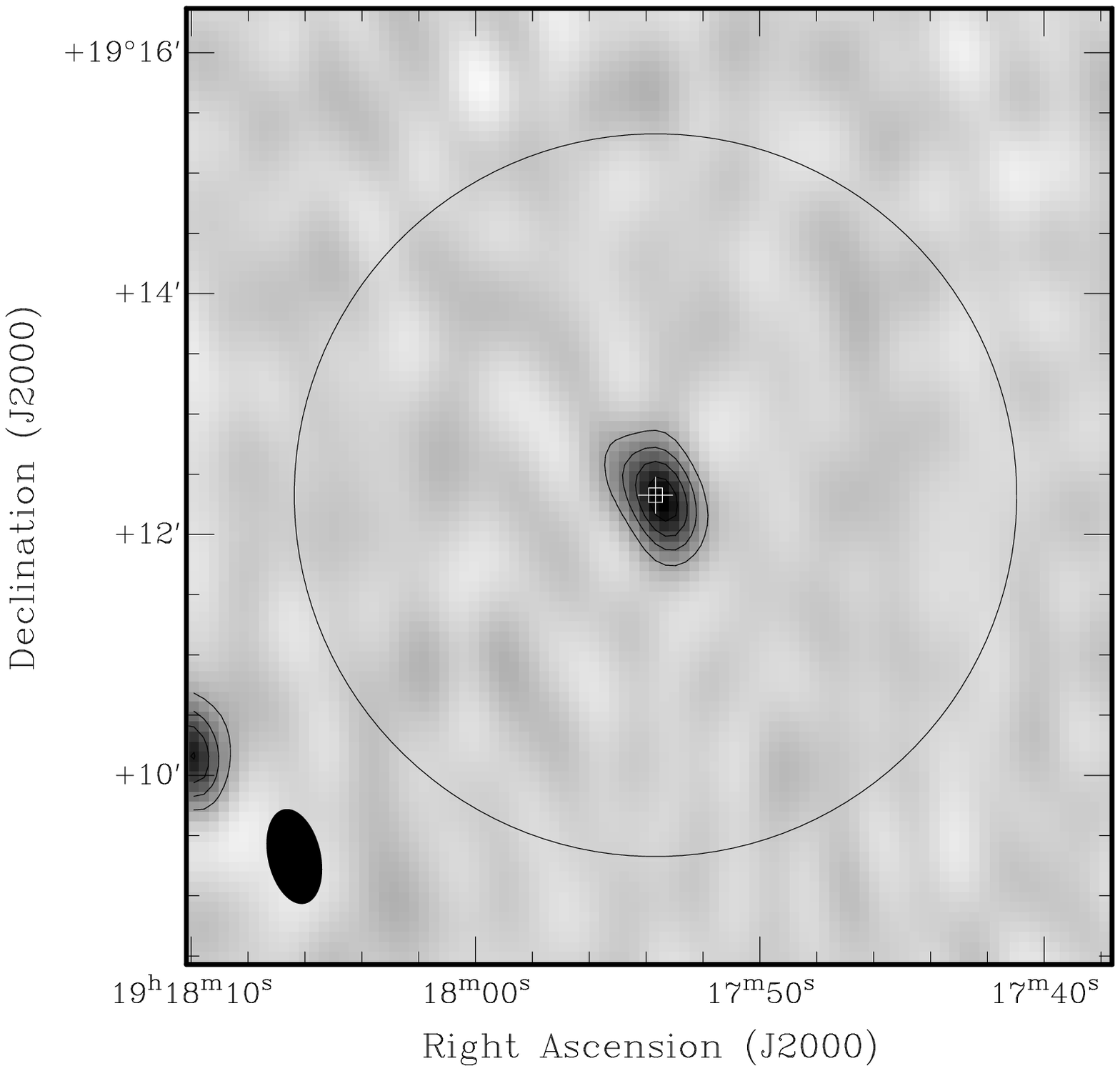}\qquad \includegraphics[width=0.4\textwidth]{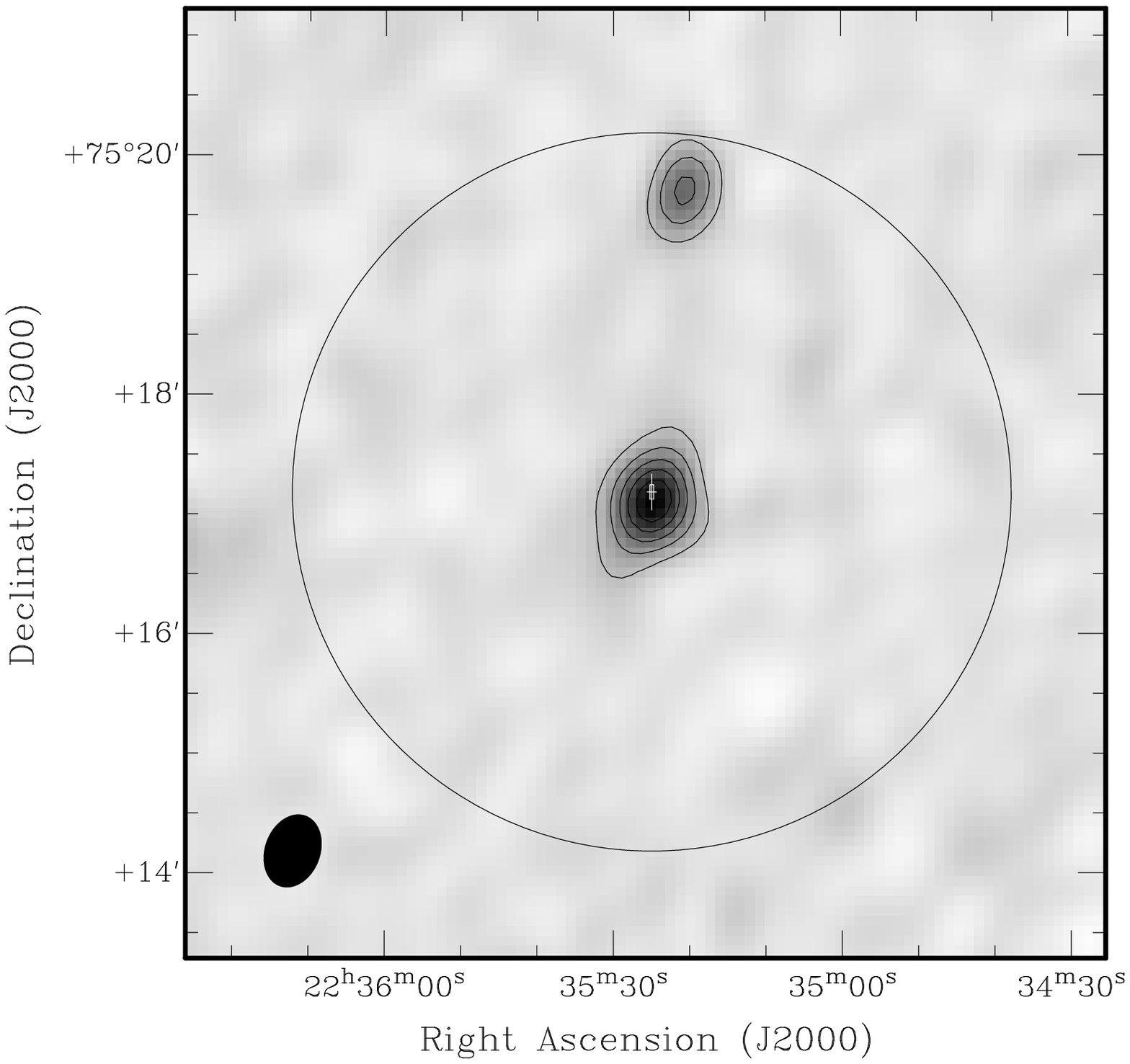}}
\centerline{L723\hspace{0.4\textwidth}L1251}
\end{figure*}

\section{Spectral Energy Distributions}

The observed radio spectra over the AMI frequency channels 3-8 combined with flux densities from the literature. Maximum likelihood results from Scenario (ii) are overlaid (see Table~\ref{tab:srcalpha2}). The list of archival data used in the spectral energy distributions can be found in Appendix~C. Only data $\nu<3$\,THz ($\lambda>100\,\mu$m) were included in the fit, but IRAS data $\nu>3$\,THz are included in the plots for illustration.

\begin{figure*}
\caption{The spectral energy distribution for each source along with the model fit. \label{fig:amised1}}
\centerline{\includegraphics[width=0.4\textwidth]{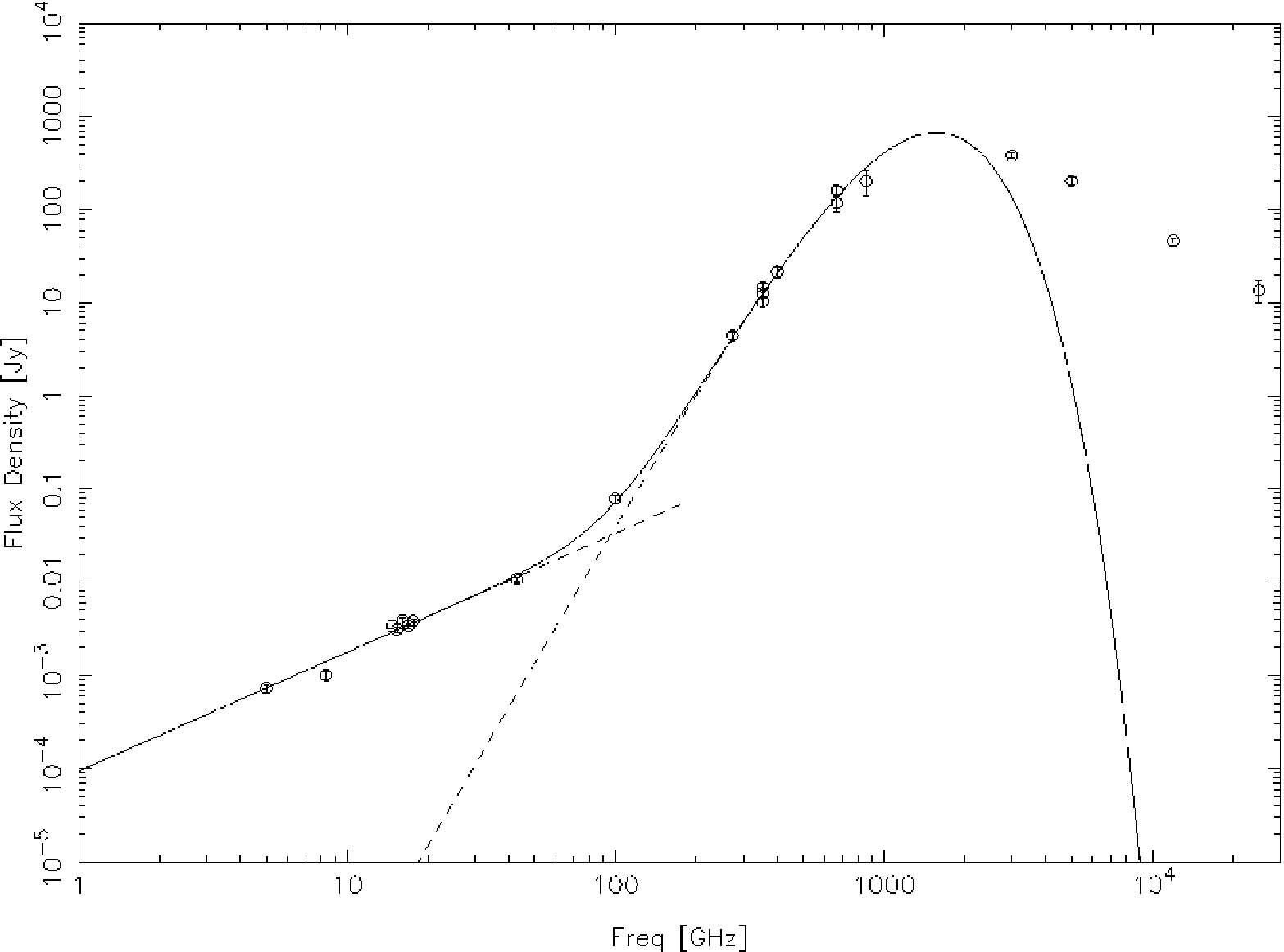}\qquad \includegraphics[width=0.4\textwidth]{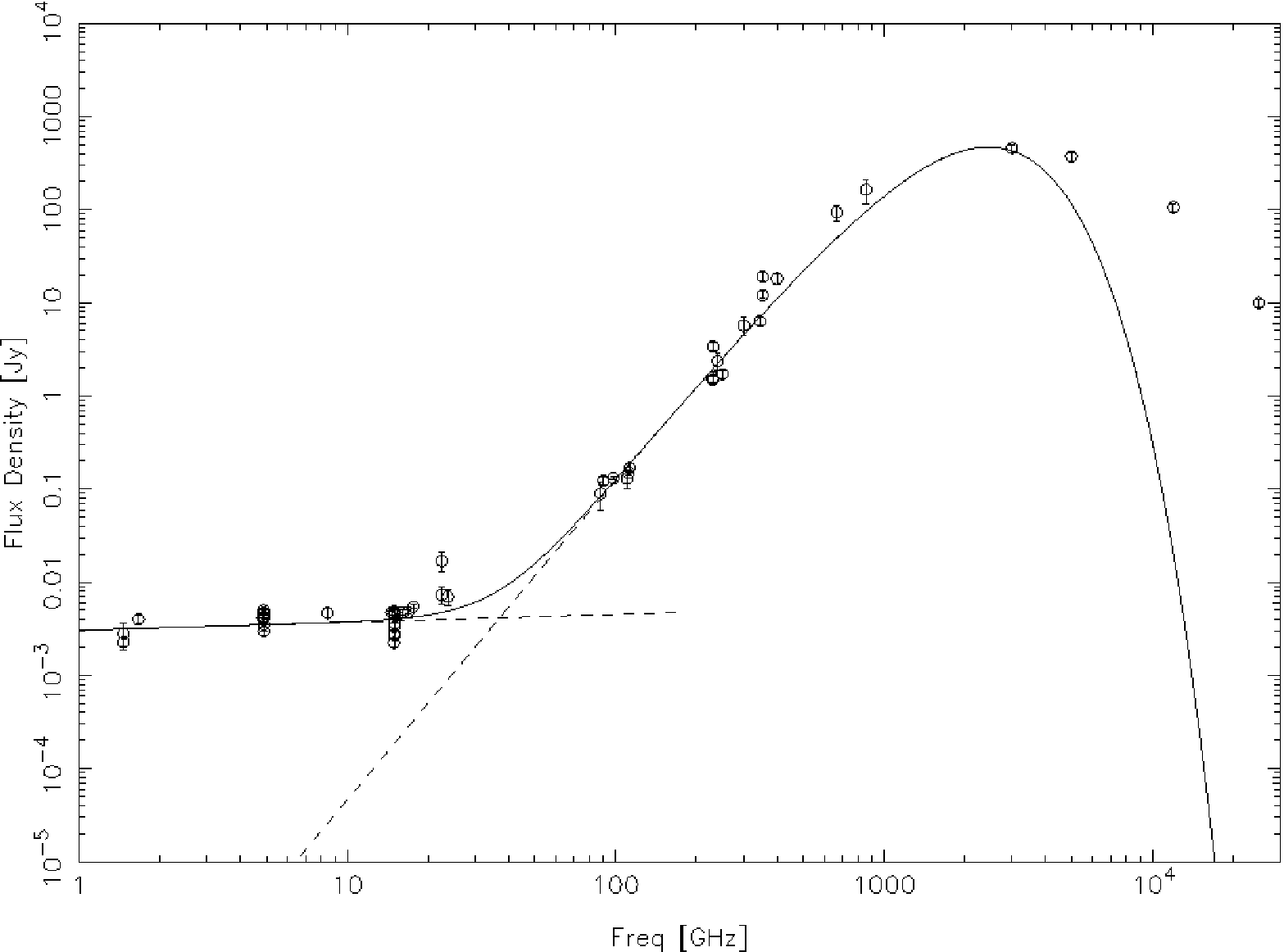}}
\centerline{HH~7-11\hspace{0.4\textwidth}L1551}
\end{figure*}

\begin{figure*}
\caption{The spectral energy distribution for each source along with the model fit - continued. \label{fig:amised1}}
\centerline{\includegraphics[width=0.4\textwidth]{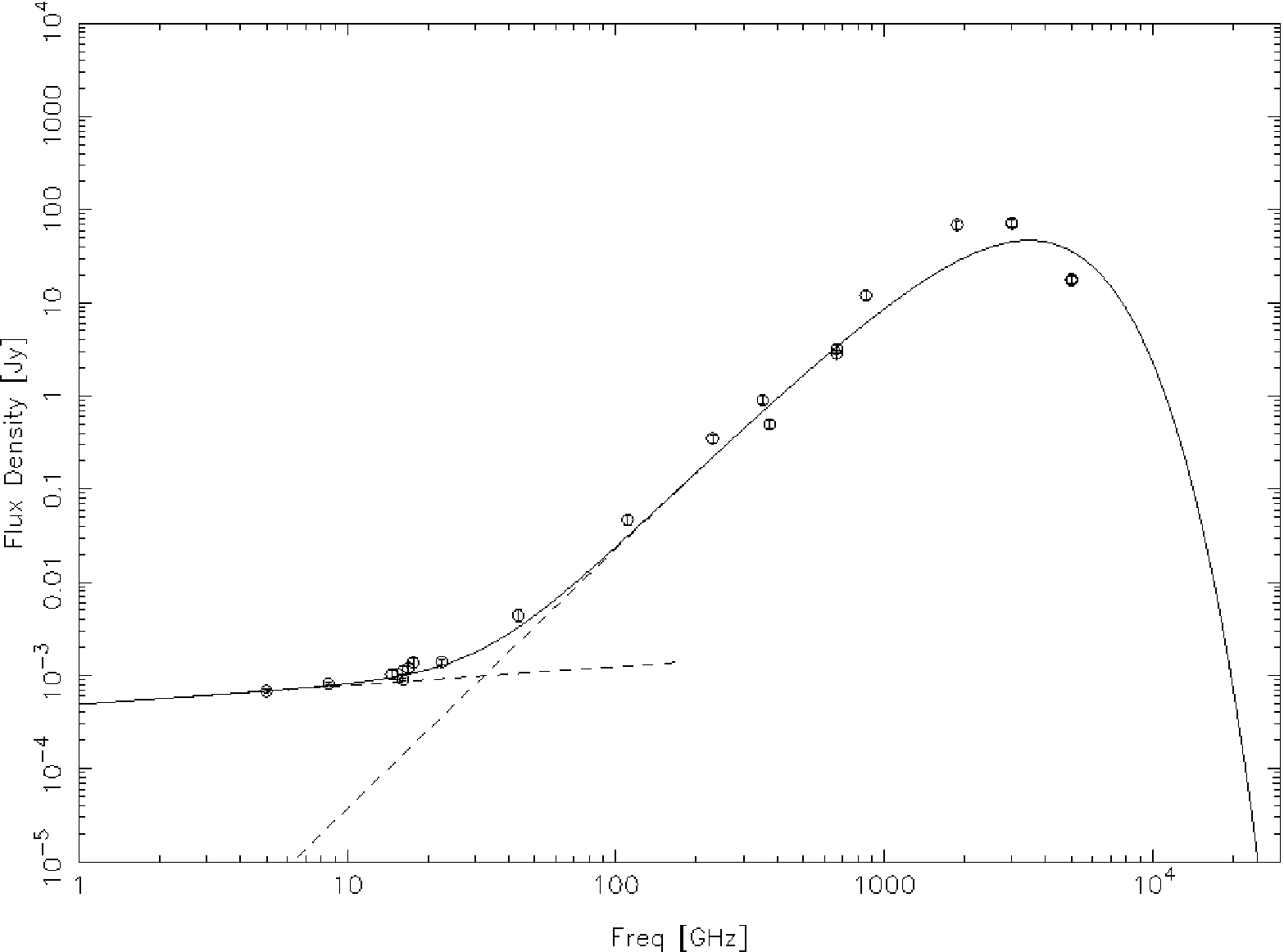}\qquad \includegraphics[width=0.4\textwidth]{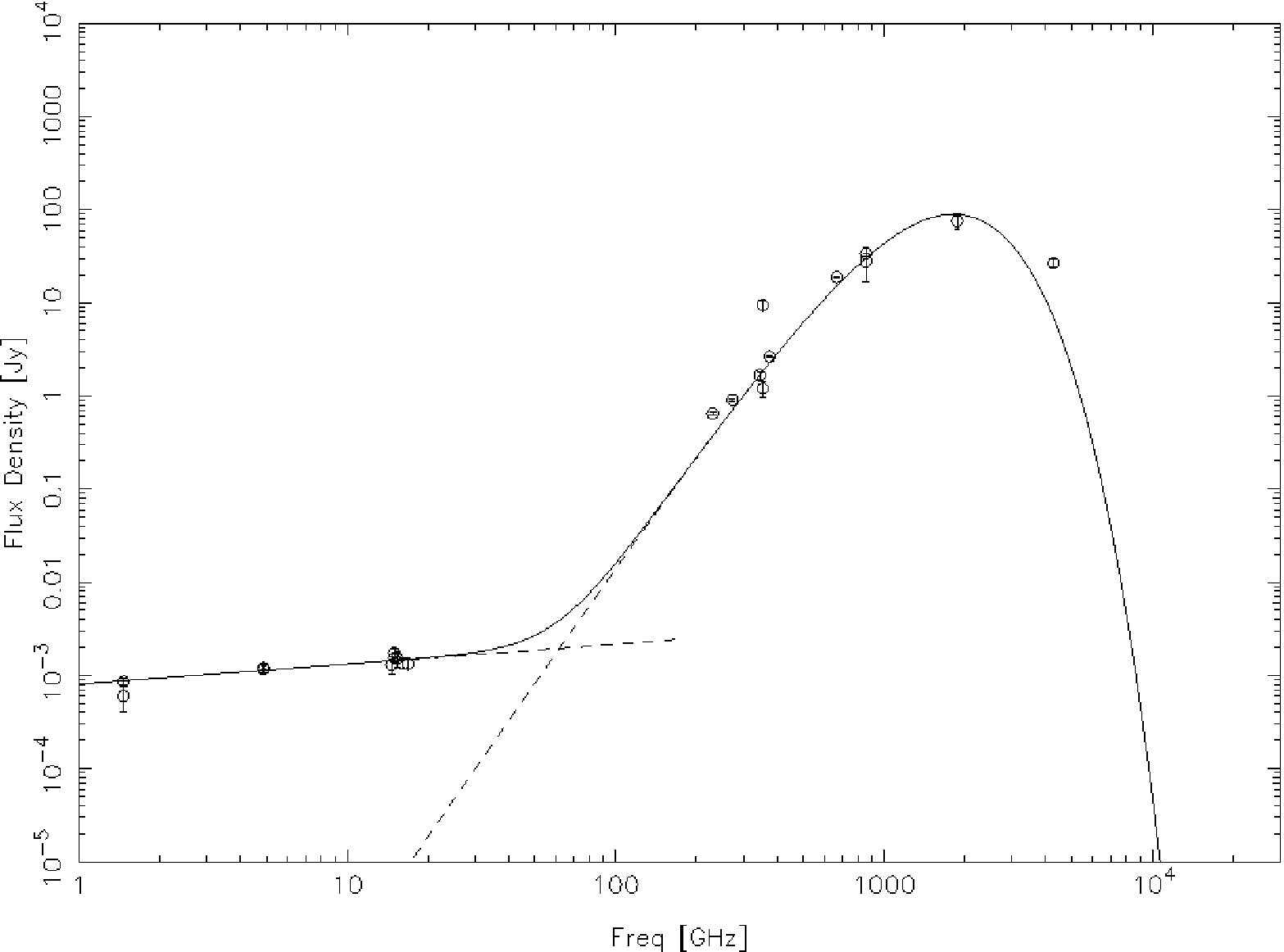}}
\centerline{L1527\hspace{0.4\textwidth}HH~1-2}
\centerline{\includegraphics[width=0.4\textwidth]{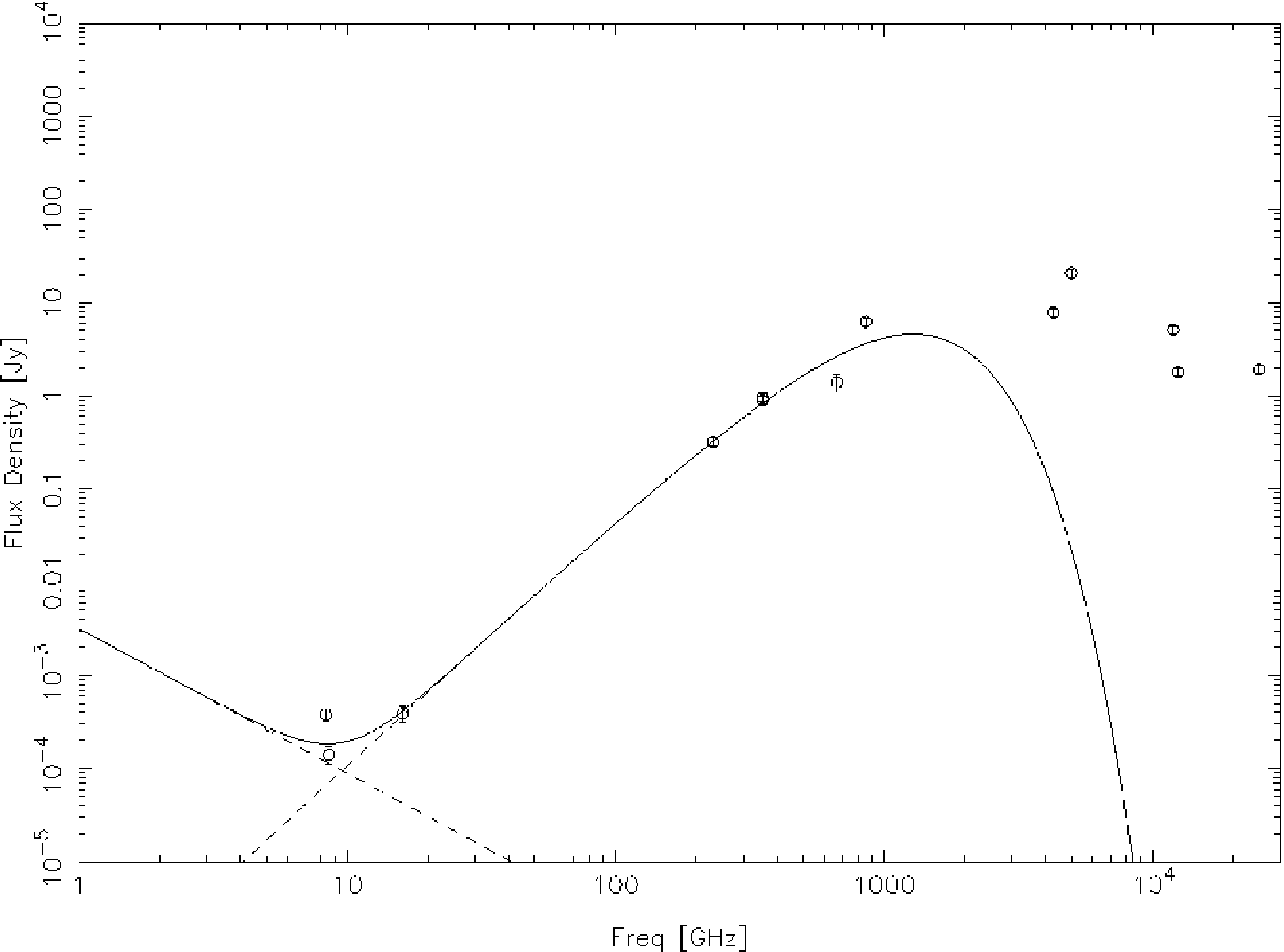}\qquad \includegraphics[width=0.4\textwidth]{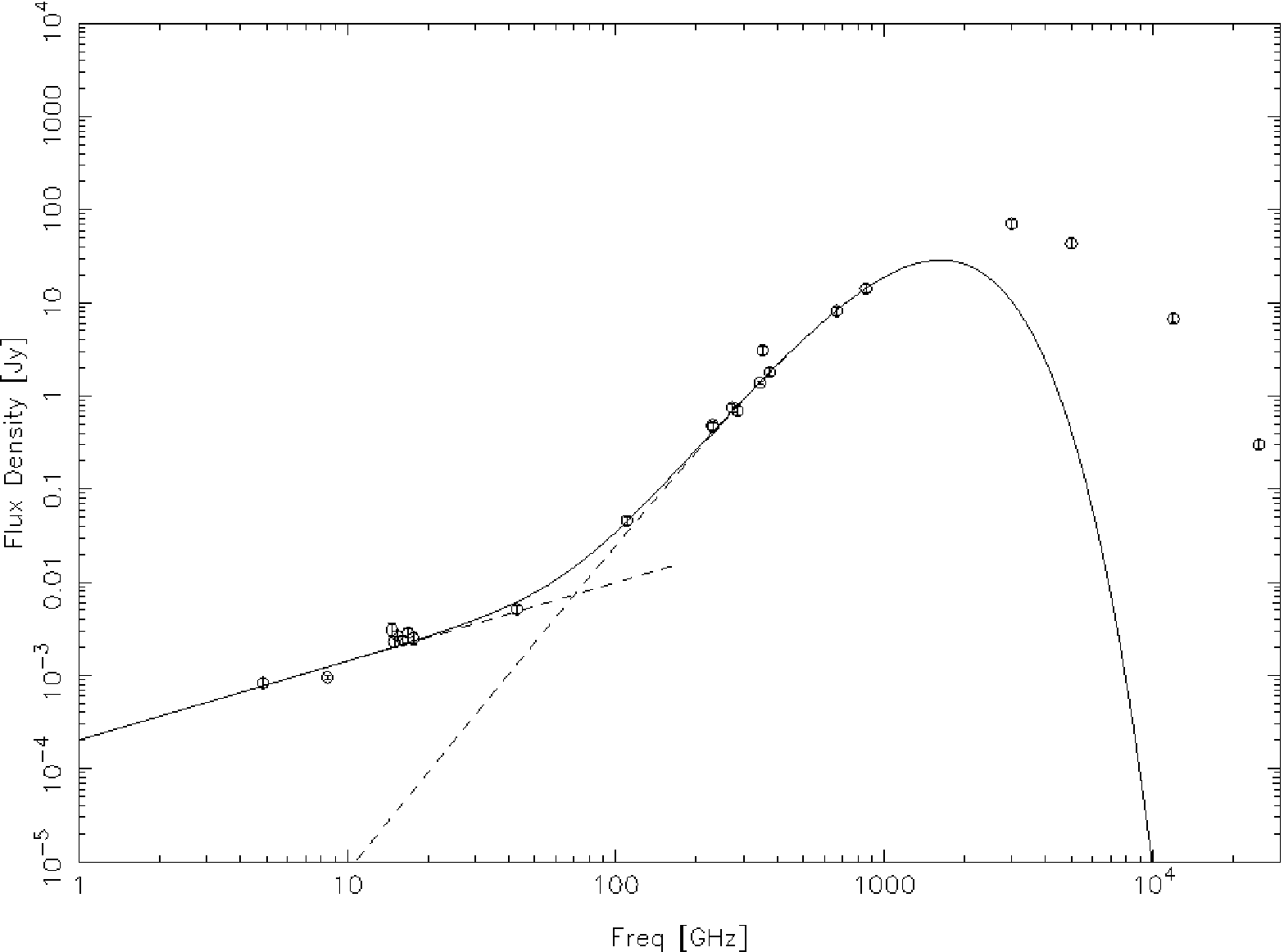}}
\centerline{HH~26~IR\hspace{0.4\textwidth}HH~111}
\centerline{\includegraphics[width=0.4\textwidth]{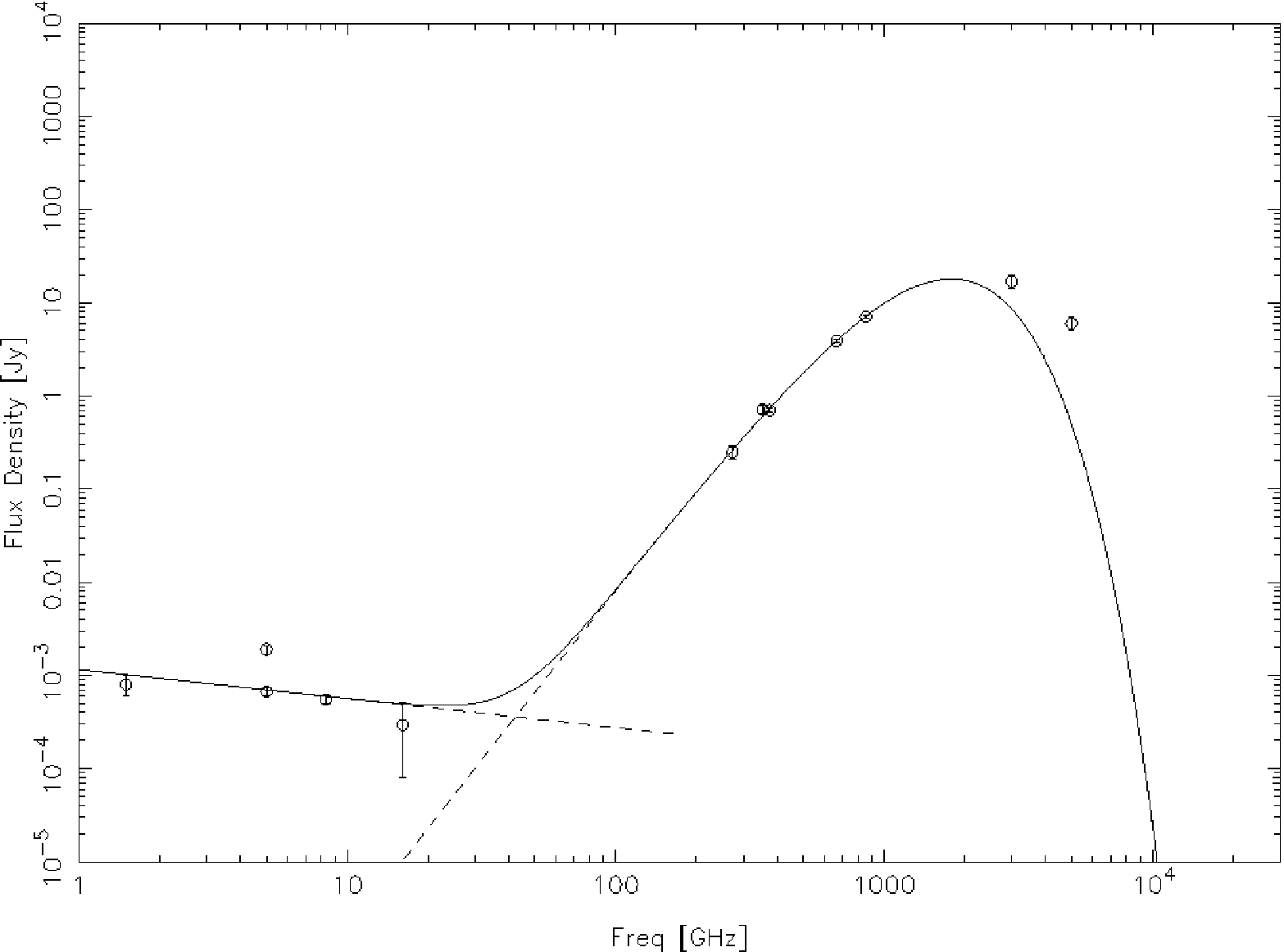}\qquad \includegraphics[width=0.4\textwidth]{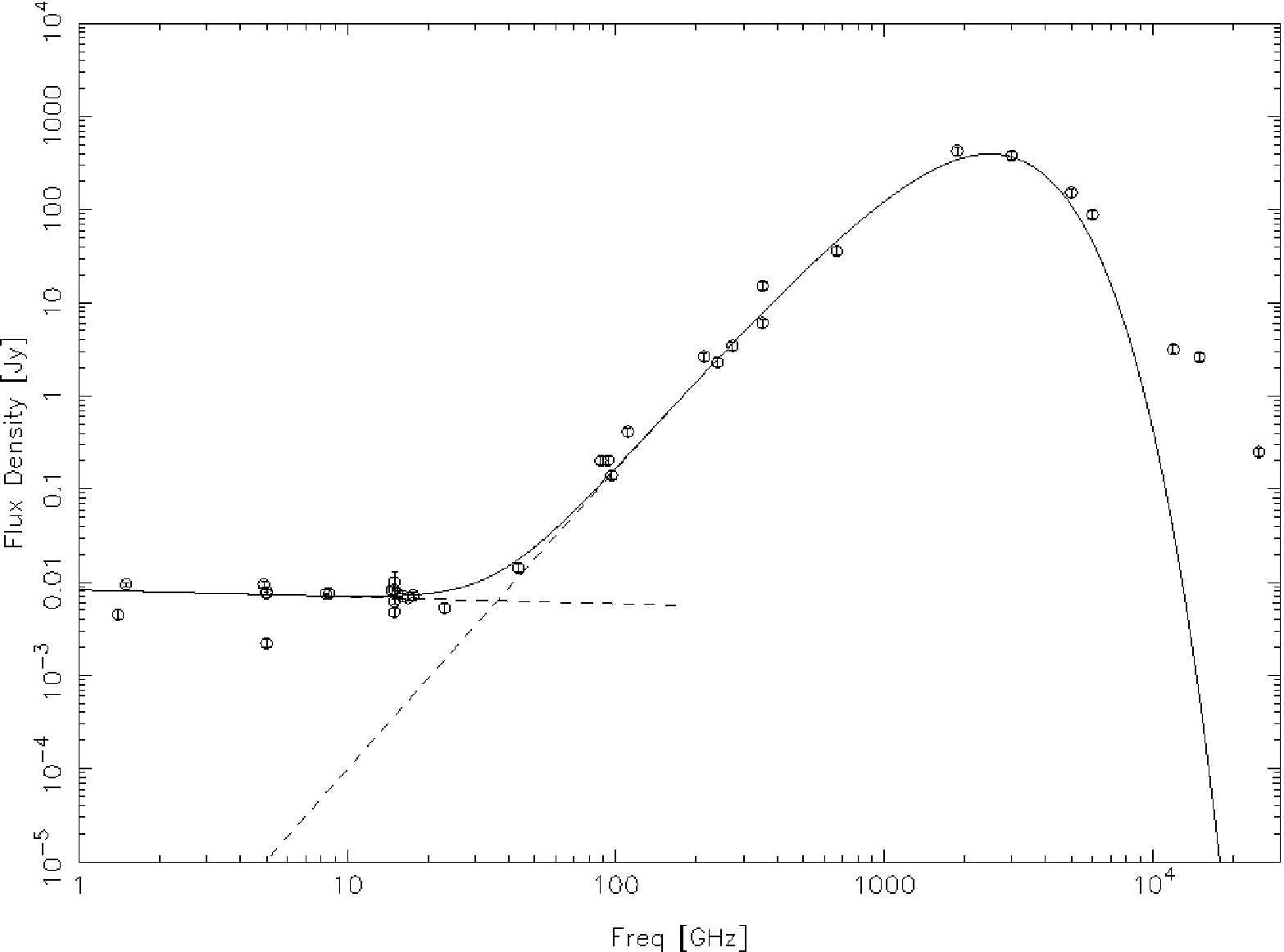}}
\centerline{NGC~2264\hspace{0.4\textwidth}Serpens}
\centerline{\includegraphics[width=0.4\textwidth]{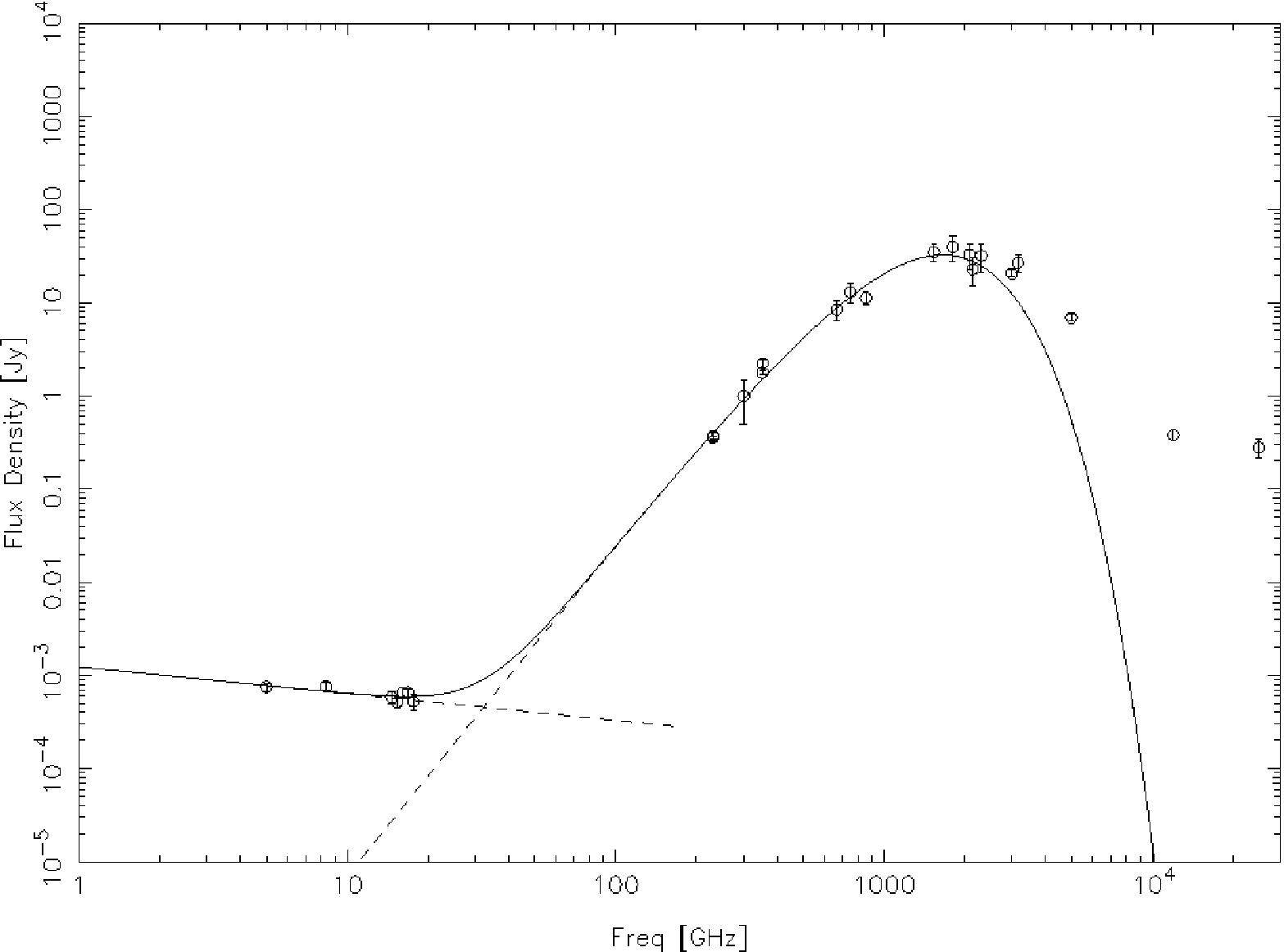}\qquad \includegraphics[width=0.4\textwidth]{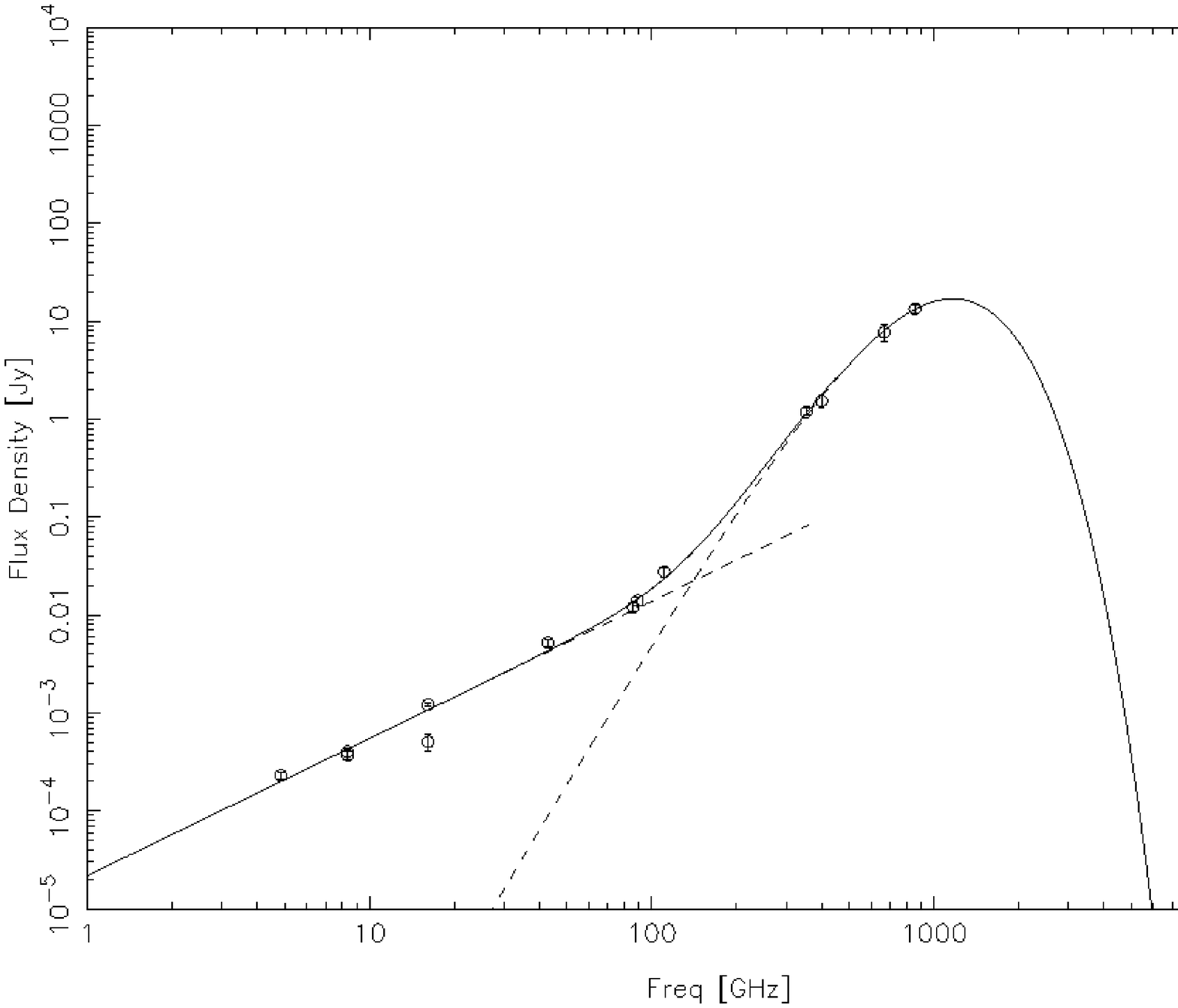}}
\centerline{L723\hspace{0.4\textwidth}L1251}
\end{figure*}

\begin{figure*}
\caption{The spectral energy distribution for each source along with the model fit - continued. \label{fig:amised1}}
\centerline{\includegraphics[width=0.4\textwidth]{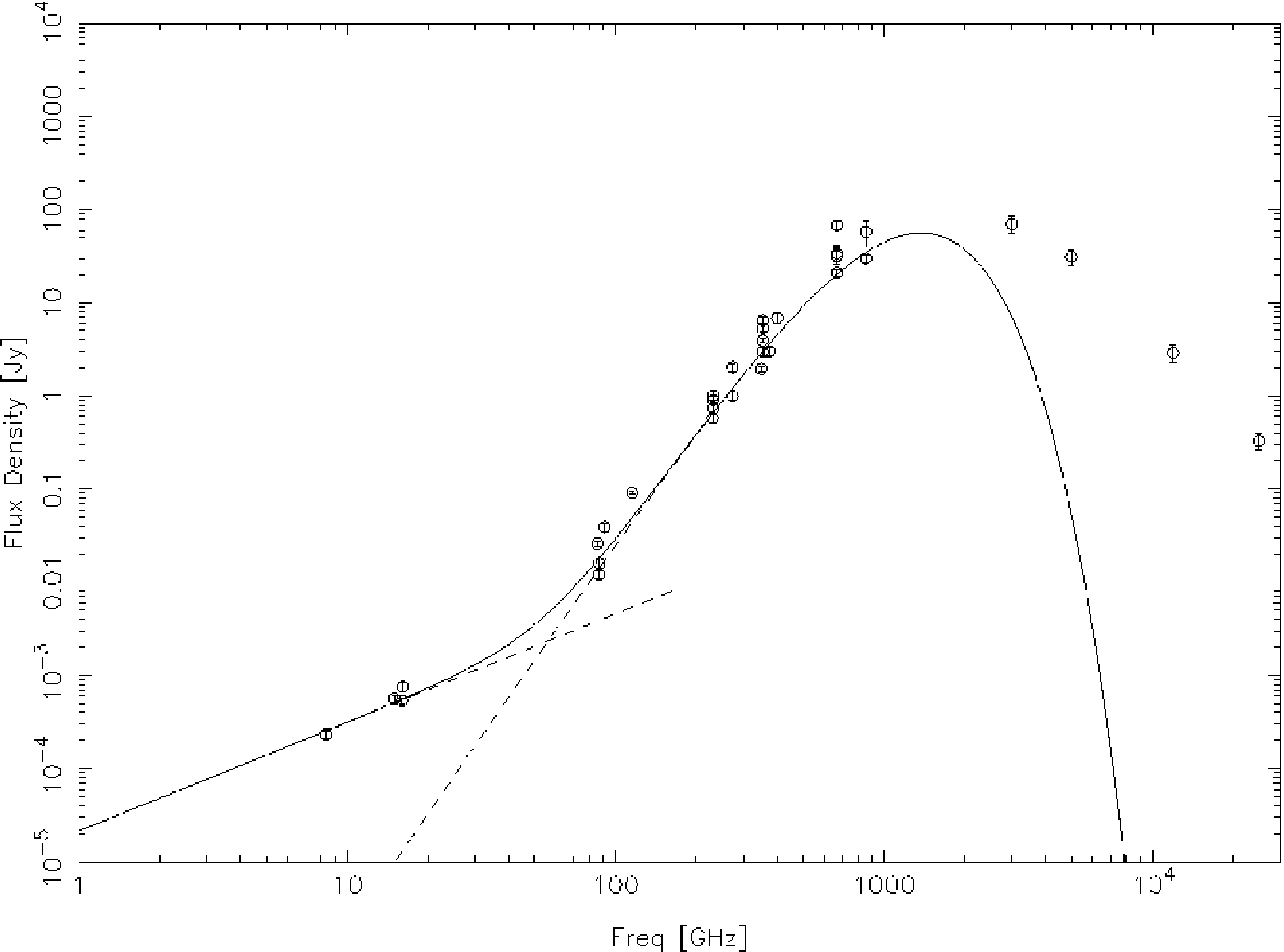}\qquad \includegraphics[width=0.4\textwidth]{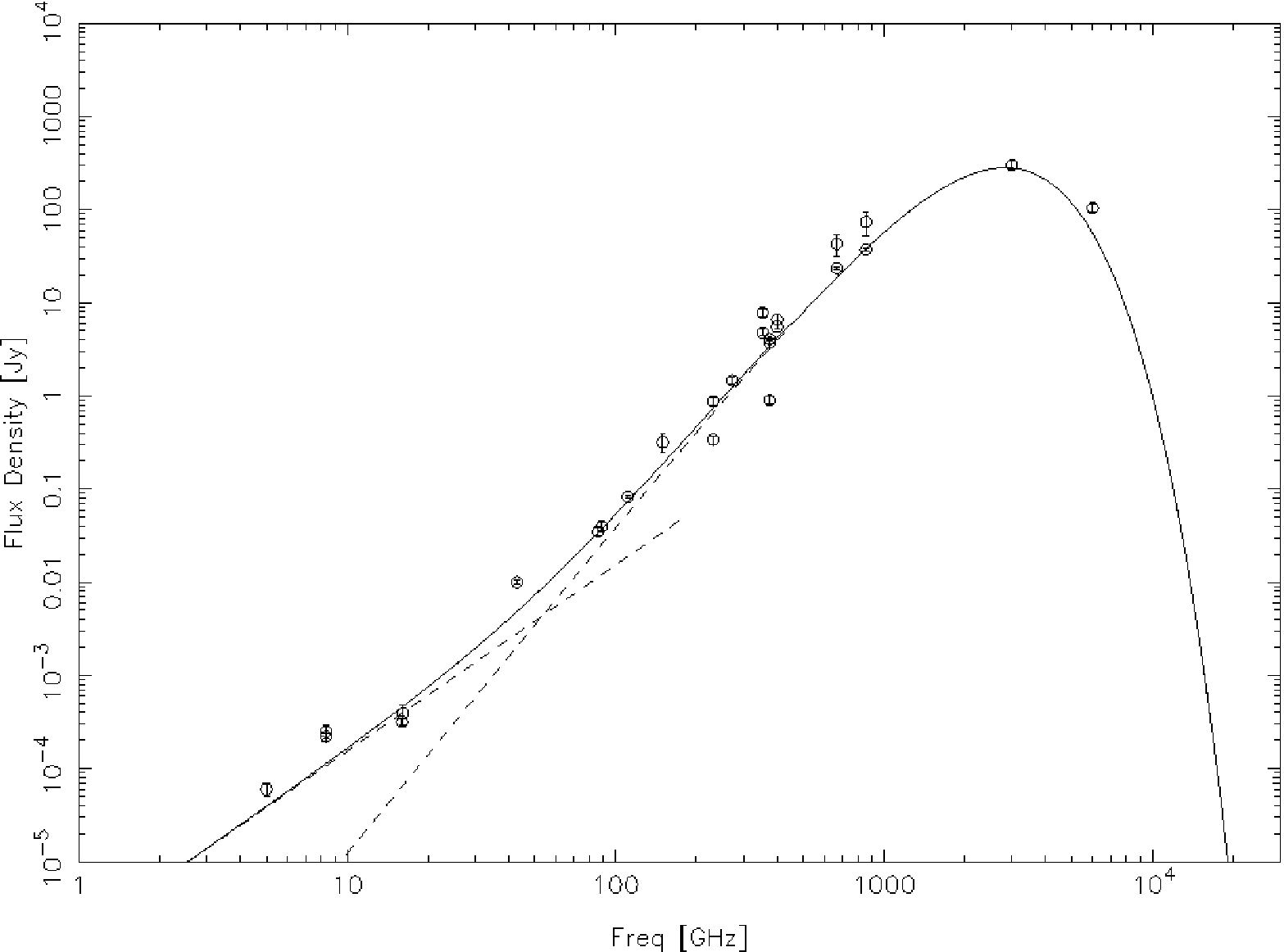}}
\centerline{L1448~C\hspace{0.4\textwidth}NGC~1333~IRAS~2A}
\centerline{\includegraphics[width=0.4\textwidth]{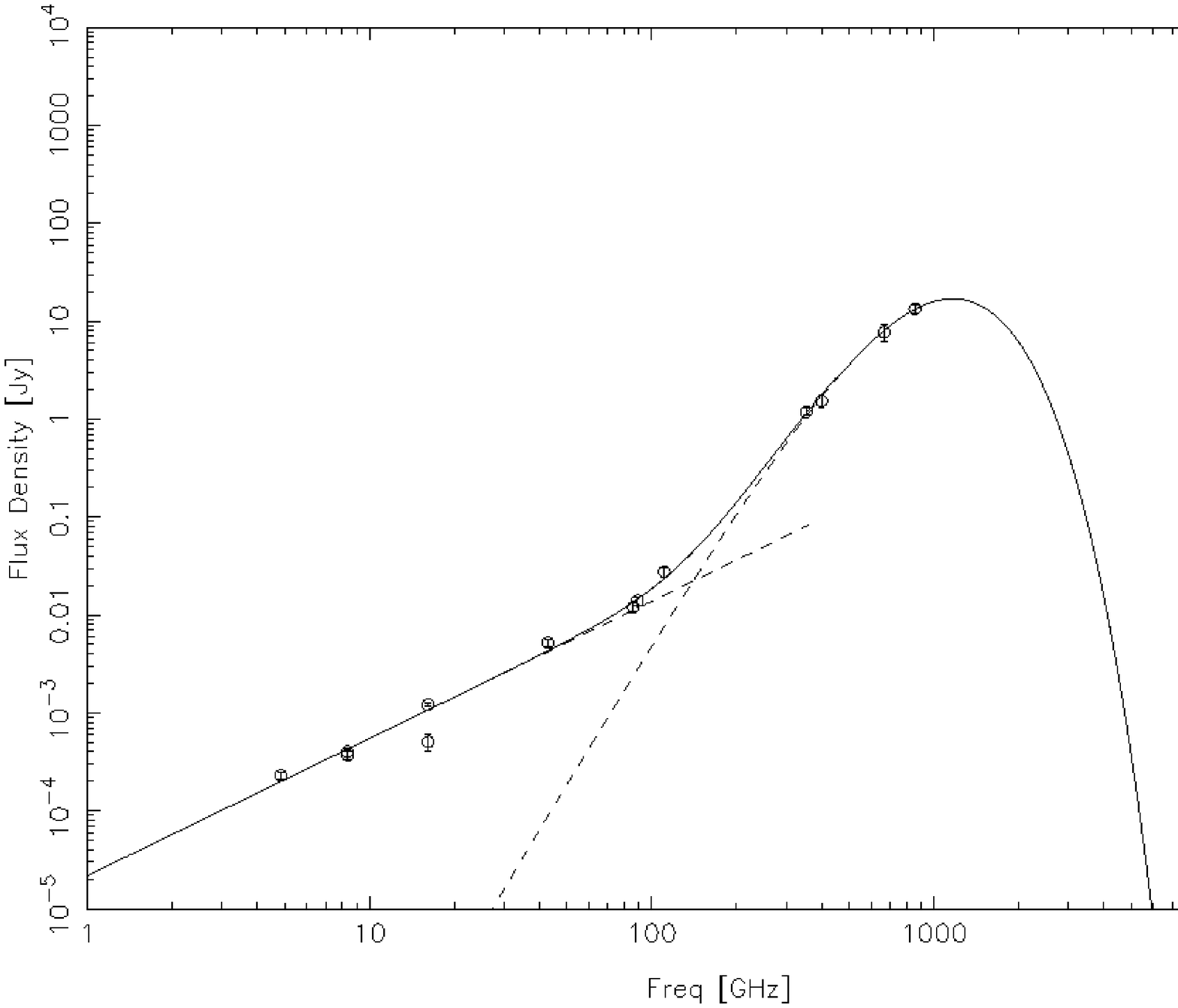}\qquad \includegraphics[width=0.4\textwidth]{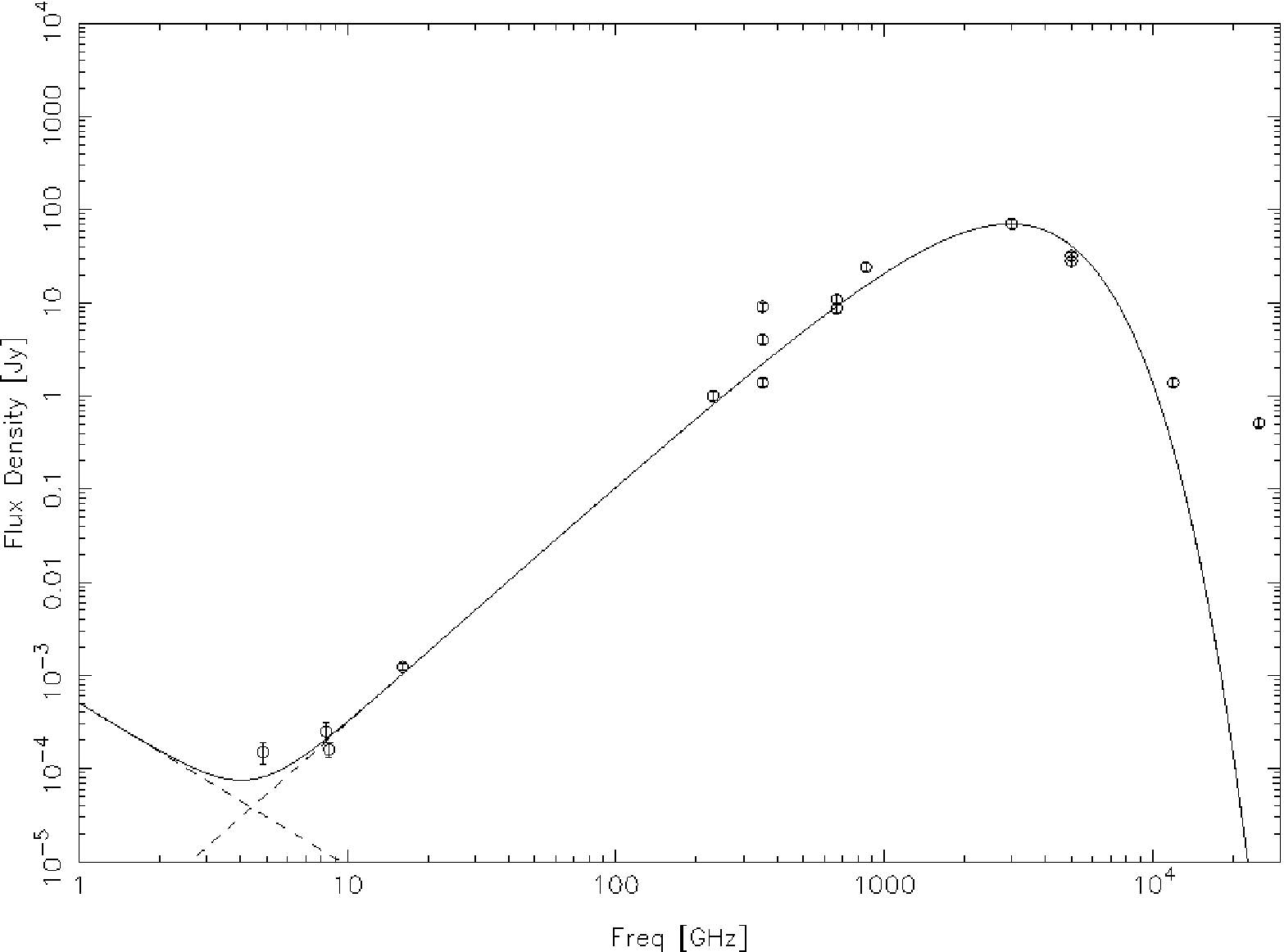}}
\centerline{NGC~1333~IRAS~2B\hspace{0.4\textwidth}HH~25~MMS}
\centerline{\includegraphics[width=0.4\textwidth]{./filler.ps}\qquad \includegraphics[width=0.4\textwidth]{./filler.ps}}
\centerline{\includegraphics[width=0.4\textwidth]{./filler.ps}\qquad \includegraphics[width=0.4\textwidth]{./filler.ps}}
\end{figure*}

\section{SED Reference Table}

An extensive literature search was conducted for unresolved, integrated flux densities to include in the spectral energy distributions. It should be noted that \citet{1995A&AS..109..177W} was a useful reference. High resolution data that were highly discrepant due to flux loss or data with high uncertainties \citep[e.g. 450\,$\mu$m data from][]{2008ApJS..175..277D} were not included. Where uncertainties were not provided, an error of 10~per~cent was used in the model fittings and this is indicated by a $^{\dag}$. 

\begin{table*}
\begin{center}
\caption{Reference list for the spectral energy distribution data.\label{tab:sedref}}
\begin{tabular}{llll}
\hline\hline
Source & $\nu$ (GHz) & $S_{\nu}$ (Jy) & Reference \\
\hline
L1448~IRS~3 & 5 & $1.15\rm{x}10^{-3\dag}$ & \citet{1990ApJ...365L..85C}\\
 & 8 & $1.6\rm{x}10^{-3\dag}$ & \citet{2002AJ....124.1045R}\\
 & 14.62 & $2.40\rm{x}10^{-3}\pm0.25$ & this work\\
 & 15 & $1.82\rm{x}10^{-3\dag}$ & \citet{1990ApJ...365L..85C}\\
 & 15.37 & $2.34\rm{x}10^{-3}\pm0.26$ & this work\\
 & 16 & $2.06\rm{x}10^{-3}\pm0.12$ & \citet{scaife2011b}\\
 & 16.12 & $2.41\rm{x}10^{-3}\pm0.37$ & this work\\
 & 16.87 & $2.26\rm{x}10^{-3}\pm0.24$ & this work\\
 & 17.62 & $2.62\rm{x}10^{-3}\pm0.29$ & this work\\
 & 91 & $0.1\pm0.01$ & \citet{2011AJ....141...39S}\\
 & 231 & $2.6\pm0.14$ & \citet{1998ApJ...509..733B}\\
 & 231 & $3^{\dag}$ & \citet{2001AA...365..440M}\\
 & 273 & $4.7\pm0.05$ & \citet{2006ApJ...638..293E}\\
 & 273 & $2.3\pm0.20$ & \citet{1998ApJ...509..733B}\\
 & 349 & $7.12\pm0.33$ & \citet{2011AJ....141...39S}\\
 & 353 & $10.18^{\dag}$ & \citet{2008ApJS..175..277D}\\
 & 353 & $16.86^{\dag}$ & \citet{hat2007a}\\
 & 353 & $8.37\pm0.91$ & \citet{2000ApJ...530..851C}\\
 & 353 & $14.7\pm0.62$ & \citet{2000ApJS..131..249S}\\
 & 375 & $7.8\pm0.64$ & \citet{1998ApJ...509..733B}\\
 & 400 & $10.9\pm1.15$ & \citet{2000ApJ...530..851C}\\
 & 666 & $56.1\pm12.28$ & \citet{2000ApJ...530..851C}\\
 & 666 & $100.3\pm12.40$ & \citet{2000ApJS..131..249S}\\
 & 857 & $91.5\pm22.78$ & \citet{2000ApJ...530..851C}\\
 & 857 & $45\pm3$ & \citet{1998ApJ...509..733B}\\
 & $3\rm{x}10^{3}$ & $112\pm20.15$ & \citet{1998ApJ...509..733B}\\
 & $5\rm{x}10^{3}$ & $32\pm6.12$ & \citet{1998ApJ...509..733B}\\
 & $1.2\rm{x}10^{4}$ & $5.75\pm1.2$ & \citet{1998ApJ...509..733B}\\
 & $2.5\rm{x}10^{4}$ & $0.69\pm0.15$ & \citet{1998ApJ...509..733B}\\
\hline
HH~7-11 & 5 & $7.3\rm{x}10^{-4\dag}$ & \citet{1999ApJS..125..427R}\\
 & 8 & $1.01\rm{x}10^{-3\dag}$ & \citet{1999ApJS..125..427R}\\
 & 14.62 & $3.43\rm{x}10^{-3}\pm0.20$ & this work\\
 & 15.37 & $3.10\rm{x}10^{-3}\pm0.18$ & this work\\
 & 16.12 & $3.91\rm{x}10^{-3}\pm0.22$ & this work\\
 & 16.87 & $3.43\rm{x}10^{-3}\pm0.19$ & this work\\
 & 17.62 & $3.83\rm{x}10^{-3}\pm0.24$ & this work\\
 & 43 & $1.08\rm{x}10^{-2}\pm0.06$ & \citet{2004ApJ...605L.137A}\\
 & 100 & $7.89\pm0.79$ & \citet{2009ApJ...691.1729C}\\
 & 273 & $4.46^{\dag}$ & \citet{2006ApJ...638..293E}\\
 & 353 & $10.3^{\dag}$ & \citet{2008ApJS..175..277D}\\
 & 353 & $12.7^{\dag}$ & \citet{hat2007a}\\
 & 353 & $14.9\pm1.2$ & \citet{2000ApJ...530..851C}\\
 & 400 & $21.5\pm2.8$ & \citet{2000ApJ...530..851C}\\
 & 666 & $119\pm24$ & \citet{2000ApJ...530..851C}\\
 & 666 & $159.9^{\dag}$ & \citet{hat2007a}\\
 & 857 & $203\pm61$ & \citet{2000ApJ...530..851C}\\
 & $3\rm{x}10^{3}$ & $381\pm23$ & IRAS\\
 & $5\rm{x}10^{3}$ & $204\pm20$ & IRAS\\
 & $1.2\rm{x}10^{4}$ & $46.5\pm2.8$ & IRAS\\
 & $2.5\rm{x}10^{4}$ & $13.6\pm3.7$ & IRAS\\
\hline
L1551~IRS~5 & 1.5 & $2.28\rm{x}10^{-3\dag}$ & \citet{1985ApJ...289L...5B}\\
 & 1.5 & $2.8\rm{x}10^{-3}\pm0.9$ & \citet{1985ApJ...290..587S}\\
 & 1.7 & $4\rm{x}10^{-3\dag}$ & \citet{1989ApJ...337..712R}\\
 & 5 & $4.2\rm{x}10^{-3\dag}$ & \citet{1987MNRAS.224..721D}\\
 & 5 & $3.5\rm{x}10^{-3}\pm0.5$ & \citet{1982ApJ...253..707C}\\
 & 5 & $3\rm{x}10^{-3\dag}$ & \citet{1984ApJ...282..699B}\\
 & 5 & $4.69\rm{x}10^{-3\dag}$ & \citet{1985ApJ...289L...5B}\\
 & 5 & $4.3\rm{x}10^{-3}\pm0.5$ & \citet{1985ApJ...290..587S}\\
 & 5 & $5\rm{x}10^{-3}\pm0.5$ & \citet{1987ApJ...320..364E}\\
 & 5 & $4.1\rm{x}10^{-3}\pm0.4$ & \citet{1990ApJ...355..635K}\\
 & 8 & $4.7\rm{x}10^{-3}\pm0.5$ & \citet{1990ApJ...355..635K}\\
 & 14.62 & $4.74\rm{x}10^{-3}\pm0.25$ & this work\\
\hline
\end{tabular}
\end{center}
\end{table*}

\begin{table*}
\begin{center}
\caption{Reference list for the spectral energy distribution data - continued.\label{tab:sedref}}
\begin{tabular}{llll}
\hline\hline
Source & $\nu$ (GHz) & $S_{\nu}$ (Jy) & Reference \\
\hline
L1551~IRS~5 cont. & 15 & $2.25\rm{x}10^{-3\dag}$ & \citet{1986ApJ...301L..25R}\\
 & 15 & $4.59\rm{x}10^{-3\dag}$ & \citet{1985ApJ...289L...5B}\\
 & 15 & $4.9\rm{x}10^{-3}\pm0.5$ & \citet{1990ApJ...355..635K}\\
 & 15 & $3.6\rm{x}10^{-3}\pm0.2$ & \citet{2003ApJ...583..330R}\\
 & 15 & $2.8\rm{x}10^{-3}\pm0.2$ & \citet{2003ApJ...583..330R}\\
 & 15 & $3.3\rm{x}10^{-3}\pm0.2$ & \citet{2003ApJ...583..330R}\\
 & 15 & $2.7\rm{x}10^{-3}\pm0.2$ & \citet{2003ApJ...583..330R}\\
 & 15.37 & $3.10\rm{x}10^{-3}\pm0.18$ & this work\\
 & 16.12 & $3.91\rm{x}10^{-3}\pm0.22$ & this work\\
 & 16.87 & $3.43\rm{x}10^{-3}\pm0.19$ & this work\\
 & 17.62 & $3.83\rm{x}10^{-3}\pm0.24$ & this work\\
 & 22.5 & $1.7\rm{x}10^{-2}\pm0.4$ & \citet{1985ApJ...288..595T}\\
 & 22.5 & $7.3\rm{x}10^{-3}\pm1.5$ & \citet{1990ApJ...355..635K}\\
 & 23.7 & $7\rm{x}10^{-3}\pm1.4$ & \citet{1993ApJ...414..333G}\\
 & 88 & $0.09\pm0.03$ & \citet{1990ApJ...355..635K}\\
 & 90 & $0.12^{\dag}$ & \citet{1994AA...281..161A}\\
 & 98 & $0.13\pm0.004$ & \citet{1991AJ....102.2054O}\\
 & 110 & $0.13\pm0.03$ & \citet{1990ApJ...355..635K}\\
 & 112 & $0.15^{\dag}$ & \citet{1989ApJ...345..257W}\\
 & 113 & $0.17^{\dag}$ & \citet{1986BAAS...18..973K}\\
 & 230 & $1.5^{\dag}$ & \citet{1991ApJ...379L..25C}\\
 & 230 & $1.57\pm0.02$ & \citet{1993AA...273..221R}\\
 & 231 & $3.4^{\dag}$ & \citet{2001AA...365..440M}\\
 & 240 & $2.37\pm0.48$ & \citet{1990ApJ...355..635K}\\
 & 250 & $1.72^{\dag}$ & \citet{1994AA...281..161A}\\
 & 300 & $5.7\pm1.3$ & \citet{1990ApJ...355..635K}\\
 & 345 & $6.36\pm0.06$ & \citet{1993AA...273..221R}\\
 & 353 & $19.01^{\dag}$ & \citet{2008ApJS..175..277D}\\
 & 353 & $12.1\pm1$ & \citet{2000ApJ...530..851C}\\
 & 400 & $18.2\pm2.4$ & \citet{2000ApJ...530..851C}\\
 & 666 & $94\pm19$ & \citet{2000ApJ...530..851C}\\
 & 857 & $164\pm49$ & \citet{2000ApJ...530..851C}\\
 & $3\rm{x}10^{3}$ & $458^{\dag}$ & \citet{1993AA...273..221R}\\
 & $5\rm{x}10^{3}$ & $373^{\dag}$ & \citet{1993AA...273..221R}\\
 & $1.2\rm{x}10^{4}$ & $106^{\dag}$ & \citet{1993AA...273..221R}\\
 & $2.5\rm{x}10^{4}$ & $10^{\dag}$ & \citet{1993AA...273..221R}\\
\hline
L1527 & 5 & $6.8\rm{x}10^{-4}\pm0.4$ & \citet{2011ApJ...739L...7M}\\
 & 8.5 & $8.10\rm{x}10^{-4}\pm0.3$ & \citet{2011ApJ...739L...7M}\\
 & 14.62 & $1.04\rm{x}10^{-3}\pm0.11$ & this work\\
 & 15.37 & $1.04\rm{x}10^{-3}\pm0.14$ & this work\\
 & 16 & $0.9\rm{x}10^{-3}\pm0.03$ & \citet{scaife2011d}\\
 & 16.12 & $1.12\rm{x}10^{-3}\pm0.13$ & this work\\
 & 16.87 & $1.20\rm{x}10^{-3}\pm0.16$ & this work\\
 & 17.62 & $1.38\rm{x}10^{-3}\pm0.15$ & this work\\
 & 22.5 & $1.4\rm{x}10^{-3}\pm0.1$ & \citet{2011ApJ...739L...7M}\\
 & 43.3 & $4.4\rm{x}10^{-3}\pm0.6$ & \citet{2011ApJ...739L...7M}\\
 & 111 & $4.7\rm{x}10^{-2}\pm0.56$ & \citet{1997ApJ...475..211O}\\
 & 230 & $1.5^{\dag}$ & \citet{2001AA...365..440M}\\
 & 230 & $0.35^{\dag}$ & \citet{2010AJ....139.2504G}\\
 & 353 & $11.78^{\dag}$ & \citet{2008ApJS..175..277D}\\
 & 353 & $0.90\pm0.01$ & \citet{2010AJ....139.2504G}\\
 & 375 & $0.50^{\dag}$ & \citet{2010AJ....139.2504G}\\
 & 666 & $2.85\pm0.22$ & \citet{2010AJ....139.2504G}\\
 & 666 & $3.21^{\dag}$ & \citet{2010AJ....139.2504G}\\
 & 857 & $12^{\dag}$ & \citet{2010AJ....139.2504G}\\
 & $1.87\rm{x}10^{3}$ & $68.8^{\dag}$ & \citet{2010AJ....139.2504G}\\
 & $3\rm{x}10^{3}$ & $71.3^{\dag}$ & \citet{2010AJ....139.2504G}\\
 & $3\rm{x}10^{3}$ & $72^{\dag}$ & \citet{2010AJ....139.2504G}\\
 & $3\rm{x}10^{3}$ & $73.3\pm11.7$ & IRAS\\
 & $5\rm{x}10^{3}$ & $17.4^{\dag}$ & \citet{2010AJ....139.2504G}\\
 & $5\rm{x}10^{3}$ & $18^{\dag}$ & \citet{2010AJ....139.2504G}\\
 & $5\rm{x}10^{3}$ & $17.8\pm1.6$ & IRAS\\
 & $1.2\rm{x}10^{3}$ & $0.74\pm0.07$ & IRAS\\
\hline
\end{tabular}
\end{center}
\end{table*}

\begin{table*}
\begin{center}
\caption{Reference list for the spectral energy distribution data - continued.\label{tab:sedref}}
\begin{tabular}{llll}
\hline\hline
Source & $\nu$ (GHz) & $S_{\nu}$ (Jy) & Reference \\
\hline
HH~1-2~MMS~1 & 1.5 & $6\rm{x}10^{-4}\pm2$ & \citet{1985ApJ...293L..35P}\\
 & 1.5 & $8.6\rm{x}10^{-4\dag}$ & \citet{1990ApJ...352..645R}\\
 & 5 & $1.19\rm{x}10^{-3}\pm1.6$ & \citet{1990ApJ...362..274M}\\
 & 5 & $1.2\rm{x}10^{-3}\pm0.04$ & \citet{1985ApJ...293L..35P}\\
 & 5 & $1.17\rm{x}10^{-3\dag}$ & \citet{1990ApJ...352..645R}\\
 & 14.62 & $1.30\rm{x}10^{-3}\pm0.28$ & this work\\
 & 15 & $1.54\rm{x}10^{-3}\pm0.18$ & \citet{1985ApJ...293L..35P}\\
 & 15 & $1.75\rm{x}10^{-3\dag}$ & \citet{1990ApJ...352..645R}\\
 & 15.37 & $\rm{x}10^{-3}\pm0.18$ & this work\\
 & 16.12 & $\rm{x}10^{-3}\pm0.22$ & this work\\
 & 16.87 & $\rm{x}10^{-3}\pm0.19$ & this work\\
 & 230 & $0.65\pm0.02$ & \citet{1993AA...273..221R}\\
 & 273 & $0.92\pm0.03$ & \citet{1998MNRAS.301.1049D}\\
 & 345 & $1.67\pm0.02$ & \citet{1993AA...273..221R}\\
 & 353 & $9.44^{\dag}$ & \citet{2008ApJS..175..277D}\\
 & 354 & $1.2\pm0.24$ & \citet{2010AA...518L.122F}\\
 & 375 & $2.65\pm0.03$ & \citet{1998MNRAS.301.1049D}\\
 & 666 & $18.9\pm0.35$ & \citet{1998MNRAS.301.1049D}\\
 & 857 & $33.95\pm0.29$ & \citet{1998MNRAS.301.1049D}\\
 & 857 & $\pm$ & \citet{2010AA...518L.122F}\\
 & $1.88\rm{x}10^{3}$ & $75.7\pm15.14$ & \citet{2010AA...518L.122F}\\
 & $4.29\rm{x}10^{3}$ & $26.6\pm2.66$ & \citet{2010AA...518L.122F}\\
\hline
HH~26~IR & 8 & $3.8\rm{x}10^{-4}\pm0.6$ & \citet{1998AJ....116.2953A}\\
 & 8 & $1.4\rm{x}10^{-4}\pm0.3$ & \citet{1999MNRAS.304....1G}\\
 & 16.12 & $3.92\rm{x}10^{-4}\pm0.76$ & this work\\
 & 231 & $0.32^{\dag}$ & \citet{1999ApJ...527..856L}\\
 & 353 & $0.97^{\dag}$ & \citet{2008ApJS..175..277D}\\
 & 353 & $0.9\pm0.09$ & \citet{2007MNRAS.374.1413N}\\
 & 666 & $1.4\pm0.28$ & \citet{2007MNRAS.374.1413N}\\
 & 857 & $6.3^{\dag}$ & \citet{1999ApJ...527..856L}\\
 & $3\rm{x}10^{3}$ & $67.93^{\dag}$ & \citet{2008AA...479..503A}\\
 & $4.29\rm{x}10^{3}$ & $7.9^{\dag}$ & \citet{2008AA...479..503A}\\
 & $5\rm{x}10^{3}$ & $20.87^{\dag}$ & \citet{2008AA...479..503A}\\
 & $1.2\rm{x}10^{4}$ & $5.13^{\dag}$ & \citet{2008AA...479..503A}\\
 & $1.25\rm{x}10^{3}$ & $1.8^{\dag}$ & \citet{2008AA...479..503A}\\
 & $2.5\rm{x}10^{4}$ & $1.93^{\dag}$ & \citet{2008AA...479..503A}\\
\hline
HH~111 & 5 & $8.3\rm{x}10^{-4}\pm0.9$ & \citet{2008AJ....136.1852R}\\
 & 8 & $9.5\rm{x}10^{-4}\pm0.4$ & \citet{1994AA...281..882R}\\
 & 14.62 & $3.11\rm{x}10^{-3}\pm0.53$ & this work\\
 & 15 & $2.3\rm{x}10^{-3}\pm0.32$ & \citet{1994AA...281..882R}\\
 & 15.37 & $2.68\rm{x}10^{-3}\pm0.28$ & this work\\
 & 16.12 & $2.39\rm{x}10^{-3}\pm0.26$ & this work\\
 & 16.87 & $2.86\rm{x}10^{-3}\pm0.30$ & this work\\
 & 17.62 & $2.51\rm{x}10^{-3}\pm0.34$ & this work\\
 & 43 & $5.15\rm{x}10^{-3}\pm0.52$ & \citet{2008AJ....136.1852R}\\
 & 110 & $4.6\rm{x}10^{-2}\pm0.46$ & \citet{1993ApJ...408..239S}\\
 & 230 & $0.49\pm0.02$ & \citet{1993AA...273..221R}\\
 & 230 & $0.47\pm0.05$ & \citet{1989ApJ...345..257W}\\
 & 273 & $0.75\pm0.01$ & \citet{1998MNRAS.301.1049D}\\
 & 285 & $0.69^{\dag}$ & \citet{1993ApJ...408..239S}\\
 & 345 & $1.39\pm0.03$ & \citet{1993AA...273..221R}\\
 & 353 & $3.09^{\dag}$ & \citet{2008ApJS..175..277D}\\
 & 375 & $1.8\pm0.02$ & \citet{1998MNRAS.301.1049D}\\
 & 666 & $8.22\pm0.26$ & \citet{1998MNRAS.301.1049D}\\
 & 857 & $14.2\pm0.9$ & \citet{1998MNRAS.301.1049D}\\
 & $3\rm{x}10^{3}$ & $71.1^{\dag}$ & \citet{1993AA...273..221R}\\
 & $5\rm{x}10^{3}$ & $43.5^{\dag}$ & \citet{1993AA...273..221R}\\
 & $1.2\rm{x}10^{4}$ & $6.8^{\dag}$ & \citet{1993AA...273..221R}\\
 & $2.5\rm{x}10^{4}$ & $0.3^{\dag}$ & \citet{1993AA...273..221R}\\
\hline
NGC~2264~G & 1.5 & $8\rm{x}10^{-4}\pm2$ & \citet{1989RMxAA..17..115R}\\
 & 5 & $1.9\rm{x}10^{-3}\pm0.2$ & \citet{1989RMxAA..17..115R}\\
 & 5 & $6.7\rm{x}10^{-4\dag}$ & \citet{1994ApJ...436..749G}\\
\hline
\end{tabular}
\end{center}
\end{table*}

\begin{table*}
\begin{center}
\caption{Reference list for the spectral energy distribution data - continued.\label{tab:sedref}}
\begin{tabular}{llll}
\hline\hline
Source & $\nu$ (GHz) & $S_{\nu}$ (Jy) & Reference \\
\hline
NGC~2264~G cont. & 8 & $5.5\rm{x}10^{-4\dag}$ & \citet{1994ApJ...436..749G}\\
 & 16.12 & $2.96\rm{x}10^{-4}\pm2.16$ & this work\\
 & 273 & $0.25\pm0.04$ & \citet{1995MNRAS.273L..25W}\\
 & 353 & $0.71^{\dag}$ & \citet{2008ApJS..175..277D}\\
 & 375 & $0.7\pm0.04$ & \citet{1995MNRAS.273L..25W}\\
 & 666 & $3.9\pm0.1$ & \citet{1995MNRAS.273L..25W}\\
 & 857 & $7.1\pm0.28$ & \citet{1995MNRAS.273L..25W}\\
 & $3\rm{x}10^{3}$ & $17\pm3$ & \citet{1995MNRAS.273L..25W}\\
 & $5\rm{x}10^{3}$ & $6\pm1$ & \citet{1995MNRAS.273L..25W}\\
\hline
Serpens~SMM~1 & 1.5 & $4.5\rm{x}10^{-3}\pm0.5$ & \citet{1998AJ....115.1693C}\\
 & 1.5 & $9.5\rm{x}10^{-3}\pm0.49$ & \citet{1993ApJ...415..191C}\\
 & 5 & $9.5\rm{x}10^{-3}\pm0.8$ & \citet{1986ApJ...303..683S}\\
 & 5 & $7.9\rm{x}10^{-3}\pm0.48$ & \citet{1989ApJ...346L..85R}\\
 & 5 & $7.6\rm{x}10^{-3}\pm0.3$ & \citet{1993ApJ...415..191C}\\
 & 5 & $2.2\rm{x}10^{-3\dag}$ & \citet{1994ApJ...424..222M}\\
 & 8 & $7.5\rm{x}10^{-3}\pm0.17$ & \citet{1993ApJ...415..191C}\\
 & 8 & $7.54\rm{x}10^{-3\dag}$ & \citet{2005AJ....130..643E}\\
 & 14.62 & $8.21\rm{x}10^{-3}\pm0.77$ & this work\\
 & 15 & $1\rm{x}10^{-2}\pm0.3$ & \citet{1986ApJ...303..683S}\\
 & 15 & $8.3\rm{x}10^{-3}\pm0.3$ & \citet{1993ApJ...415..191C}\\
 & 15 & $6.2\rm{x}10^{-3}\pm0.0.44$ & \citet{1989ApJ...346L..85R}\\
 & 15 & $4.8\rm{x}10^{-3\dag}$ & \citet{1994ApJ...424..222M}\\
 & 15.37 & $7.66\rm{x}10^{-3}\pm0.0.55$ & this work\\
 & 16 & $4.74\pm0.24$ & \citet{scaife2011c}\\
 & 16.12 & $7.04\rm{x}10^{-3}\pm0.43$ & this work\\
 & 16.87 & $6.77\rm{x}10^{-3}\pm0.44$ & this work\\
 & 17.62 & $7.30\rm{x}10^{-3}\pm0.54$ & this work\\
 & 23 & $5.3\rm{x}10^{-3\dag}$ & \citet{1994ApJ...424..222M}\\
 & 43 & $1.42\rm{x}10^{-2}\pm0.04$ & \citet{2009ApJ...705.1730C}\\
 & 88 & $0.2^{\dag}$ & \citet{1999ApJ...513..350H}\\
 & 94 & $0.2^{\dag}$ & \citet{1999ApJ...513..350H}\\
 & 97 & $0.14^{\dag}$ & \citet{1994ApJ...424..222M}\\
 & 111 & $0.41^{\dag}$ & \citet{1999ApJ...513..350H}\\
 & 214 & $2.65^{\dag}$ & \citet{1999ApJ...513..350H}\\
 & 240 & $2.3^{\dag}$ & \citet{1994ApJ...424..222M}\\
 & 273 & $3.47\pm0.1$ & \citet{1993AA...275..195C}\\
 & 353 & $15.23{\dag}$ & \citet{2008ApJS..175..277D}\\
 & 353 & $6.1^{\dag}$ & \citet{1999MNRAS.309..141D}\\
 & 666 & $35.7^{\dag}$ & \citet{1999MNRAS.309..141D}\\
 & $1.88\rm{x}10^{3}$ & $430^{\dag}$ & \citet{1994ApJ...424..222M}\\
 & $3\rm{x}10^{3}$ & $380^{\dag}$ & \citet{1994ApJ...424..222M}\\
 & $5\rm{x}10^{3}$ & $152.92^{\dag}$ & \citet{1994ApJ...424..222M}\\
 & $6\rm{x}10^{3}$ & $88.6^{\dag}$ & \citet{1994ApJ...424..222M}\\
 & $1.2\rm{x}10^{4}$ & $3.17^{\dag}$ & \citet{1994ApJ...424..222M}\\
 & $1.5\rm{x}10^{4}$ & $2.6^{\dag}$ & \citet{1994ApJ...424..222M}\\
 & $2.5\rm{x}10^{4}$ & $0.25^{\dag}$ & \citet{1994ApJ...424..222M}\\
\hline
L723 & 5 & $7.5\rm{x}10^{-4\dag}$ & \citet{1996ApJ...473L.123A}\\
 & 8 & $7.6\rm{x}10^{-4\dag}$ & \citet{1996ApJ...473L.123A}\\
 & 14.62 & $5.82\rm{x}10^{-4}\pm0.86$ & this work\\
 & 15.37 & $5.16\rm{x}10^{-4}\pm0.64$ & this work\\
 & 16.12 & $6.53\rm{x}10^{-4}\pm0.78$ & this work\\
 & 16.87 & $6.55\rm{x}10^{-4}\pm0.69$ & this work\\
 & 17.62 & $5.22\rm{x}10^{-4}\pm1.00$ & this work\\
 & 230 & $0.37^{\dag}$ & \citet{2001AA...365..440M}\\
 & 230 & $0.36\pm0.02$ & \citet{1993AA...273..221R}\\
 & 300 & $1\pm0.5$ & \citet{1993AA...273..221R}\\
 & 353 & $1.79\pm0.11$ & \citet{2000ApJS..131..249S}\\
 & 353 & $2.24^{\dag}$ & \citet{2008ApJS..175..277D}\\
 & 666 & $8.5\pm2.1$ & \citet{2000ApJS..131..249S}\\
 & 750 & $13\pm3$ & \citet{1993AA...273..221R}\\
 & 857 & $11.3\pm1.8$ & \citet{2007AJ....133.1560W}\\
 & $1.54\rm{x}10^{3}$ & $35\pm7$ & \citet{1993AA...273..221R}\\
 & $1.81\rm{x}10^{3}$ & $40\pm12$ & \citet{1993AA...273..221R}\\
\hline
\end{tabular}
\end{center}
\end{table*}

\begin{table*}
\begin{center}
\caption{Reference list for the spectral energy distribution data - continued.\label{tab:sedref}}
\begin{tabular}{llll}
\hline\hline
Source & $\nu$ (GHz) & $S_{\nu}$ (Jy) & Reference \\
\hline
L723 cont. & $2.08\rm{x}10^{3}$ & $33\pm10$ & \citet{1993AA...273..221R}\\
 & $2.14\rm{x}10^{3}$ & $23\pm8$ & \citet{1993AA...273..221R}\\
 & $2.31\rm{x}10^{3}$ & $32\pm11$ & \citet{1993AA...273..221R}\\
 & $3\rm{x}10^{3}$ & $20.7\pm1.7$ & \citet{2000ApJS..131..249S}\\
 & $3.16\rm{x}10^{3}$ & $27\pm6$ & \citet{2000ApJS..131..249S}\\
 & $5\rm{x}10^{3}$ & $6.93\pm0.62$ & \citet{2000ApJS..131..249S}\\
 & $1.2\rm{x}10^{4}$ & $0.38\pm0.05$ & \citet{2000ApJS..131..249S}\\
 & $2.5\rm{x}10^{4}$ & $0.28\pm0.06$ & \citet{2000ApJS..131..249S}\\
\hline
L1251~A & 5 & $1.7\rm{x}10^{-4}\pm0.03$ & \citet{2001AJ....121.1556B}\\
 & 8 & $2.9\rm{x}10^{-4\dag}$ & \citet{2004AJ...127.1736R}\\
 & 8 & $4.7\rm{x}10^{-4}\pm0.03$ & \citet{2001AJ....121.1556B}\\
 & 14.62 & $8.57\rm{x}10^{-4}\pm1.77$ & this work\\
 & 15.37 & $9.85\rm{x}10^{-4}\pm1.36$ & this work\\
 & 16.12 & $1.00\rm{x}10^{-3}\pm0.20$ & this work\\
 & 16.87 & $7.61\rm{x}10^{-4}\pm1.58$ & this work\\
 & 17.62 & $1.10\rm{x}10^{-3}\pm0.23$ & this work\\
 & 230 & $0.23\pm0.02$ & \citet{1995PASP..107...49R}\\
 & 273 & $0.38\pm0.03$ & \citet{1995PASP..107...49R}\\
 & 353 & $5.68^{\dag}$ & \citet{2008ApJS..175..277D}\\
 & 375 & $0.71\pm0.11$ & \citet{1995PASP..107...49R}\\
 & 857 & $20.8\pm3.2$ & \citet{2007AJ....133.1560W}\\
 & $3\rm{x}10^{3}$ & $78.5\pm12.6$ & \citet{1995PASP..107...49R}\\
 & $3\rm{x}10^{3}$ & $79^{\dag}$ & \citet{1989ApJ...343..773S}\\
 & $5\rm{x}10^{3}$ & $67.7\pm6.1$ & \citet{1995PASP..107...49R}\\
 & $5\rm{x}10^{3}$ & $66^{\dag}$ & \citet{1989ApJ...343..773S}\\
 & $1.2\rm{x}10^{4}$ & $28.3\pm1.4$ & \citet{1995PASP..107...49R}\\
 & $1.2\rm{x}10^{4}$ & $26^{\dag}$ & \citet{1989ApJ...343..773S}\\
 & $2.5\rm{x}10^{4}$ & $6.2\pm0.3$ & \citet{1995PASP..107...49R}\\
 & $2.5\rm{x}10^{4}$ & $5^{\dag}$ & \citet{1989ApJ...343..773S}\\
\hline
L1448~C & 8 & $2.3\rm{x}10^{-4\dag}$ & \citet{2002AJ....124.1045R}\\
 & 15 & $5.6\rm{x}10^{-4}\pm0.5$ & \citet{1990ApJ...365L..85C}\\
 & 16 & $5.37\rm{x}10^{-4}\pm0.33$ & \citet{scaife2011b}\\
 & 16.12 & $7.58\rm{x}10^{-4}\pm0.94$ & this work\\
 & 86 & $2.6\rm{x}10^{-2}\pm0.2$ & \citet{2000ApJS..131..249S}\\
 & 87 & $1.2\rm{x}10^{-2\dag}$ & \citet{1991AA...241L..43B}\\
 & 87 & $1.6\rm{x}10^{-2\dag}$ & \citet{1992AA...265L..49G}\\
 & 91 & $3.92\rm{x}10^{-2}\pm0.39$ & \citet{2011AJ....141...39S}\\
 & 115 & $9.1\rm{x}10^{-2}\pm0.2$ & \citet{2000ApJS..131..249S}\\
 & 230 & $1\pm0.1$ & \citet{1998ApJ...509..733B}\\
 & 230 & $0.74\pm0.11\pm$ & \citet{2000ApJS..131..249S}\\
 & 230 & $0.91^{\dag}\pm$ & \citet{2001AA...365..440M}\\
 & 230 & $0.58^{\dag}$ & \citet{1991AA...241L..43B}\\
 & 273 & $2.04^{\dag}$ & \citet{2006ApJ...638..293E}\\
 & 273 & $1\pm0.1$ & \citet{1998ApJ...509..733B}\\
 & 349 & $1.97\pm0.13$ & \citet{2011AJ....141...39S}\\
 & 353 & $2.98^{\dag}m$ & \citet{2008ApJS..175..277D}\\
 & 353 & $6.51^{\dag}$ & \citet{hat2007a}\\
 & 353 & $3.95\pm0.24$ & \citet{2000ApJS..131..249S}\\
 & 353 & $5.34\pm0.43$ & \citet{2000ApJ...530..851C}\\
 & 375 & $3\pm0.3$ & \citet{1998ApJ...509..733B}\\
 & 400 & $6.91\pm0.91$ & \citet{2000ApJ...530..851C}\\
 & 666 & $68.4^{\dag}$ & \citet{hat2007a}\\
 & 666 & $21\pm2$ & \citet{1998ApJ...509..733B}\\
 & 666 & $31.8\pm5.5$ & \citet{2000ApJS..131..249S}\\
 & 666 & $34.2\pm6.9$ & \citet{2000ApJ...530..851C}\\
 & 857 & $58\pm18$ & \citet{2000ApJ...530..851C}\\
 & 857 & $30\pm3$ & \citet{1998ApJ...509..733B}\\
 & $3\rm{x}10^{3}$ & $70.3\pm14.8$ & \citet{1998ApJ...509..733B}\\
 & $5\rm{x}10^{3}$ & $31.2\pm6.5$ & \citet{1998ApJ...509..733B}\\
 & $1.2\rm{x}10^{4}$ & $2.9\pm0.6$ & \citet{1998ApJ...509..733B}\\
 & $2.5\rm{x}10^{4}$ & $0.33\pm0.07$ & \citet{1998ApJ...509..733B}\\
\hline
NGC~1333~IRAS~2A & 5 & $6\rm{x}10^{-5}\pm1$ & \citet{1999ApJS..125..427R}\\
\hline
\end{tabular}
\end{center}
\end{table*}

\begin{table*}
\begin{center}
\caption{Reference list for the spectral energy distribution data - continued.\label{tab:sedref}}
\begin{tabular}{llll}
\hline\hline
Source & $\nu$ (GHz) & $S_{\nu}$ (Jy) & Reference \\
\hline
NGC~1333~IRAS~2A cont. & 8 & $2.2\rm{x}10^{-4\dag}$ & \citet{2002AJ....124.1045R}\\
 & 8 & $2.5\rm{x}10^{-4}\pm0.4$ & \citet{1999ApJS..125..427R}\\
 & 16 & $3.2\rm{x}10^{-4}\pm0.25$ & \citet{scaife2011b}\\
 & 16.12 & $3.92\rm{x}10^{-4}\pm0.96$ & this work\\
 & 43 & $1\rm{x}10^{-2}\pm0.5$ & \citet{2004ApJ...605L.137A}\\
 & 86 & $3.5\rm{x}10^{-2\dag}$ & \citet{2004AA...413..993J}\\
 & 89 & $4\rm{x}10^{-2\dag}$ & \citet{2004AA...413..993J}\\
 & 111 & $8.28\rm{x}10^{-2}\pm0.4$ & \citet{2000ApJ...529..477L}\\
 & 150 & $0.32\pm0.07$ & \citet{1994AA...285L...1S}\\
 & 230 & $0.88^{\dag}$ & \citet{2001AA...365..440M}\\
 & 230 & $0.88\pm0.12$ & \citet{1994AA...285L...1S}\\
 & 273 & $1.46\pm0.12$ & \citet{1994AA...285L...1S}\\
 & 353 & $7.78^{\dag}$ & \citet{2007ApJ...668.1042K}\\
 & 353 & $4.79\pm0.39$ & \citet{2000ApJ...530..851C}\\
 & 375 & $4.08\pm0.05$ & \citet{1994AA...285L...1S}\\
 & 375 & $3.75\pm0.15$ & \citet{1994AA...285L...1S}\\
 & 400 & $5.54\pm0.21$ & \citet{1994AA...285L...1S}\\
 & 400 & $6.61\pm0.86$ & \citet{2000ApJ...530..851C}\\
 & 666 & $43\pm11$ & \citet{2000ApJ...530..851C}\\
 & 666 & $23.7\pm0.9$ & \citet{1994AA...285L...1S}\\
 & 857 & $37.2\pm1.2$ & \citet{1994AA...285L...1S}\\
 & 857 & $74\pm22$ & \citet{2000ApJ...530..851C}\\
 & $3\rm{x}10^{3}$ & $300^{\dag}$ & \citet{1987MNRAS.226..461J}\\
 & $6\rm{x}10^{3}$ & $104^{\dag}$ & \citet{1987MNRAS.226..461J}\\
\hline
NGC~1333~IRAS~2B & 5 & $2.3\rm{x}10^{-4}\pm0.1$ & \citet{1999ApJS..125..427R}\\
 & 8 & $4\rm{x}10^{-4}\pm0.2$ & \citet{1999ApJS..125..427R}\\
 & 8 & $3.7\rm{x}10^{-4}\pm0.37$ & \citet{2002AJ....124.1045R}\\
 & 16 & $1.21\rm{x}10^{-3}\pm0.05$ & \citet{scaife2011b}\\
 & 16.12 & $5.05\rm{x}10^{-4}\pm1.01$ & this work\\
 & 43 & $5.2\rm{x}10^{-3}\pm0.3$ & \citet{2004ApJ...605L.137A}\\
 & 86 & $1.2\rm{x}10^{-2\dag}$ & \citet{2004AA...413..993J}\\
 & 89 & $1.4\rm{x}10^{-2\dag}$ & \citet{2004AA...413..993J}\\
 & 111 & $2.77\rm{x}10^{-2}\pm0.32$ & \citet{2000ApJ...529..477L}\\
 & 353 & $1.19\pm0.1$ & \citet{2000ApJ...530..851C}\\
 & 400 & $1.53\pm0.21$ & \citet{2000ApJ...530..851C}\\
 & 666 & $7.7\pm1.6$ & \citet{2000ApJ...530..851C}\\
 & 857 & $13.4\pm1.3$ & \citet{2000ApJ...530..851C}\\
\hline
L1551~NE & 8 & $6.6\rm{x}10^{-4\dag}$ & \citet{2002AJ....124.1045R}\\
 & 16.12 & $7.43\rm{x}10^{-4}\pm0.92$ & this work\\
 & 230 & $1.5^{\dag}$ & \citet{2001AA...365..440M}\\
 & 230 & $0.85\pm0.01$ & \citet{2005ApJ...631.1134A}\\
 & 230 & $0.85\pm0.08$ & \citet{2000ApJ...533L.143M}\\
 & 238 & $0.8\pm0.12$ & \citet{1993ApJ...406L..71B}\\
 & 272 & $1.07\pm0.11$ & \citet{1993ApJ...406L..71B}\\
 & 353 & $2.78^{\dag}$ & \citet{2006ApJ...645..357M}\\
 & 353 & $10.5^{\dag}$ & \citet{2008ApJS..175..277D}\\
 & 379 & $2.22\pm0.37$ & \citet{1993ApJ...406L..71B}\\
 & 666 & $8.33^{\dag}$ & \citet{2006ApJ...645..357M}\\
 & 677 & $16.2\pm4.6$ & \citet{1993ApJ...406L..71B}\\
 & 857 & $22.83\pm0.72$ & \citet{2005ApJ...631.1134A}\\
 & $3\rm{x}10^{3}$ & $130^{\dag}$ & \citet{1984ApJ...278L..49E}\\
 & $5\rm{x}10^{3}$ & $80^{\dag}$ & \citet{1984ApJ...278L..49E}\\
 & $1.2\rm{x}10^{4}$ & $16^{\dag}$ & \citet{1984ApJ...278L..49E}\\
 & $2.5\rm{x}10^{4}$ & $1.2^{\dag}$ & \citet{1984ApJ...278L..49E}\\
\hline
HH~25~MMS & 5 & $1.5\rm{x}10^{-4}\pm0.4$ & \citet{1995AA...297...98B}\\
 & 8 & $2.5\rm{x}10^{-4}\pm0.6$ & \citet{1995AA...297...98B}\\
 & 8 & $1.6\rm{x}10^{-4}\pm0.3$ & \citet{1999MNRAS.304....1G}\\
 & 16.12 & $1.23\rm{x}10^{-3}\pm0.10$ & this work\\
 & 231 & $1^{\dag}$ & \citet{1999ApJ...527..856L}\\
 & 353 & $9.11^{\dag}$ & \citet{2008ApJS..175..277D}\\
 & 353 & $4^{\dag}$ & \citet{2007MNRAS.374.1413N}\\
 & 666 & $10.8^{\dag}$ & \citet{2007MNRAS.374.1413N}\\
\hline
\end{tabular}
\end{center}
\end{table*}

\begin{table*}
\begin{center}
\caption{Reference list for the spectral energy distribution data - continued.\label{tab:sedref}}
\begin{tabular}{llll}
\hline\hline
Source & $\nu$ (GHz) & $S_{\nu}$ (Jy) & Reference \\
\hline
HH~25~MMS cont. & 666 & $8.7^{\dag}$ & \citet{2001MNRAS.326..927P}\\
 & 353 & $1.4^{\dag}$ & \citet{2001MNRAS.326..927P}\\
 & 857 & $24^{\dag}$ & \citet{1999ApJ...527..856L}\\
 & $3\rm{x}10^{3}$ & $70.66^{\dag}$ & IRAS\\
 & $5\rm{x}10^{3}$ & $28.04^{\dag}$ & IRAS\\
 & $5\rm{x}10^{3}$ & $32^{\dag}$ & IRAS\\
 & $1.2\rm{x}10^{4}$ & $1.39^{\dag}$ & \citet{1999yCat.2225....0G}\\
 & $2.5\rm{x}10^{4}$ & $0.51^{\dag}$ & \citet{1999yCat.2225....0G}\\
\hline
\end{tabular}
\end{center}
\end{table*}

\end{document}